\documentclass[11pt,twocolumn]{article}
\usepackage[authoryear,sectionbib]{natbib}
\usepackage[english]{babel}
\usepackage{latexsym}
\usepackage{exscale}
\usepackage{psfig}

\newcommand{\mbb}[1]{\mbox{\boldmath $#1$}}
\newcommand{\gsim}{\raisebox{-.5ex}{$\sim$}\raisebox{.5ex}%
{\hspace{-1.75ex}$>$}}

\begin{document}
\setcounter{tocdepth}{2}

\renewcommand{\thefootnote}{\fnsymbol{footnote}}
\setcounter{footnote}{1}
%\begin{titlepage}
\hspace*{-2ex}\parbox{\textwidth}{ 
\begin{center}
{\Large \bf
       Spontaneous Decay in the Presence of Absorbing Media
}
\\[4ex]
%\large
       Ho Trung Dung\footnotemark, 
       Ludwig Kn\"{o}ll, and Dirk-Gunnar Welsch\\[2ex]
{\it 
       Theoretisch-Physikalisches Institut,
       Friedrich-Schiller-Universit\"{a}t Jena,\\
       Max-Wien-Platz 1, D-07743 Jena, Germany}
\\[1ex] 
(April 11, 2001)
\end{center}
}

\footnotetext{On leave from the Institute of
Physics, National Center for Sciences and Technology,
1 Mac Dinh Chi Street, District 1, Ho Chi Minh City, Vietnam.}
\setcounter{footnote}{0}
%\end{titlepage}

\vspace*{-1cm}    

\section*{ABSTRACT}
After giving a summary of the basic-theoretical concept
of quantization of the electromagnetic field in the presence
of dispersing and absorbing (macroscopic) bodies,
their effect on spontaneous decay of an excited atom is studied.
Various configurations such as bulk material,  planar half space 
media, spherical cavities, and microspheres are considered.
In particular, the influence of  material absorption
on the local-field correction, the decay rate, the
line shift, and the emission pattern are examined.
Further, the interplay between
radiative losses and losses due to material absorption is analyzed.
Finally, the possibility of generating entangled states of two atoms 
coupled by a microsphere-assisted field is discussed.

\renewcommand{\thefootnote}{\arabic{footnote}}
\renewcommand{\theequation}{\arabic{section}.\arabic{equation}}
\setcounter{equation}{0}
\section{INTRODUCTION}
\label{sec:Intro}

Spontaneous emission of an excited atom is not an immutable
property of the atom, but it sensitively depends on the photonic
spectral density of states that are involved in the atomic transition
at the chosen location of the atom. Already  
\citet{Purcell46} pointed out that spontaneous emission 
can be enhanced when the atom is inside a cavity and
its transition is in resonance with a cavity mode.
The opposite case of inhibition of spontaneous emission
is also possible [\citet{Kleppner81}]. It is further well 
known that the decay process can even be reversed 
by strongly coupling the atom to a sufficiently sharp 
cavity-field mode, so that the emitted photon can be
\pagebreak
\vspace*{4.5cm}
\\
reabsorbed and reemitted.
Obviously, the photonic density of states can be modified
by the presence of macroscopic bodies, which, in the
simplest case, change the boundary conditions for the
electromagnetic field. For the last years, engineering periodic
dielectric structures (photonic crystals)
has been of increasing interest 
[\citet{John87,Yablonovich87,John94,Kofman94,Joannopoulous95,
Soukoulis96,
Woldeyohannes99,Nikolopoulos00,Zhu00,Schriemer01}]. 

It is worth noting that spontaneous emission may be regarded
as being a basic process in the rapidly  growing field of 
cavity quantum electrodynamics (QED), where strong (resonant)
interactions of a single or a few atoms with a single or a few
radiation-field modes formed by material bodies are studied.
Cavity QED itself has offered novel possibilities of testing
fundamental aspects of quantum physics, such as quantum
nondemolition measurement, complementarity, and
entanglement [for reviews, see, e.g., \citet{Hinds91,Haroche92,Meschede92,
Meystre92,
Berman94,Haroche98,Kimble98,Walther98}].

Spontaneous emission in the presence of material bodies
is not only interesting from the point of view
of fundamental research, but it has also offered a number
of interesting applications. It can  provide a reliable
and efficient single-photon source to be used in quantum
information processing [\citet{DeMartini96}, \citet{Kitson98}].
The sensitivity to the ambient medium of resonance 
fluorescence is crucial in scanning near-field optical 
microscopy [\citet{Betzig93,Kopelman93,Bian95,Henkel98,Gersen00}].
Another important potential application is the so-called
thresholdless laser [\citet{DeMartini88,Yamamoto93,Protsenko99}].
In a conventional laser, only a small portion of the spontaneous emission 
is channeled into the lasing mode formed by the cavity mirrors, 
the rest being lost to the free space modes. In a microcavity, 
due to strongly modified emission pattern and enhanced 
emission rate, a large portion of spontaneously emitted 
photons is stored in the cavity resonance mode. Losses
due to excitation of free-space modes are thus drastically
reduced and ultralow threshold lasing
can be achieved. 

Controlling of the spontaneous decay also plays an important role
in solid-state systems, where instead of atoms, quantum 
well or quantum dot excitons play the role of the emitters
[\citet{Yokoyama95,Khitrova99,Yamamoto00}].
So, the improved directionality of the spontaneously
emitted light could have dramatic impact on manufacture
of high-efficient light-emitting diodes and displays
[\citet{Yamamoto93}], and the spectral
narrowing could help to increase the transmission capacity of
optical fiber systems where chromatic dispersion is the limiting 
factor [\citet{Hunt93}].

Although certain properties of spontaneous emission
such as the decay rate can be described classically,
using the model of a classically oscillating dipole interacting
with its own radiation field [\citet{Chance78,Wylie85,Haroche92}], 
spontaneous emission is an intrinsically quantum mechanical 
process. Its proper description requires quantization
of both the atom and the radiation field. Obviously, in the
presence of material bodies the medium-assisted electromagnetic
field must be quantized. As long as the medium can be
regarded as being nondispersing and nonabsorbing,
whose (real) permittivity changes with space in general,
electromagnetic-field quantization can be performed, using,
e.g., generalized orthogonal-mode expansion [\citet{Knoell92}]. 
However, the concept fails when material absorption is included 
and the (spatially varying) permittivity becomes a complex function
of frequency. The systematic study of the problem during the
last years has generalized earlier results and offered  
powerful methods to deal with the special requirements of quantizing
the electromagnetic field in absorbing media [for a review, 
see \citet{Knoell01}]. 

There are many reasons why inclusion of material absorption 
in the study of spontaneous emission is desired. One might ask
what would happen when the atomic transition frequency becomes
close to a medium resonance, where absorption is strong.
In particular the question of the effect of absorption on
spontaneous emission in the presence of band-gap material arises.   
Obviously, spontaneous decay must not necessarily be accompanied
by a really observable photon, if the atom is near an
absorbing body, and the question is of what is the (average)
fraction of emitted light. Another question is of how can
absorption modify the local field felt by the atom.
A rigorous approach to the problem has acquired even 
more significance with the recent progress in designing certain 
types of microcavities (e.g., microspheres), where the 
ultimate quality level determined by intrinsic material 
losses has been achieved \ [\citet{Gorodetsky96}].

Although there has been a large body of theoretical work on
medium-assisted spontaneous emission, material absorption has
been ignored usually. Roughly speaking, there have been two 
concepts to treat absorption, namely the microscopic and the
macroscopic approach.  The microscopic approach
starts from a microscopic model of the medium
[\citet{Lee95,Yeung96}; \citet{Juzeliunas97,Fleischhauer99};
\citet{Crenshaw00a,Crenshaw00b,Wubs01}].
Accordingly, the underlying total Hamiltonian typically consists
of the Hamiltonians of the free atom, the free radiation field,
the atomic systems of the medium, and all the  mutual interactions.  
The resulting equations of motion for the coupled system
are then tried to rewrite in order to eliminate,
on applying various approximation schemes, the medium variables
and to obtain closed equations of motion for the atom-field
system only. In this way, the life time of an excited atom
in absorbing bulk material [\citet{Lee95}; \citet{Juzeliunas97};
\citet{Fleischhauer99,Crenshaw00a,Crenshaw00b,Wubs01}], the
initial transient regime [\citet{Wubs01}], and the problem of local 
field corrections [\citet{Juzeliunas97,Fleischhauer99,Crenshaw00a,
Crenshaw00b,Wubs01}] have been studied, and the problem of
spontaneous emission of an excited atom near an absorbing
interface has been considered [\citet{Yeung96}].
The concepts typically borrow, at some stage of
calculation, from macroscopic electrodynamics, e.g., when
a (model-specific) permittivity is introduced, boundary conditions
at surfaces of discontinuity are set or local-field corrections
within the framework of cavity models are considered.
Apart from the fact that the (simplified) microscopic models do not
yield, in general, the exact permittivities, the calculations
can become rather involved, particularly when surfaces
of discontinuity are taken into account [\citet{Yeung96}].
Further, the elimination of the medium variables must
be done very carefully in order to ensure that the
equal-time commutation relations are preserved. If
this is not the case [\citet{Crenshaw00a,Crenshaw00b}],
the results are questionable.

In the macroscopic approach, the medium is
described, from the very beginning, in terms of a
spatially varying permittivity, which is a complex function
of frequency, $\varepsilon({\bf r},\omega)$, that satisfies
the Kramers--Kronig relations.
This approach has -- similar to classical optics -- the
benefit of being universally valid, because it uses
only general physical properties, without the need of
involved {\em ab initio} calculations. Clearly, this concept
is valid only to some approximately fixed length scale 
which exceeds the average distance of two atoms.
With regard to the calculation of the lifetimes
and line shifts, the macroscopic approach is simple.
It is well known that, according to Fermi's golden
rule, the rate of spontaneous decay $\Gamma$ of an excited
atom [position ${\bf r}_A$, (real) transition dipole moment ${\bf d}$, 
transition frequency $\omega_A$] can be expressed in
terms of an electric-field correlation function as
follows [see, e.g., \citet{Loudon83}]:
\begin{equation}
\label{1.1}
\Gamma = \frac{2\pi}{\hbar^2}\!\!
\int_0^\infty\!\!{\rm d}\omega\,
\langle 0|{\bf d}\hat{\bf E}({\bf r}_{\rm A},\omega)\!\otimes\!
\hat{\bf E}^\dagger({\bf r}_{\rm A},\omega_{\rm A}){\bf d}|0 \rangle 
\end{equation}
[cf. Eqs.~(\ref{e2.1}) -- (\ref{e2.2})]. It is also well known
[see, e.g., \citet{Abrikosov75}] that, in agreement with the
dissipation-fluctuation theorem, the relation
\begin{eqnarray}
\label{1.2}
\lefteqn{
\langle 0|\hat{\mbb{E}}({\bf r},\omega)\otimes
\hat{\mbb{E}}{^\dagger}({\bf r}',\omega') | 0 \rangle
}
\nonumber\\&&
= \frac{\hbar\omega^2}{\pi\epsilon_0c^2}
\,{\rm Im}\,\mbb{G}({\bf r},{\bf r}',\omega) \,\delta(\omega-\omega')
\qquad
\end{eqnarray}
($c$, vacuum velocity of light)
is valid, where $\mbb{G}({\bf r},{\bf r}',\omega)$ is the
Green tensor of the classical, macroscopic Maxwell equations.
Combining Eqs.~(\ref{1.1}) and (\ref{1.2}) yields
\begin{equation}
\label{e3.12}
\Gamma = \frac{2\omega_{\rm A}^2}{\hbar\varepsilon_0 c^2}
\,{\bf d}\,{\rm Im}\,
\mbb{G}({\bf r}_{\rm A},{\bf r}_{\rm A},\omega_{\rm A})\,{\bf d}
\end{equation}
[see also \citet{Agarwal75,Wylie84,Wylie85}]. The line shift
can be calculated in a similar way to obtain
\begin{equation}
\label{e3.13}
\delta\omega_A = {{\cal P} \over \pi\hbar\varepsilon_0}
   \!\int_0^\infty \!\!{\rm d}\omega\, 
   {\omega^2\over c^2}
   \frac{{\bf d}\,{\rm Im}\,\mbb{G}({\bf r}_A,{\bf r}_A,\omega)\,{\bf d}}
   {\omega-\omega_A} 
\end{equation}
[${\cal P}$, principal value; for a more rigorous
derivation of Eqs.~(\ref{e3.12}) and
(\ref{e3.13}), see Section \ref{sec3.1.1}].

Hence, knowing the Green tensor of the classical problem for given
complex permittivity, the decay rate and the line shift are known
as well. Equations (\ref{e3.12}) and (\ref{e3.13}) were used 
in order to calculate
decay rates and line shifts for an excited atom near a realistic
(i.e., absorbing) metallic sphere [\citet{Ruppin82,Agarwal83}],
near an absorbing interface [\citet{Agarwal75,Agarwal77,
Wylie84,Wylie85}], and in
a planar cavity filled with an absorbing medium
[\citet{Tomas99}]. Based on Eq.~(\ref{e3.12}), spontaneous
emission of an excited atom in absorbing bulk material 
was studied [\citet{Barnett92}], including local
field corrections [\citet{Barnett96}]. The associated line
shift (without local-field correction) was considered in
Welton's interpretation [\citet{Matloob00a}]. 
Spontaneous decay of an atom at the center of an absorbing 
sphere has been calculated in \citet{Tomas01} (with local 
field correction).

In what follows we restrict our attention to the
macroscopic approach that is based on QED
in dispersing and absorbing media, within the framework of
a source-quantity representation of the electromagnetic
field in terms of the (classical) Green tensor of
the macroscopic Maxwell equations and appropriately
chosen fundamental bosonic fields  
[\citet{Huttner92,Ho93,Matloob95,Gruner96,Ho98,Scheel98,Knoell01}].
The quantization scheme is outlined in Section \ref{sec:quan}.
In Section \ref{sec:Gen_Theor}, the basic formulas for studying
the spontaneous decay of an excited atom are given, which cover
both the strong- and the weak-coupling regime. 
Section \ref{sec:realcavity} is devoted to the
spontaneous decay in bulk material, with special emphasis
on local-field effects.
The problem of spontaneous decay near a planar interface
is considered in Section \ref{sec:planar}, and
Sections \ref{sec:beyond} and \ref{sec:microsphere},
respectively, present results for an atom in a spherical
cavity and near a microsphere. 
In Section \ref{sec:2atoms} a system 
of two atoms coupled to a microsphere is analyzed,
with special emphasis on entangled-state preparation.  
Finally, a summary is given in Section \ref{sec:concl}.

%%%%%%%%%%%%%%%%%%%%%%%%%%%%%%%%%%%%%%%%%%%%%%%%%%%%%%%%%%
%%%%%%%%%%%%%%%%%%%%%%%%%%%%%%%%%%%%%%%%%%%%%%%%%%%%%%%%%%

\setcounter{equation}{0}
\section{\hspace{-0ex}QUANTIZATION SCHEME}
\label{sec:quan}

Following \citet{Ho98,Scheel98,Knoell01},
we first consider the electromagnetic field
in the presence of dispersing and absorbing macroscopic
bodies in the case where no additional atomic sources
are present. The electric-field operator $\hat{\bf E}$ can
be represented in the form of
\begin{eqnarray}
\label{e2.1}
&\displaystyle
   \hat{\bf E}({\bf r}) = \hat{\bf E}^{(+)}({\bf r})
   + \hat{\bf E}^{(-)}({\bf r}),
\\[.5ex]
\label{e2.1a}
&\displaystyle
\hat{\bf E}^{(-)}({\bf r})
   = \big[\hat{\bf E}^{(+)}({\bf r})\big]^\dagger ,&
\\[.5ex]
\label{e2.2}
&\displaystyle
  \hat{\bf E}^{(+)}({\bf r}) 
   = \int_{0}^{\infty} {\rm d}\omega\,
   \underline{\hat{\bf E}}({\bf r},\omega) ,&
\end{eqnarray}
and the induction-field operator $\hat{\bf B}$ accordingly. 
The fields $\underline{\hat{\bf E}}$ and $\underline{\hat{\bf B}}$
then satisfy the macroscopic Maxwell equations 
\begin{equation} 
\label{e2.3}
\mbb{\nabla} {} \underline{\hat{\bf B}}({\bf r},\omega) = 0,
\end{equation}
\begin{equation}        
\label{e2.4}
\mbb{\nabla} {} \left[ \varepsilon_0 \varepsilon({\bf r},\omega)
   \underline{\hat{\bf E}}({\bf r},\omega)\right] =
   \underline{\hat{\rho}}_{\rm N}({\bf r},\omega) ,
\end{equation}
\begin{equation}         
\label{e2.5}
\mbb{\nabla} \times \underline{\hat{\bf E}}({\bf r},\omega) = i\omega
   \underline{\hat{\bf B}}({\bf r},\omega) ,
\end{equation}
\begin{equation}
\label{e2.6}
\mbb{\nabla} \!\times\! \underline{\hat{\bf B}}({\bf r},\omega)
   \!=\! \mu_0 \underline{\hat{\bf j}}_{\rm N}({\bf r},\omega)
   \!-\!\frac{i\omega}{c^2}\,\varepsilon({\bf r},\omega) 
       \underline{\hat{\bf E}}({\bf r},\omega).
\end{equation}
As already mentioned, the real part $\varepsilon_{R}$
and the imaginary part $\varepsilon_{I}$ of the
complex (relative) permittivity $\varepsilon({\bf r},\omega)$ 
satisfy (for any ${\bf r}$) the Kramers--Kronig relations. 
The  operator noise charge and current densities  
$\underline{\hat{\rho}}_{\rm N}({\bf r},\omega)$ and 
$\underline{\hat{\bf j}}_{\rm N}({\bf r},\omega)$ respectively,
which are associated with absorption, are related to the operator
noise polarization $\underline{\hat{\bf P}}_{\rm N}({\bf r},\omega)$ as
\begin{eqnarray}
\label{e2.7}
        &&\underline{\hat{\rho}}_{\rm N}({\bf r},\omega) = 
         - \mbb{\nabla} {} 
         \underline{\hat{\bf P}}_{\rm N}({\bf r},\omega) , \qquad
\\[.5ex]
\label{e2.8}
         &&\underline{\hat{\bf j}}_{\rm N}({\bf r},\omega) =
         -i\omega \underline{\hat{\bf P}}_{\rm N}({\bf r},\omega) , \qquad
\end{eqnarray}
where 
\begin{equation} 
\label{e2.9}
       \underline{\hat{\bf P}}_{\rm N}({\bf r},\omega) 
       = i \sqrt{\frac{\hbar
       \varepsilon_0}{\pi} \varepsilon_{I}({\bf r},\omega)} 
       \,\hat{\bf f}({\bf r},\omega).
\end{equation}
Here, $\hat{\bf f}({\bf r},\omega)$ and
$\hat{\bf f}^\dagger({\bf r},\omega)$ are bosonic fields
which play the role of the fundamental variables of the
composed system (electromagnetic field and medium
including a dissipative system),
\begin{equation}
\label{e2.10}
   \big[ \hat{f}_i({\bf r},\omega),
   \hat{f}^{\dagger}_j({\bf r}',\omega')\big] =         
       \delta_{ij}
       \delta({\bf r}\!-\!{\bf r}') \delta(\omega\!-\!\omega') ,
\end{equation}
\begin{equation}       
\label{e2.11}
      \big[ \hat{f}_i({\bf r},\omega),
       \hat{f}_j({\bf r}',\omega')\big] = 0
        . 
\end{equation}
   
{F}rom Eqs.~(\ref{e2.3}) -- (\ref{e2.9}) it follows that
$\underline{\hat{\bf E}}$ can be written in the form
\begin{eqnarray}
\label{e2.12}
\lefteqn{
\hat{\underline{\bf E}}({\bf r},\omega)
   = i \sqrt{\frac{\hbar}{\pi\varepsilon_0}}\,\frac{\omega^2}{c^2}
}
\nonumber\\[.5ex]&&\hspace{-2.5ex}\times\!
   \!\!\int \!\!{\rm d}^3{\bf r}'
   \sqrt{\varepsilon_{I}({\bf r}',\omega)}
   \,\mbb{G}({\bf r},{\bf r}',\omega)   
   {}\hat{\bf f}({\bf r}',\omega),
\qquad    
\end{eqnarray}
and $\hat{\underline{\bf B}}$ $\!=$
$\!(i\omega)^{-1} \mbb{\nabla}\times\hat{\underline{\bf E}}$ accordingly,
where
$\mbb{G}({\bf r},{\bf r}',\omega)$
is the classical Green tensor satisfying the equation
\begin{eqnarray}
\label{e2.13}
      \left[
      \frac{\omega^2}{c^2}\,\varepsilon({\bf r},\omega)
      \!-\!\mbb{\nabla}\!\times\!\mbb{\nabla}\!\times\!
      \right]
      \mbb{G}({\bf r},{\bf r}',\omega)
     \!=\! -\mbb{\delta}({\bf r}\!-\!{\bf r}') \ \ 
\end{eqnarray}
together with the boundary condition at infinity
[$\mbb{\delta}({\bf r})$, dyadic $\delta$-function].
In this way, the electric field and the induction field are 
expressed in terms of a continuum set of the bosonic fields
$\hat{\bf f}({\bf r},\omega)$ [and $\hat{\bf f}^\dagger
({\bf r},\omega)$], and the Hamiltonian
of the composed system reads (without the infinite ground-state
energy)
\begin{equation}
\label{e2.14}
\hat{H} =  \int\! {\rm d}^3{\bf r}
          \! \int_0^\infty \!{\rm d}\omega \,\hbar\omega
   \,\hat{\bf f}^\dagger({\bf r},\omega){}\hat{\bf f}({\bf r},\omega).
\end{equation}  

Using Eq.~(\ref{e2.12}) [together with Eqs.~(\ref{e2.1}) and
(\ref{e2.2})], one can introduce scalar and vector potentials
$\hat{\varphi}$ and $\hat{\bf A}$, respectively, and express them
in terms of the fundamental bosonic fields. In particular,
in the Coulomb gauge one obtains
\begin{eqnarray} 
\label{e2.15}
         &\displaystyle 
         -\mbb{\nabla} \hat{\varphi}({\bf r}) 
         = \hat{\bf E}^\parallel({\bf r}) , &
\\[.5ex]
\label{e2.16}
         &\displaystyle \hspace{-5ex}
         \hat{\bf A}({\bf r}) \!=\! 
         \!\int_0^\infty\!\! {\rm d}\omega \, (i\omega)^{-1} 
         \underline{\hat{\bf E}}^\perp({\bf r},\omega) 
         + {\rm H.c.}, &
\end{eqnarray}
where
\begin{equation}
\label{e2.17}
\hat{\bf E}^{\perp(\parallel)}({\bf r})
= \int \!{\rm d}^3{\bf r}' \,
   \mbox{\boldmath $\delta$}^{\perp(\parallel)}
   ({\bf r}\!-\!{\bf r}')
         {} \, \hat{\bf E}({\bf r}'),
\end{equation}
with $\mbox{\boldmath $\delta$}^\perp({\bf r})$ 
and $\mbox{\boldmath $\delta$}^\parallel({\bf r})$ being
the dyadic transverse and longitudinal $\delta$-functions,
respectively.

We now consider the interaction of the medium-assisted
electromagnetic field with additional point charges $q_\alpha$.
Applying the minimal-coupling scheme, we may write the
complete Hamiltonian in the form of
\begin{eqnarray}
\label{e2.18}
\lefteqn{
     \hat{H} = \!\int\! {\rm d}^3{\bf r}
     \!\int_0^\infty\! {\rm d}\omega \,\hbar\omega
     \,\hat{\bf f}^\dagger({\bf r},\omega){}\hat{\bf f}({\bf r},\omega)
}
\nonumber\\&&\hspace{2ex}   
     + \sum_\alpha {1\over 2m_\alpha} \left[ \hat{\bf p}_\alpha 
     - q_\alpha \hat{\bf A}(\hat{\bf r}_\alpha) \right]^2
\nonumber\\&&\hspace{-10ex}
     +\,{\textstyle{1\over 2}} \int {\rm d}^3{\bf r}\, 
     \hat{\rho}_A({\bf r}) \hat{\varphi}_A({\bf r})
     + \int {\rm d}^3{\bf r}\, 
     \hat{\rho}_A({\bf r}) \hat{\varphi}({\bf r}) ,
%\nonumber\\
\end{eqnarray}
where $\hat{\bf r}_\alpha$ is the position operator
and $\hat{\bf p}_\alpha$ is the canonical momentum operator
of the $\alpha$th charged particle of mass $m_\alpha$. The
Hamiltonian (\ref{e2.18}) consists of four terms. The first 
term is the energy observed when the particles are absent
[cf. Eq.~(\ref{e2.14})]. The second term is the 
kinetic energy of the particles, and the third and fourth terms  
are their Coulomb energies, where the potential $\hat{\varphi}_A$ 
can be given by
\begin{equation}
\label{e2.19}
         \hat{\varphi}_A({\bf r}) = 
         \int {\rm d}^3{\bf r}' \,\frac {\hat{\rho}_A({\bf r}')}
         {4\pi\varepsilon_0|{\bf r}-{\bf r}'|} \,,
\end{equation}
with
\begin{equation}
\label{e2.20}
         \hat{\rho}_A({\bf r}) = 
         \sum_\alpha q_\alpha 
         \delta({\bf r}-\hat{\bf r}_\alpha) 
\end{equation}
being the charge density. Obviously, the last term
in Eq.~(\ref{e2.18}) is the Coulomb energy 
of interaction of the particles with the medium. Note
that all terms are expressed in terms of the
dynamical variables $\hat{\bf f}({\bf r},\omega)$, 
$\hat{\bf f}^\dagger({\bf r},\omega)$, $\hat{\bf r}_\alpha$, and
$\hat{\bf p}_\alpha$. It is worth noting the quantization scheme 
is fully equivalent to the so-called auxiliary-field scheme
introduced \ by \ \citet{Tip98} [for details, see \citet{Tip01}]. 

%%%%%%%%%%%%%%%%%%%%%%%%%%%%%%%%%%%%%%%%%%%%%%%%%%%%%%%%%%%%%%%%%%%
%%%%%%%%%%%%%%%%%%%%%%%%%%%%%%%%%%%%%%%%%%%%%%%%%%%%%%%%%%%%%%%%%%%

\setcounter{equation}{0}
\section{SPONTANEOUS DECAY:\\ GENERAL FORMALISM}
\label{sec:Gen_Theor}

%%%%%%%%%%%%%%%%%%%%%%%%%%%%%%%%%%%%%%%%%%%%%%%%%%%%%%%%%%%%%%%%%%

\subsection{Basic equations}
\label{subsec:basic}

Let us consider $N$ two-level atoms [positions ${\bf r}_A$, transition
frequencies $\omega_A$ ($A$ $\!=$ $\!1,2,...,N$)] 
that resonantly interact with radiation
via electric-dipole transition (dipole moments ${\bf d}_A$).
Let us further assume that the atoms are sufficiently far from each
other, so that the interatom Coulomb interaction can be ignored.
In this case, the electric-dipole approximation and the rotating
wave approximation apply, and the minimal-coupling 
Hamiltonian takes the form of [\citet{Ho00,Knoell01}]
\begin{eqnarray}
\label{e3.1}
\lefteqn{
   \hat{H} = \int \!{\rm d}^3{\bf r}
   \! \int_0^\infty \!\!{\rm d}\omega \,\hbar\omega
   \,\hat{\bf f}^\dagger({\bf r},\omega){}\hat{\bf f}({\bf r},\omega)
}
\nonumber\\[.5ex]&& \hspace{-5ex}
    + \sum_{A} {\textstyle{1\over 2}}\hbar\omega_A \hat{\sigma}_{Az}
%\nonumber\\[.5ex]&&
    - \sum_{A} \big[
    \hat{\sigma}_A^\dagger
    \hat{\bf E}^{(+)}({\bf r}_A){}{\bf d}_A  
    + {\rm H.c.}\big]. 
\qquad
\end{eqnarray}
Here and in the following, the two-level atoms are described
in terms of the Pauli operators $\hat{\sigma}_A$,
$\hat{\sigma}_A^\dagger$, and $\hat{\sigma}_{Az}$. 

For a single-quantum excitation of the system, the system wave 
function at time $t$ can be written as
\begin{eqnarray}
\label{e3.2}
\lefteqn{
    |\psi(t)\rangle = \sum_{A} C_{U_A}(t)
    e^{-i(\omega_A-\bar{\omega})t}
    |U_A\rangle |\{0\}\rangle
}
\nonumber\\[.5ex]&&\hspace{0ex} 
     +\!\int\! {\rm d}^3{\bf r} \int_0^\infty\!\! {\rm d}\omega\, 
     \Bigl[ C_{Li}({\bf r},\omega,t)
\nonumber\\[.5ex]&&\hspace{2ex} \times \,
     e^{-i (\omega-\bar{\omega})t}
     |L\rangle |\{1_i({\bf r},\omega)\} \rangle \Bigr]
\qquad
\end{eqnarray}
($\bar{\omega}$ $\!=$ $\!$ $\!\frac{1}{2}\sum_A\omega_A$).
Here, $|U_A\rangle$ is the excited atomic state, where
the $A$th atom is in the upper state and all the other
atoms are in the lower state, and $|L\rangle$ is the
atomic state, where all atoms are in the lower state.
Accordingly, $|\{0\}\rangle$ is the 
vacuum state of the rest of the system, and
$|\{1_i({\bf r},\omega)\}\rangle$ is the state,
where it is excited in a single-quantum Fock state.
The Schr\"odinger equation yields
\begin{eqnarray}
\label{e3.3}
\lefteqn{\hspace{-1ex}
   \dot{C}_{U_A}(t) =
   \frac{-1}{\sqrt{\pi\varepsilon_0\hbar}}           
   \!\int_0^\infty \!\!{\rm d}\omega
   \!\int\!{\rm d}^3{\bf r}
   \bigg\{
   \frac{\omega^2}{c^2}\,e^{-i(\omega-\omega_A)t}  
}
\nonumber \\[0ex]&& \hspace{-5ex}\times
   \Bigl[ \sqrt{\varepsilon_{I}({\bf r},\omega)}
   \,{\bf d}_A\mbb{G}({\bf r}_A,{\bf r},\omega)
   {\bf C}_{L}({\bf r},\omega,t) \Bigr]
   \bigg\},
\end{eqnarray}
\begin{eqnarray}
\label{e3.4}
\lefteqn{
   \dot{\bf C}_{L}({\bf r},\omega,t) = \sum_{A} 
   \frac{1}{\sqrt{\pi\varepsilon_0\hbar}}
   \,\frac{\omega^2}{c^2}
   \,e^{i(\omega-\omega_A)t}
}
\nonumber \\[.5ex]&&\hspace{-2ex}\times\,
   \sqrt{\varepsilon_{I}({\bf r},\omega)}          
   {\bf d}_A\mbb{G}^\ast({\bf r}_A,{\bf r},\omega)\, C_{U_A}(t).
\quad   
\end{eqnarray}
We now substitute the result of formal integration of 
Eq.~(\ref{e3.4}) into Eq.~(\ref{e3.3}). 
Making use of the relationship 
\begin{eqnarray}
\label{e3.5}
\lefteqn{
   {\rm Im}\,G_{ij}({\bf r},{\bf r'},\omega) =
   \int {\rm d}^3{\bf s}\,
   \bigg[
   \frac{\omega^2}{c^2}\, \varepsilon_{I}({\bf s},\omega)
}\nonumber\\[.5ex]&& \hspace{4ex}\times\,
    G_{ik}({\bf r},{\bf s},\omega) 
    G^\ast_{jk}({\bf r'},{\bf s},\omega)
    \bigg],
\qquad    
\end{eqnarray}
we obtain the following (closed) system of
integro-differential equations:
\begin{eqnarray}
\label{e8.5}
\lefteqn{\hspace{0ex}
        \dot{C}_{U_A}(t) = \sum_{A'}
        \int_0^t {\rm d}t'\, K_{AA'}(t,t')\, C_{U_{A'}}(t')
}
\nonumber\\[.5ex]&&\hspace{-3ex}
   -\frac{1}{\sqrt{\pi\varepsilon_0\hbar}}           
   \!\int_0^\infty \!\!{\rm d}\omega
   \!\int\!{\rm d}^3{\bf r}
   \bigg\{
   \frac{\omega^2}{c^2}\,e^{-i(\omega-\omega_A)t}  
\nonumber\\[.5ex]&& \hspace{-5ex}\times
   \Big[ \sqrt{\varepsilon_{I}({\bf r},\omega)}
   \,{\bf d}_A\mbb{G}({\bf r}_A,{\bf r},\omega)
   {\bf C}_{L}({\bf r},\omega,0) \Big]
   \bigg\},
%\nonumber\\&&
\end{eqnarray}
\begin{eqnarray}
\label{e8.6} 
\lefteqn{\hspace{0ex}
        K_{AA'}(t,t') =
        \frac{-1 }
          {\hbar\pi\varepsilon_0}
        \int_0^\infty {\rm d}\omega \biggl[
        {\omega^2\over c^2}
        e^{-i(\omega-\omega_A)t}
} 
\nonumber\\[.5ex]&&\hspace{-1ex}\times\,
        e^{i(\omega-\omega_{A'})t'}         
       {\bf d}_A {\rm Im}\,\mbb{G}({\bf r}_A,{\bf r}_{A'},\omega)
       {\bf d}_{A'} \biggr].
\qquad
\end{eqnarray}
Note that
\begin{equation}
\label{e8.6a}
K_{AA'}(t,t')=K_{A'A}^\ast(t',t),
\end{equation}
because of the reciprocity theorem.

The excitation can initially reside in either 
an atom or the medium-assisted electromagnetic field.
The latter case, i.e., 
\mbox{${\bf C}_L({\bf r},\omega,0)$ $\!\neq$
$\!0$} in Eq.~(\ref{e8.5}), could be realized, for example, 
by coupling the field first to an excited atom {$D$}
in a time interval $\Delta t$ such that, according to Eq.~(\ref{e3.4}),
${\bf C}_L({\bf r},\omega,0)$ reads
\begin{eqnarray}
\label{e3.35}
\lefteqn{       
   {\bf C}_{L}({\bf r},\omega,0) =  \int_{-\Delta t}^0 {\rm d}t'
   \frac{1}{\sqrt{\pi\varepsilon_0\hbar}}
   \,\frac{\omega^2}{c^2}
   \,e^{i(\omega-\omega_D)t'}
}
\nonumber \\[.5ex]&&\hspace{-2ex}\times\,
   \sqrt{\varepsilon_{I}({\bf r},\omega)}          
   {\bf d}_D\mbb{G}^\ast({\bf r}_D,{\bf r},\omega)\, C_{U_D}(t'),
\qquad   
\end{eqnarray}
where
$C_{U_D}(t)$
describes the single-atom decay according to
Eq.~(\ref{e3.8}) given below. Substitution of the
expression (\ref{e3.35}) into Eq.~(\ref{e8.5}) then yields
\begin{eqnarray}
\label{e8.6b}
\lefteqn{
        \dot{C}_{U_A}(t) = \sum_{A'}
        \int_0^t {\rm d}t'\, K_{AA'}(t,t')\,C_{U_{A'}}(t')
}
\nonumber\\[.5ex]&&\hspace{2ex}
        + \int_{-\Delta t}^0 {\rm d}t'\, K_{AD}(t,t')\,C_{U_D}(t').
\qquad
\end{eqnarray}

%%%%%%%%%%%%%%%%%%%%%%%%%%%%%%%%%%%%%%%%%%%%%%%%%%%%%%%%%%%%%%%%%%%

\subsection{Single-Atom Decay}
\label{sec3.1}

%%%%%%%%%%%%%%%%%%%%%%%%%%%%%%%%%%%%%%%%%%%%%%%%%%%%%%%%%%%%%%%%%%%

\subsubsection{Atomic Dynamics}
\label{sec3.1.1}

Let us restrict our attention to the spontaneous decay 
of a single atom. For the initial condition 
${\bf C}_{L}({\bf r},\omega,0)$ $\!=$ $\!0$,
Eq.~(\ref{e8.5}) becomes
($C_{U}$ $\!\equiv$ $\!C_{U_A}$, \mbox{${\bf d}$ $\!\equiv$ $\!{\bf d}_A$})
\begin{equation}
\label{e3.6}
        \dot{C}_{U}(t) =\int_0^t {\rm d}t'\, K(t-t') C_{U}(t'),
\end{equation}
where
\begin{equation}
\label{e3.7}
K(t-t') = K_{AA}(t,t').
\end{equation}
Integrating both sides of Eq.~(\ref{e3.6}) with respect to time, 
we easily derive, on changing the order of integrations
on the right-hand side, a Volterra integral equation of
the second kind,
\begin{equation}
\label{e3.8}
   C_{U}(t) =\int_0^t {\rm d}t'\, \bar{K}(t-t') C_{U}(t') + 1
\end{equation}
[$C_{U}(0)$ $=$ $\!1$], where, according to Eqs.~(\ref{e3.7})
and (\ref{e8.6}), 
\begin{eqnarray}
\label{e3.9}
\lefteqn{\hspace{-2ex}              
   \bar{K}(t\!-\!t') =
   \frac{-1}{\hbar\pi\varepsilon_0}
   \!\int_0^\infty \!\!{\rm d}\omega
   \bigg\{\!\Big[1\!
    -\!e^{-i(\omega-\omega_A)(t-t')}\Big]
}
\nonumber\\[.5ex]&&\hspace{4ex}
   \times\,{\omega^2\over c^2}\,
   \frac{{\bf d}{\rm Im}\,\mbb{G}({\bf r}_A,{\bf  r}_A,\omega)
   {\bf d}}{i(\omega-\omega_A)}\bigg\}.
\qquad 
\end{eqnarray}

It is worth noting
that Eqs.~(\ref{e3.6}) and (\ref{e3.8})
apply to the spontaneous decay of an atom in the presence of an 
arbitrary configuration of dispersing and absorbing macroscopic 
bodies. All the matter parameters that are relevant for  
the atomic evolution are contained, via the 
Green tensor, in the kernel functions (\ref{e3.7})
and (\ref{e3.9}). In particular when absorption is
disregarded and the permittivity is regarded as being 
a real, frequency-independent quantity (which of 
course can change with space), then the formalism
yields the results of standard mode decomposition,
obtained by Laplace transform techniques
[\citet{Lewenstein88a,Lewenstein88b,John94}] and
delay-differential-equation techniques
[\citet{Cook87,Feng89,Ho99}].
It should be pointed out that the Green tensor has been available 
for a large variety of configurations such as planarly, spherically, 
and cylindrically multilayered media [\citet{Tai94,Chew95}].  

In order to study the
case where the atom is surrounded by matter,
the atom should be assumed to be localized in some 
small free-space region, so that the Green tensor at 
the position of the atom reads 
\begin{equation}
\label{e3.16}
   \mbb{G}({\bf r}_A,{\bf r}_A,\omega)
    \!=\!\mbb{G}^V({\bf r}_A,{\bf r}_A,\omega)
    \!+\! \mbb{G}^R({\bf r}_A,{\bf r}_A,\omega),
\end{equation}
where 
$\mbb{G}^V$ is the vacuum Green tensor with
\begin{equation}
\label{e3.17}
         {\rm Im}\,\mbb{G}^V({\bf r}_A,{\bf r}_A,\omega)
         = {\omega \over 6\pi c} \, \mbb{I}
\end{equation}
(Appendix \ref{app:bulk}), and 
$\mbb{G}^R$ describes the effects of reflections at the 
(surface of discontinuity of the) surrounding medium.
The contribution of $\mbb{G}^V$ to $\bar{K}$ in Eq.~(\ref{e3.9})
can be treated in the Markov approximation (see below), thus
\begin{eqnarray}
\label{e3.19}
\lefteqn{
   \bar{K}(t-t') = - {\textstyle\frac{1}{2}}\Gamma_0 + 
   \frac{1}{\hbar\pi\varepsilon_0}
   \int_0^\infty \!\!{\rm d}\omega\,
   {\omega^2\over c^2}
}
\nonumber\\&&\hspace{-2ex}\times\,         
   \frac{{\bf d}{\rm Im}\,\mbb{G}^R({\bf r}_A,
   {\bf r}_A,\omega){\bf d}}
   {i(\omega-\omega_A)}
   \Big[e^{-i(\omega-\omega_A)(t-t')}\!-\!1 \Big],
\nonumber\\
\end{eqnarray}
where $\Gamma_0$ is the well-known decay rate in free space,
\begin{equation}
\label{e3.18}
        \Gamma_0= {\omega_A^3d^2 \over 3\hbar\pi\varepsilon_0c^3}\,.
\end{equation}
The integro-differential equation (\ref{e3.6})
[or the integral equation (\ref{e3.8})] together with the kernel
function (\ref{e3.7}) [or  (\ref{e3.19})] can be regarded as
the basic equation for studying the influence of an arbitrary
configuration of dispersing and absorbing matter on the
spontaneous decay of an excited atom. 

%%%%%%%%%%%%%%%%%%%%%%%%%%%%%%%%%%%%%%%%%%%%%%%%%%%%%%%%%%%%%%%%%%%%%%

\paragraph{Weak Coupling}

When the Markov approximation applies, i.e.,  when
in a coarse-grained description of the atomic motion
memory effects are disregarded, then we may let
\begin{equation}
\label{e3.10}
\frac{e^{i(\omega_A-\omega)(t-t')}-1}
{i(\omega_A-\omega)}
\quad\to\quad
\zeta(\omega_A-\omega)
\end{equation}
in Eq.~(\ref{e3.9}) [$\zeta(x)$ $\!=$ $\!\pi\delta(x)$ $\!+$ 
$\!i{\cal P}/x$],
and thus
\begin{equation}
\label{e3.11}
\bar{K}(t-t') = - {\textstyle\frac{1}{2}} \Gamma
    + i \delta\omega_A,
\end{equation}
where
$\Gamma$ and $\delta\omega_A$ are respectively given by
Eqs.~(\ref{e3.12}) and (\ref{e3.13}).
Substitution of the expression (\ref{e3.11})
into Eq.~(\ref{e3.8}) for the kernel function
yields the familiar (weak-coupling) result that
\begin{equation}
\label{e3.15}
C_{U}(t) = \exp\!\left[\left(-{\textstyle\frac{1}{2}}
   \Gamma+i\delta\omega_A\right)t\right].
\end{equation}
Application of Eqs.~(\ref{e3.12}), (\ref{e3.13}), (\ref{e3.16}), and 
(\ref{e3.17}) yields
\begin{equation}
\label{e3.17a}
   \Gamma \!=\! \Gamma_0  \!+\!  
   {2\omega_A^2\over \hbar\varepsilon_0c^2}\, 
   {\bf d}{\rm Im}\,\mbb{G}^R({\bf r}_A,{\bf r}_A,\omega_A){\bf d},
\end{equation}
\begin{eqnarray}
\label{e3.14}
\lefteqn{\hspace{-5ex} 
\delta\omega_A =
   {\omega_A^2\over \hbar\varepsilon_0c^2}\,
   \biggl[ 
   {\bf d}{\rm Re}\,\mbb{G}^R({\bf r}_A,{\bf r}_A,\omega_A){\bf d}
}
\nonumber\\[.5ex]&& \hspace{-8ex} 
   - {1\over \pi}\!\int_0^\infty \!\!{\rm d}\omega\, 
   {\omega^2\over \omega_A^2} 
   \frac{{\bf d}{\rm Im}\,\mbb{G}^R({\bf r}_A,{\bf
   r}_A,\omega){\bf d}}
   {\omega+\omega_A}
   \biggr].
\end{eqnarray}
In Eq.~(\ref{e3.14}), the Kramers--Kronig relations have 
been used and the divergent contribution of the vacuum to the 
line shift is thought of as being included in the atomic 
transition frequency $\omega_A$. It is not difficult to see 
that in Eq.~(\ref{e3.14}) the second term, which is only weakly
sensitive to the atomic transition frequency, is small compared
to the first one and can therefore be neglected in general. 

%%%%%%%%%%%%%%%%%%%%%%%%%%%%%%%%%%%%%%%%%%%%%%%%%%%%%%%%%%%%%%%%%%

\paragraph{Strong Coupling}

When the atomic transition frequency
approaches a resonance frequency of a resonator-like
equipment of macroscopic bodies, then the
strength of the coupling between the atom and the
electromagnetic field can increase to such an extent that the
weak-coupling approximation fails and the integral equation
(\ref{e3.6}) must be considered. 
Let us assume, for simplicity, that only a single (field-)resonance
line of Lorentzian shape  is involved in the atom--field interaction.
In this case, the kernel function (\ref{e3.7}) may be approximated by
\begin{eqnarray}
\label{e6.2}
\lefteqn{\hspace{-.5ex}
      K(t-t') \simeq -\frac{\Gamma_C(\Delta\omega_C)^2}{2\pi}
     {\rm e}^{-i(\omega_C-\omega_A)(t-t')}
}
\nonumber\\[.5ex]&&\hspace{0ex}
     \times
     \int_{-\infty}^\infty \!{\rm d}\omega \,
     \frac{{\rm e}^{-i(\omega-\omega_C)(t-t')}}{(\omega-\omega_C)^2
     +\left(\Delta\omega_C \right)^2}
\nonumber \\[.5ex]&&\hspace{-4ex}
=\!{\textstyle\frac{1}{2}}\Gamma_C \Delta\omega_C
{\rm e}^{-i(\omega_C-\omega_A)(t-t')}
{\rm e}^{-\Delta\omega_C|t-t'|},
\qquad  
\end{eqnarray}
and thus the integral equation (\ref{e3.6}) corresponds to
the differential equation [\citet{Ho00}]
\begin{eqnarray}
\label{e6.3}
\lefteqn{\hspace{-2ex}
   \ddot{C}_{U}(t) +\left[i(\omega_C-\omega_A)
   +\Delta\omega_C \right] \dot{C}_{U}(t) 
}
\nonumber\\ && \hspace{8ex}
   +\,{\textstyle\frac{1}{2}}\Gamma_C
   \Delta\omega_C C_{U}(t) = 0.
\quad   
\end{eqnarray}
Here, $\omega_C$ and $\Delta\omega_C$ are respectively
the mid-frequency and the
line width of the
field resonance associated with the bodies, 
and $\Gamma_C$ is the (weak-coupling) decay rate at $\omega_C$.

Equation (\ref{e6.3}) typically applies
to the strong-coupling regime for an arbitrary resonator
configuration, provided that the field that effectively
interacts with the atom can be regarded as being a
single-resonance field of
Lorentzian shape.\footnote{Equations of
   the type (\ref{e6.3}) can also be obtained within the
   framework of standard (Markovian) quantum noise theory,
   where an appropriately chosen undamped mode is coupled to a
   two-level atom and some reservoir
   [\citet{Sachdev84}].\label{fnote1}}   
In particular, when material absorption is disregarded,
then the line broadening
solely results from the radiative losses due
to the input-output coupling
[\citet{Cook87,Lai88,Feng89}].

Equation (\ref{e6.3}) reveals that
the upper-state probability amplitude of the atom 
obeys the equation
of motion for a damped harmonic oscillator.
In the strong-coupling regime, where \mbox{$\omega_A$ $\!=$
$\!\omega_C$} and \mbox{$\Omega$ $\!\gg$
$\!\Delta\omega_C$}, damped Rabi oscillations are
observed:
\begin{equation}
\label{e6.5}
     C_{U}(t) = {\rm e}^{-\Delta\omega_C t/2} 
     \cos(\Omega t/2),
\end{equation}
where the Rabi frequency $\Omega$ reads
\begin{equation}
\label{e6.6}
\Omega = \sqrt{2\Gamma_C \Delta\omega_C}\,.
\end{equation}

%%%%%%%%%%%%%%%%%%%%%%%%%%%%%%%%%%%%%%%%%%%%%%%%%%%%%%%%%%%%%%%%%%%%

\subsubsection{Emitted-Light Intensity}
\label{sec3.1.2}

It is well known that the intensity of light registered by a point-like
photodetector at position ${\bf r}$ and time $t$ is given by
\begin{equation}
\label{e3.20}
           I({\bf r},t) \equiv 
           \langle \psi(t) | 
           \hat{\bf E}^{(-)}({\bf r}){}\hat{\bf E}^{(+)}({\bf r})
           | \psi(t) \rangle.
\end{equation}
The emitted-light intensity associated with the spontaneous decay 
of an excited atom in the presence of dispersing and absorbing 
matter can be obtained by combining Eqs. (\ref{e2.1}) -- (\ref{e2.2}),
(\ref{e2.12}), (\ref{e3.2}), and (\ref{e3.5}). The result is
\begin{eqnarray}
\label{e3.21}
\lefteqn{               
          I({\bf r},t)  =
          \biggl| {\omega_A^2 \over \pi\varepsilon_0c^2}
          \int_0^t {\rm d}t' 
          \int_0^\infty\!\! {\rm d}\omega\,
          \Big[ C_{U}(t')
}          
\nonumber \\ &&         
          \times\,
          e^{-i(\omega-\omega_A)(t-t')}
          {\rm Im}\, \mbb{G}({\bf r},{\bf r}_A,\omega)\,{\bf d}
          \Big]\biggr|^2\!,
\qquad\quad
\end{eqnarray}
where, in the spirit of the rotating wave approximation
used, \mbox{$\omega^2\!=\!\omega_A^2$} has been set
in the frequency integral. Again, all relevant matter
parameters are contained in the Green tensor.
In contrast to Eq.~(\ref{e3.8}) [together with the kernel
function (\ref{e3.19})], Eq.~(\ref{e3.21}) requires information 
about the Green tensor at different space points.
In particular, its dependence on space and frequency 
essentially determines the retardation effects.

In the simplest case of free space we have
\begin{eqnarray}
\label{e3.22}
\lefteqn{          
         {\rm Im}\,\mbb{G}^V({\bf r},{\bf r}_A,\omega) 
         = {1\over 8i\pi\rho} \left( \mbb{I}
         - \frac{\mbb{\rho}\otimes\mbb{\rho}}
         {\rho^2} \right)
}         
\nonumber\\[.5ex]&&\hspace{0ex}
         \times
         \left(e^{i\omega\rho/c} - e^{-i\omega\rho/c} \right) 
         + {\cal O}\!\left(\rho^{-2}\right)
\qquad         
\end{eqnarray}
($\mbb{\rho}$ $\!=$ $\!{\bf r}$ $\!-$ $\!{\bf r}_A$;
Appendix \ref{app:bulk}). We substitute Eqs.~(\ref{e3.15}) 
($\Gamma$ $\!=$ $\!\Gamma_0$)
and (\ref{e3.22}) into Eq.~(\ref{e3.21}), calculate the time integral, 
and extend the lower limit in the frequency integral to $-\infty$, 
\begin{eqnarray}
\label{e3.23}
\lefteqn{\hspace{-5ex}
         \int_{-\infty}^\infty \!\!{\rm d}\omega 
         \left(e^{i\omega\rho/c}\! -\! e^{-i\omega\rho/c} \right)
         \frac
         {e^{-(\Gamma_0/2+i\omega_A')t}
         \!-\! e^{-i\omega t} }
         {i[\omega \!-\! (\omega_A'\!-\! i\Gamma_0/2)]}        
}
\nonumber\\[.5ex] &&   \hspace{-8ex}
         =\! - 2\pi \exp\!\left[
         \left(-{\textstyle\frac{1}{2}}\Gamma_0  
         \!-\!i\omega_A'\right)\!\left(\!t\!
         -\!{\rho\over c}\right)\!\right]
         \Theta\!\left(\!t\!-\!{\rho\over c}\right)
%\nonumber\\
\end{eqnarray}
[$\Theta(x)$, unit step function], where
\begin{eqnarray}         
\label{e3.24}
         \omega_A' = \omega_A
         - \delta\omega_A\,. 
\end{eqnarray}
Thus, the well-known (far-field) result 
\begin{equation}
\label{e3.25}
          I({\bf r},t)  =
          \left( {\omega_A^2d\sin\theta \over 
          4\pi\varepsilon_0c^2\rho} \right)^2 
          \!e^{-\Gamma_0(t-\rho/c)} \, \Theta(t\!-\!\rho/c)
\end{equation}
is recognized ($\theta$, angle between $\mbb{\rho}$ and
$\mbb{d}$).

It should be noted that the general expression (\ref{e3.21}) is
valid for an arbitrary coupling regime. In particular, 
in the weak-coupling regime the Markov approximation applies,
and $C_{U}(t')$ can be taken at \mbox{$t'$ $\!=$ $\!t$} and
put in front of the time integral in Eq.~(\ref{e3.21}),
with $C_{U}(t)$ being simply the exponential (\ref{e3.15}).
Equation (\ref{e3.21}) thus simplifies to 
\begin{equation}
\label{e3.26}
          I({\bf r},t)  \simeq 
          |{\bf F}({\bf r},{\bf r}_A,\omega_A)|^2 
          e^{-\Gamma t} ,
\end{equation}
where
\begin{eqnarray}
\label{e3.27}
\lefteqn{ 
        {\bf F}({\bf r},{\bf r}_A,\omega_A) = 
        -{i\omega_A^2 \over \varepsilon_0c^2}
        \biggl[ 
        \mbb{G}({\bf r},{\bf r}_A,\omega_A){\bf d}
}
\nonumber\\[.5ex]&& 
        -\, {1\over\pi} \int_0^\infty {\rm d}\omega\, 
        \frac{{\rm Im}\,
        \mbb{G}({\bf r},{\bf r}_A,\omega){\bf d}}
        {\omega+\omega_A}
        \biggr].
\qquad        
\end{eqnarray}
Since the second term on the right-hand side of  
Eq.~(\ref{e3.27}) is small compared to the first one,
it can be omitted, and the spatial distribution
of the emitted-light intensity (emission pattern) can
be given by, on disregarding transit time delay,  
\begin{equation}
\label{e3.28}
      |{\bf F}({\bf r},{\bf r}_A,\omega_A)|^2
      \simeq 
      \biggl| {\omega_A^2 \over \varepsilon_0c^2}\, 
      \mbb{G}({\bf r},{\bf r}_A,\omega_A){\bf d} \biggr|^2\!.
\end{equation}

Material absorption gives rise to nonradiative decay.
The fraction of really emitted radiation energy can be obtained
by integration of the Pointing-vector expectation value with 
respect to time and over the surface of a sphere
whose radius $r$ is much larger than the extension of the
system consisting of the macroscopic bodies and the atom,
\begin{equation}
\label{e3.29}
W = 2c\varepsilon_0\!\int_0^\infty \!\!{\rm d}t
   \!\int_0^{2\pi} \!\!d\phi
   \!\int_0^\pi \!\!d\theta\,  r^2 \sin\theta \,I({\bf r},t).
\end{equation}
The ratio $W/W_0$ (\mbox{$W_0$ $\!=$ $\!\hbar\omega_A$})
then gives us a measure of the emitted radiation energy, and
accordingly, \mbox{$1$ $\!-$ $\!W/W_0$}
measures the energy absorbed by the bodies.

%%%%%%%%%%%%%%%%%%%%%%%%%%%%%%%%%%%%%%%%%%%%%%%%%%%%%%%%%%%%%%%

\subsubsection{Emitted-Light Spectrum}
\label{sec3.1.3}

Next, let us consider the time-dependent power spectrum of 
the emitted light, which for sufficiently small passband width
of the spectral apparatus can be given by [see, e.g., \citet{Vogel94}]
\begin{eqnarray}
\label{e3.30}
\lefteqn{      
S({\bf r},\omega_{S},T) = 
   \int_0^T \!\!{\rm d}t_2
   \!\int_0^T \!\!{\rm d}t_1 \Big[e^{-i\omega_S(t_2-t_1)}
}
\nonumber\\&&
   \hspace{4ex}
   \times\,          
   \big\langle \hat{\bf E}^{(-)}({\bf r},t_2){} 
   \hat{\bf E}^{(+)}({\bf r},t_1) \big\rangle\Big],
\qquad    
\end{eqnarray}
where $\omega_{S}$ is the setting frequency of the spectral 
apparatus and $T$ is the operating-time interval of the detector.
In close analogy to the derivation of Eq.~(\ref{e3.21}),
combination of Eqs.~(\ref{e2.1}) -- (\ref{e2.2}), (\ref{e2.12}),  
(\ref{e3.2}), and (\ref{e3.5}) leads to 
\begin{eqnarray}
\label{e3.31}
\lefteqn{
          S({\bf r},\omega_{S},T) = 
          \biggl| {\omega_A^2 \over \pi\varepsilon_0c^2}
          \int_0^T {\rm d}t_1
          \Big[e^{i(\omega_{S}-\omega_A) t_1}
}
\nonumber\\[.5ex]&&\times
          \int_0^{t_1}{\rm d}t'\,C_{U}(t') 
          \int_0^\infty{\rm d}\omega\, 
          e^{-i(\omega-\omega_A) (t_1-t')}          
\nonumber\\[.5ex]&&\hspace{6ex}\times\,
          {\rm Im}\, \mbb{G}({\bf r},{\bf r}_A,\omega){\bf d}
          \Big]\biggr|^2 .
\qquad          
\end{eqnarray}
Further calculation again requires
knowledge of the Green tensor of the problem.

Let us use Eq.~(\ref{e3.31}) to recover the free-space result.
Following the line that has led from Eq.~(\ref{e3.21}) to
Eq.~(\ref{e3.25}), we find that 
\begin{eqnarray}
\label{e3.32}
\lefteqn{
          S({\bf r},\omega_{S},T) = 
          \left( {\omega_A^2 d \sin\theta \over
          4\pi \varepsilon_0c^2\rho} \right)^2
          \Theta\!\left(T\!-\!\rho/c\right)
}          
\nonumber\\&&\hspace{2ex}\times\,
          \left| \frac
          {e^{[-\Gamma_0/2+i(\omega_{S}-\omega_A')](T-\rho/c)} -1}
          {\omega_{S}- \omega_A'+i\Gamma_0/2} \right|^2\!.
\qquad\quad          
\end{eqnarray}
In particular for $T$ $\!\rightarrow$ $\!\infty$, we recognize
the well-known Lorentzian:
\begin{eqnarray}
\label{e3.33}
\lefteqn{
          \lim_{T\to\infty} S({\bf r},\omega_{S},T) = 
          \left( {\omega_A^2 d \sin\theta \over
          4\pi \varepsilon_0c^2\rho} \right)^2
}         
\nonumber\\[.5ex]&&\hspace{2ex}\times\,
          {1\over 
          (\omega_{S} - \omega_A')^2
          +\Gamma_0^2/4}\,.
\qquad          
\end{eqnarray}

If retardation is ignored and the Markov approximation applies,
Eq.~(\ref{e3.31}) can be simplified in a similar
way as Eq.~(\ref{e3.21}). In close analogy to the derivation of 
Eq.~(\ref{e3.26}) we may write
\begin{eqnarray}
\label{e3.34}
\lefteqn{
          S({\bf r},\omega_{S},T) = 
          |{\bf F}({\bf r},{\bf r}_A,\omega_A)|^2
}          
\nonumber\\ &&\hspace{2ex}\times\, 
          \left| \frac
      {e^{[-\Gamma/2+i(\omega_S-\omega_A')] T} -1} 
          {\omega_{S}-\omega_A'
          + i\Gamma/2} \right|^2\!,
\qquad
\end{eqnarray}
with ${\bf F}({\bf r},{\bf r}_A,\omega_A)$ from
Eq.~(\ref{e3.27}). 

%%%%%%%%%%%%%%%%%%%%%%%%%%%%%%%%%%%%%%%%%%%%%%%%%%%%%%%%%%%%%%%%%%%

\subsection{Two-Atom Coupling}
\label{sec3.2}

We now turn to the problem of two atoms
(denoted by A and B) coupled through a 
medium-assisted electromagnetic field in the case 
of single-quantum excitation. For simplicity,
let us consider 
atoms with equal transition frequencies, so that
\begin{equation}
\label{e8.20b}
       K_{AA'}(t,t') \equiv K_{AA'}(t-t')
\end{equation}
($A'$ $\!=$ $\!B,D$) and
\begin{equation}
\label{e8.20c}
       K_{AB}(t-t') = K_{BA}(t-t'),
\end{equation}
and assume that the isolated atoms 
undergo the same decay law,
\begin{equation}
\label{e8.20a}
   K_{AA}(t,t') =  K_{BB}(t,t') \equiv K(t-t'). 
\end{equation}
Introducing the new variables
\begin{equation}
\label{e8.12}
      C_\pm (t) = 2^{-1/2}
      \Bigl[ C_{U_A}(t) \pm C_{U_B}(t) \Bigr],
\end{equation}
it is not difficult to prove that the integro-differential 
equations (\ref{e8.6b}) decouple as follows:
\begin{eqnarray}
\label{e8.17}
\lefteqn{       
       \dot{C}_\pm(t) = \int_{0}^t {\rm d}t'\, K_\pm(t-t')\,C_\pm(t')
}
\nonumber\\[.5ex]&& \hspace{-0ex} 
       +\,2^{-1/2} \int_{-\Delta t}^0 {\rm d}t'\,
       \left[K_{AD}(t-t')
       \right.
\nonumber\\[.5ex]&&\hspace{6ex} 
       \left.
       \pm\,K_{BD}(t-t')\right]C_{U_D}(t') ,
\qquad       
\end{eqnarray}
where
\begin{equation}
\label{e8.18}
       K_\pm(t-t') = K(t-t') \pm  K_{AB}(t-t').
\end{equation}
Obviously, the $C_{\pm}$ are the expansion coefficients
of the wave function with respect to the (atomic) basis 
\begin{equation}
\label{e8.14}
      |\pm \rangle = 2^{-1/2} 
      \left( |U_A\rangle \pm |U_B\rangle \right),
\end{equation}
and $|L\rangle$ (instead of the basis $|U_A\rangle$,
$|U_B\rangle$, and $|L\rangle$). Thus, they 
are the probability amplitudes of finding the total system in the 
states  $|+\rangle |\{0\}\rangle$ and $|-\rangle |\{0\}\rangle$, 
respectively. In the further treatment of Eq.~(\ref{e8.17})
one can again distinguish between the weak- and the strong-coupling
regime.

%%%%%%%%%%%%%%%%%%%%%%%%%%%%%%%%%%%%%%%%%%%%%%%%%%%%%%%%%%%%%%%%

\paragraph{Weak Coupling}

In the weak-coupling regime, the Markov approximation applies,
and in Eq.~(\ref{e8.17}) $C_\pm(t')$ 
can be replaced with $C_\pm(t)$,
with the time integrals being $\zeta$-functions.
In particular, when the field is initially not excited,
then the second term on the right-hand side of Eq.~(\ref{e8.17})
vanishes and we are left with a homogeneous first-order differential 
equation, whose solution is, in analogy to Eq.~(\ref{e3.15}),
\begin{equation}
\label{e8.13}
      C_\pm (t)
       = e^{(-\Gamma_\pm/2 +i \delta_\pm) t} C_\pm(0),
\end{equation}
where ($\Gamma$ $\!\equiv$ $\Gamma_{AA}$,
$\delta$ $\!\equiv$ $\delta_{AA}$)
\begin{equation}
\label{e8.9}
      \Gamma_\pm = \Gamma \pm \Gamma_{AB}\,,
\end{equation}
\begin{equation}
\label{e8.9a}
      \delta_\pm = \delta \pm \delta_{AB}\,, 
\end{equation}
\begin{equation}
\label{e8.10} 
     \Gamma_{AB} = {2\omega_A^2\over \hbar\varepsilon_0c^2}\, 
     {\bf d}_A {\rm Im}\,\mbb{G}({\bf r}_A,{\bf r}_B,\omega_A) {\bf d}_B \,,
\end{equation}
\begin{equation}
\label{e8.11} 
   \delta_{AB} \!=\! {{\cal P}\over \pi\hbar\varepsilon_0}\,
   \!\int_0^\infty \!\!{\rm d}\omega\,
   {\omega^2\over c^2} 
   \frac{{\bf d}_A {\rm Im}\,\mbb{G}({\bf r}_A,{\bf r}_B,\omega) {\bf d}_B}
   {\omega-\omega_A}\,. 
\end{equation}
Clearly, $\Gamma_\pm$
are the decay rates of the states $|\pm\rangle$, and  
the assumption (\ref{e8.20a}) means
that the two atoms are positioned in such a way that they have 
equal single-atom decay rates and line shifts.
Note that the values of $\Gamma_+$ and $\Gamma_-$ can
substantially differ from each other, because of the
interference term $\Gamma_{AB}$ (of positive or negative sign).   

%%%%%%%%%%%%%%%%%%%%%%%%%%%%%%%%%%%%%%%%%%%%%%%%%%%%%%%%%%%%%%%%%%%%

\paragraph{Strong Coupling}

In the strong-coupling regime, the atoms
are predominantly coupled (in a resonator-like equipment) 
to a sharp field resonance, whose mid-frequency approximately
equals the atomic transition frequency. As a result, the atomic probability
amplitudes in Eq.~(\ref{e8.17}) must not necessarily
be slowly varying compared with the kernel functions and
the Markov approximation thus fails in general.  
Regarding the line shape of the field resonance as being
a Lorentzian, one can of course approximate the kernels
$K(t-t')$, $K_{AB}(t-t')$ [and $K_{AD}(t-t')$ and $K_{BD}(t-t')$]
in a similar way as done in Eq.~(\ref{e6.2}) for a single atom.

Equation (\ref{e8.17}) reveals that the motion of the
states $|\pm\rangle$ defined by Eq.~(\ref{e8.14})
is governed by the kernel functions
$K_\pm(t-t')$, and it may happen that one of them becomes
very small, because of destructive interference 
[cf. Eq.~(\ref{e8.18})]. In that case, either
$|+\rangle$ or $|-\rangle$ is weakly coupled to the
field, and thus  
the strong-coupling regime cannot be realized for
both of these states simultaneously.

%%%%%%%%%%%%%%%%%%%%%%%%%%%%%%%%%%%%%%%%%%%%%%%%%%%%%%%%%%%%%%%%%%%
%%%%%%%%%%%%%%%%%%%%%%%%%%%%%%%%%%%%%%%%%%%%%%%%%%%%%%%%%%%%%%%%%%%

\setcounter{equation}{0}
\section{BULK MEDIUM}
\label{sec:realcavity}

The formalism outlined in Section \ref{sec:Gen_Theor}
requires knowledge of the permittivity as a function
of space and frequency. The spatial variation
is typically determined by some arrangement of macroscopic
bodies, each of which being characterized by a
permittivity that is a function of frequency only. 
The frequency response of a dielectric body depends on
its atomic structure and can be measured with high
precision. For theoretical studies it may be useful
to have some analytical expression at hand.     

%%%%%%%%%%%%%%%%%%%%%%%%%%%%%%%%%%%%%%%%%%%%%%%%%%%%%%%%%%%%%%%%%%%

\subsection{Drude--Lorentz Model}
\label{subsec:DLmodel}

\begin{figure}[ht]
\begin{center}
\mbox{\psfig{file=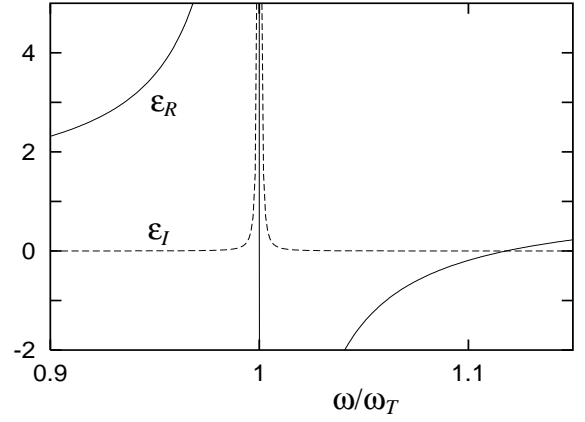,width=8.5cm}}
\end{center} 
\caption{\label{f1a}
Real and imaginary parts of the permittivity of a 
single-resonance Drude--Lorentz-type dielectric
for \mbox{$\omega_P$ $\!=$ $\!0.5\,\omega_T$} and
\mbox{$\gamma$ $\!=$ $\!10^{-4}\,\omega_T$}.
The band gap covers the interval from $\omega_T$ to 
\mbox{$\omega_L$ $\!\simeq$ $\!1.12\,\omega_T$}.
}
\end{figure}

In the Drude--Lorentz model, which is widely used in practice,
the permittivity is given by
\begin{equation}
\label{e4.5}
        \varepsilon(\omega) = 1 + \sum_\alpha
        {\omega_{P\alpha}^2 \over 
        \omega_{T\alpha}^2 - \omega^2 - i\omega \gamma_\alpha}\,,
\end{equation}
where $\omega_{T\alpha}$ and $\gamma_\alpha$ are the medium
oscillation frequencies and linewidths, respectively, and
$\omega_{P\alpha}$ correspond to the coupling constants. 
It is worth noting that the Drude--Lorentz model 
covers both dielectric (\mbox{$\omega_{T\alpha}$ $\!\neq$ $\! 0$}) and
metallic (\mbox{$\omega_{T\alpha}$ $\!=$ $\!0$}) matter.  
An example of the permittivity for a (single-resonance)
dielectric as a function of frequency is shown in Fig.~\ref{f1a}.
%%%%%%%%%%%%%%%%%%%%%%%%%%%%%%%%%%%%%%%%%%%%%%%%%%%%%
%%%%%%%%%%%%%%%%%%%%%%%%%%%%%%%%%%%%%%%%%%%%%%%%%%%%%
{F}rom the permittivity, the refractive index can
be obtained according to the relations 
\begin{equation}
\label{e4.5.1}
     n(\omega)=\sqrt{\varepsilon(\omega)}=n_R(\omega)
     + in_I(\omega),
\end{equation}
\begin{eqnarray}
\label{e4.5.1a}
\lefteqn{\hspace{-4ex}
n_{R(I)}(\omega) =
} 
\nonumber\\[.5ex]&&\hspace{-6ex}
     \sqrt{{1\over2}\left[
     \sqrt{\varepsilon_{R}^2(\omega)+\varepsilon_{I}^2(\omega)}
     +(-)\, \varepsilon_{R} (\omega)\right]} .
\end{eqnarray}

The Drude--Lorentz model
features band gaps between the transverse 
frequencies $\omega_{T\alpha}$ and the 
longitudinal frequencies \mbox{$\omega_{L\alpha}$ $\!=$ 
$\!\sqrt{\omega_{T\alpha}^2\!+\!\omega_{P\alpha}^2}$}.
Far from a medium resonance, we typically observe that
\begin{equation}
\label{e4.5.2}
\varepsilon_{I}(\omega) \ll |\varepsilon_{R}(\omega)|.
\end{equation}
For $\omega$ $\!<$ $\!\omega_{T\alpha}$ (outside a
band gap) we have
\begin{eqnarray}
\label{e4.5.3}
     &\varepsilon_{R}(\omega)>1,
\\&\displaystyle \hspace{-2ex}
     n_R(\omega)\simeq\sqrt{\varepsilon_{R}(\omega)}\gg
     n_I(\omega)\simeq {\varepsilon_{I}(\omega)\over
     2\sqrt{\varepsilon_{R}(\omega)}}\,,  
\nonumber\\
\label{e4.5.3a}
\end{eqnarray}
and for $\omega_{T\alpha}$ $\!<$ $\!\omega$ $\!<\omega_{L\alpha}$
(inside a band gap) 
\begin{eqnarray}
\label{e4.5.4}
     &\varepsilon_{R}(\omega)<0,
\\&\displaystyle \hspace{-1ex}
     n_R(\omega)\simeq {\varepsilon_{I}(\omega)\over
     2\sqrt{|\varepsilon_{R}(\omega)|}}\ll
     n_I(\omega)\simeq \sqrt{|\varepsilon_{R}(\omega)|}\,.  
\nonumber\\
\label{e4.5.5}
\end{eqnarray}

When (inside a band gap) 
\begin{equation}
\label{e4.5.6}
\varepsilon_{R}(\omega)<-1
\end{equation}
is valid, which, in view of Eq. (\ref{e4.5}), leads to
\begin{equation}
\label{e4.5.7}
     \omega<
     \sqrt{\omega_{T\alpha}^2
     +{\textstyle\frac{1}{2}}\,\omega_{P\alpha}^2} \,,
\end{equation}
then the Drude--Lorentz model also incorporates  
surface-guided waves
[see, e.g., \citet{Raether88,Ho01}],
which are observed in the presence of an interface.
These waves are bound to the interface, with the amplitudes 
being damped into either of the neighboring media. Typical
examples are surface phonon polaritons for dielectrics and
surface plasmon  polaritons for metals. 
Note that in any case \mbox{$\varepsilon_{I}(\omega)$ $\!>$ $\!0$}
is valid. 

%%%%%%%%%%%%%%%%%%%%%%%%%%%%%%%%%%%%%%%%%%%%%%%%%%%%%%%%%%%%%%%%%%%

\subsection{Local-Field Correction}
\label{subsec:LFcorrection}

{F}rom simple arguments
based on the change of the mode density, it was suggested that the
spontaneous emission rate of an atom inside a nonabsorbing medium
should be modified according to \mbox{$\Gamma$ $\!=$ $\!n\Gamma_0$},
where $n$ is the (real) refractive index
of the medium and $\Gamma_0$ is given by Eq.~(\ref{e3.18})
[\citet{Dexter56,diBartolo68,Yariv75,Nienhuis76}].
In this formula, it is assumed that the local field the atom
interacts with
is the medium-assisted electromagnetic field
obtained by averaging over a region which contains a great
number of medium constituents. In reality, the atom is located
in a small region of free space, and the field therein
differs from the averaged field. This effect is usually taken
into consideration by introduction of a local-field correction
factor $\xi$, thus
\begin{equation}
\label{e4.1b}
\Gamma  = n \xi \Gamma_0\,.
\end{equation}
In the (Clausius--Mosotti) virtual-cavity model it
is given by [\citet{Knoester89,Milonni95}]
\begin{equation}
\label{e4.1a}
      \xi  = \left( \frac{n^2+2}{3} \right)^2,
\end{equation}
and in the (Onsager) real-cavity model by [\citet{Glauber91}]
\begin{equation}
\label{e4.1}
      \xi  = \left( \frac{3n^2}{2n^2+1} \right)^2. 
\end{equation}

For absorbing media it was suggested that the 
index of refraction should be replaced by its real part and
the square of the correction factor in Eqs.~(\ref{e4.1a}) 
and  (\ref{e4.1}) by the absolute square
[\citet{Barnett92,Barnett96,Juzeliunas97}].
Later it was found, on the basis of the quantization scheme
outlined in Section \ref{sec:quan}, that a proper inclusion of
the (quantum) noise polarization leads to a more 
complicated form of the local-field correction
[\citet{Scheel99a,Scheel99b}], which (for the virtual-cavity
model) was confirmed by an alternative, microscopic 
approach [\citet{Fleischhauer99}].

In contrast to the virtual cavity model where the modification
of the field outside the cavity is disregarded,
in the real-cavity model the mutual modification of the field
outside and inside the cavity are taken into account in a
consistent way. Experiments suggested that the real-cavity model 
is suitable for describing the decay of 
substitutional guest atoms different from the constituents 
of the medium [\citet{Rikken95,deVries98,Schuurmans98}]. 

Let us consider an excited atom placed at the center of an empty,
spherical cavity (embedded in an otherwise homogeneous medium).
According to Eq.~(\ref{e3.17a}) and the Green tensor given in 
Appendix \ref{app:spherical}, the
decay rate can be given in the form of [\citet{Scheel99b}]    
\begin{equation}
\label{e4.2}
\Gamma = \Gamma_0 \left[ 1+{\rm Re}\, C_1^N(\omega_A) \right],
\end{equation}
where the generalized reflection coefficient $C_1^N(\omega)$
reads [$\tilde{k}$ $\!=$ $\!\tilde{k}(\omega)$ $\!=$ $R\omega/c$] 
\begin{eqnarray}
\label{e4.3}
\lefteqn{\hspace{-1ex}
    C_1^N(\omega) = e^{i\tilde{k}}
    \bigl\{i+\tilde{k}[n(\omega)\!+\!1]-i\tilde{k}^2n(\omega)
}
\nonumber \\ &&\hspace{-2ex}
    -\,\tilde{k}^3n^2(\omega)/[n(\omega)\!+\!1] \bigr\}
    \bigl\{ \sin \tilde{k}
    \!-\! \tilde{k}[\cos \tilde{k}
\nonumber \\ &&\hspace{-2ex}  
    +\,in(\omega)\sin \tilde{k}]
    \!+\!i\tilde{k}^2n(\omega)\cos \tilde{k}
    \!-\!\tilde{k}^3[\cos \tilde{k}
\nonumber \\ &&\hspace{-2ex} 
   -\,in(\omega)\sin \tilde{k}] 
    n^2(\omega)/[n^2(\omega)\!-\!1] \bigr\}^{-1}\!\!.
\qquad    
\end{eqnarray}
As long as the surrounding medium can be treated as a continuum,
Eq.~(\ref{e4.2}) [together with Eq.~(\ref{e4.3})] 
is exact. It is valid for arbitrary cavity radius and arbitrary
complex refractive index, without restriction to transition
frequencies far from medium resonances.

When the cavity radius is much smaller than the wavelength
of the atomic transition, i.e., $R\omega_A/c$ $\!=$
$\!\tilde{k}(\omega_A)$ $\!\ll$ $\!1$, then the real-cavity
model of local-field correction is realized. In this case,
$C_1^N(\omega_A)$ can be expanded in powers of
$\tilde{k}(\omega_A)$ to obtain [\citet{Scheel99b,Tomas01}]
\begin{eqnarray}
\label{e4.4}
\lefteqn{
     \Gamma = \Gamma_0 
     \left|\frac{3\varepsilon}
     {2\varepsilon\!+\!1}\right|^2
     \Bigg\{ n_R
}
\nonumber \\[.5ex] && \hspace{-3ex}
     \!+\frac{\varepsilon_{I}}
     {|\varepsilon|^2}
     \biggl[
     \left(\!\frac{c}{\omega_A R}\!\right)^3
     \!+\!\frac{28|\varepsilon|^2 
     \!+\!16\varepsilon_{R}\!+\!1}
     {5|2\varepsilon+1|^2}
     \!\left(\!\frac{c}{\omega_A R}\!\right)
\nonumber \\[.5ex] && \hspace{-3ex}
     -\frac{2}{|2\varepsilon\!+\!1|^2}
     \bigl( 2n_I|\varepsilon|^2
     +\,n_I\varepsilon_{R}
     \!+\!n_R \varepsilon_{I}\bigr)
     \biggr]\Biggr\} 
\nonumber \\[.5ex] && 
     +\,O(\omega_AR/c),
\end{eqnarray}
where the dependence of the permittivity $\varepsilon$ and
the refractive index $n$ on $\omega_A$ has been 
suppressed.
For $\varepsilon_{I}(\omega_A)$ $\!=$ $\!0$, i.e, when 
material absorption is fully disregarded, Eq. (\ref{e4.4})
reproduces exactly local-field correction factor 
(\ref{e4.1}). The second term in the curly
brackets essentially results from absorption. 
It is seen that material absorption gives   
rise to a strong dependence of the decay rate on the cavity
radius. In particular, the leading term proportional to $R^{-3}$ 
can be regarded as corresponding to nonradiative
energy transfer from the atom to the medium.
%%%%%%%%%%%%%%%%%%%%%%%%%%%%%%%%%%%%%%%%
\begin{figure}[ht]
\begin{center}
\mbox{\psfig{file=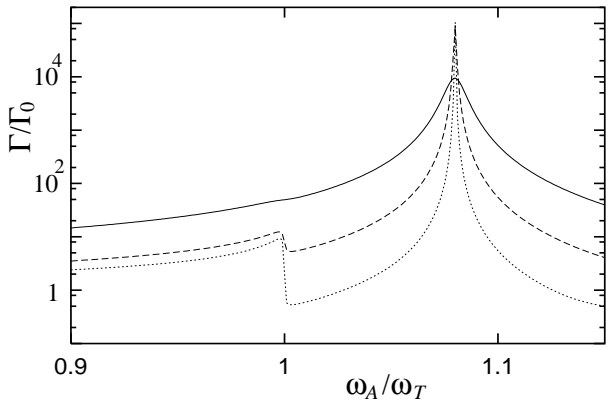,width=8.5cm}} 
\end{center} 
\caption{\label{fig:se}
Spontaneous decay rate
[Eq.~(\protect\ref{e4.2})]
of an atom embedded in a dielectric medium 
as a function of the atomic transition frequency 
near a medium resonance for
a single-resonance Drude--Lorentz-type dielectric
[\mbox{$\omega_{\rm P}$ $\!=$ $\!0.5\,\omega_T$}; 
$\gamma/\omega_T\!=\!$
$10^{-2}$ (solid line), $10^{-3}$ (dashed line), and 
$10^{-4}$ (dotted line)].
The (real-) cavity radius is $R\!=\!0.02\,\lambda_T$.
}
\end{figure}
%%%%%%%%%%%%%%%%%%%%%%%%%%%%%%%%%%%%%%%%
Examples of the dependence of rate of spontaneous decay on the atomic
transition frequency are plotted in Fig.~\ref{fig:se} for a  
Drude--Lorentz-type dielectric medium. It is seen that
in the band-gap region (where for a nonabsorbing medium
spontaneous emission would be inhibited) the decay rate can
drastically increase, because of the non-radiative decay channel
associated with absorption. Note that the strongest
enhancement of spontaneous decay is observed at
$\omega_A$ $\!\simeq$ $\!\sqrt{\omega_T^2\!+\!
\frac{3}{2}\omega_P^2}$, which [for small values of
$\varepsilon_I(\omega_A)$] corresponds to
$2\varepsilon(\omega_A)$ $\!+$ $\!1$ $\!\simeq$ $\!0$.

%%%%%%%%%%%%%%%%%%%%%%%%%%%%%%%%%%%%%%%%%%%%%%%%%%%%%%%%%%%%%
%%%%%%%%%%%%%%%%%%%%%%%%%%%%%%%%%%%%%%%%%%%%%%%%%%%%%%%%%%%%%%

\setcounter{equation}{0}
\section{PLANAR SURFACE}
\label{sec:planar}

Let us turn to the problem of
spontaneous decay of an excited atom located
near the surface of a half-space medium.
For real permittivity,
configurations of that type have been 
studied extensively in connection with Casimir and
van der Waals forces [see, e.g., \citet{Meschede90,Fichet95}
and references therein] and
with regard to scanning near-field optical
microscopy [see, e.g., \citet{Henkel98}].

To be more specific, let us consider two infinite half-spaces
such that
\begin{equation}
\label{e5.1a}
\varepsilon({\bf r},\omega) = \left\{
\begin{array}{l@{\quad \rm if \quad }l}
\varepsilon(\omega) & z \leq 0\\
1 & z > 0
\end{array}
\right..
\end{equation}
For $z$ $\!>$ $\!0$, the 
reflection part of the Green tensor
reads [\citet{Maradudin75,Mills75,Tomas95,Ho98}] 
\begin{eqnarray}
\label{e5.1}
\lefteqn{\hspace{-1ex}
     G^R_{xx}(z,z,\omega)
     = -\frac{i}{8\pi k^2}\!\int_0^\infty
     \!\!{\rm d}k_\| \,k_\|\beta\, {\rm e}^{2i\beta z} r^p(k_\|)
}
\nonumber\\&& \hspace{6ex}
    +\,\frac{i}{8\pi} \!\int_0^\infty \!\!{\rm d}k_\| \,
    \frac{k_\|{\rm e}^{2i\beta z}}{\beta} \,r^s(k_\|),
\qquad    
\end{eqnarray}
\begin{equation}
\label{e5.2}
     G^R_{yy}(z,z,\omega)= G^R_{xx}(z,z,\omega),
\end{equation}
\begin{equation}
\label{e5.3}
     G^R_{zz}(z,z,\omega) = 
     \frac{i}{4\pi k^2} \int_0^\infty \!{\rm d}k_\| \, k_\|^3
     \frac{{\rm e}^{2i\beta z}}{\beta} \,r^p(k_\|)
\end{equation}
[$k$ $\!=$ $\!\omega/c$, $\beta$ $\!=$ $\!(k^2$ $\!-$
$\!k_\|^2)^{1/2}$], where $r^p(k_\|)$ and $r^s(k_\|)$
are respectively the familiar Fresnel reflection coefficients
for $p$- (TM) and $s$- (TE) polarized waves. Substitution
of these expressions
($\omega$ $\!=$ $\!\omega_A$) into Eqs.~(\ref{e3.17a}) and
(\ref{e3.14}) yields
the decay rate and the line shift of an atom at a distance
$z$ from the surface [\citet{Agarwal75,Agarwal77,
Scheel99c}], which are in agreement
with classical results [see, e.g., \citet{Chance78} and
references therein].

When the distance of the atom from the surface is small
compared to the wavelength, $kz$ $\!\ll$ $\!1$, then
the integrals in Eqs. (\ref{e5.1}) -- (\ref{e5.3}) can be evaluated
asymptotically to give [\citet{Scheel99c}]
\begin{eqnarray}
\label{e5.4}
\lefteqn{\hspace{-3ex}    
     G^R_{zz}(z,z,\omega)
     =\frac{1}{16\pi k^2z^3}\,\frac{n^2(\omega)-1}{n^2(\omega)+1}
}
\nonumber\\[.5ex]&& \hspace{-6ex}
     +\,\frac{1}{8\pi z}\,\frac{[n(\omega)-1]^2}{n(\omega)[n(\omega)+1]}
\nonumber\\[.5ex]&& \hspace{-6ex}
     +\,\frac{ik}{12\pi}\frac{[n(\omega)-1][2n(\omega)-1]}
     {n(\omega)[n(\omega)+1]}
          + O(kz),  \hspace{2ex}
\end{eqnarray}
\begin{eqnarray}
\label{e5.5} 
\lefteqn{  \hspace{-8ex}     
      G^R_{xx}(z,z,\omega)\!=\!{\textstyle\frac{1}{2}}G^R_{zz}(z,z,\omega)
      \!-\!\frac{1}{16\pi z}\,\frac{n^2(\omega)\!-\!1}{n^2(\omega)\!+\!1}
}
\nonumber\\[.5ex]&&
      -\frac{ik}{3\pi}\frac{n(\omega)-1}{n(\omega)+1} + O(kz),  
\end{eqnarray}
\begin{equation}
\label{e5.6}
     G^R_{yy}(z,z,\omega)= G^R_{xx}(z,z,\omega).
\end{equation}
Inserting Eqs.~(\ref{e5.4}) -- (\ref{e5.6}) into
Eq.~(\ref{e3.17a})
yields 
\begin{eqnarray}
\label{e5.7}
\lefteqn{
     \Gamma = {3\Gamma_0\over 8}
     \left(1+{d_z^2\over d^2} \right)
     \left({c\over \omega_Az}\right)^3
}
\nonumber\\[.5ex]&&\times\,  
     {\varepsilon_{I}(\omega_A)
     \over |\varepsilon(\omega_A)+1|^2}  
     + O\!\left({c \over \omega_Az }\right)\!.
\end{eqnarray}
The leading (\mbox{$\sim$ $\!z^{-3}$}) term
is the same as in the
microscopic approach by \citet{Yeung96}.
This term, which is proportional to $\varepsilon_{I}(\omega_A)$,
is closely related to nonradiative decay, i.e.,
energy transfer from the atom to the medium.
Obviously, a change of $\varepsilon_{I}(\omega_A)$ 
mostly affects the near-surface behavior of the decay rate.
Note that the distance of the
atom from the surface must not
be smaller than interatomic distances
in the medium (otherwise
an interface cannot be defined).
%%%%%%%%%%%%%%%%%%%%%%%%%%%%%%%%%%%%%%%%%%%%%%%%%%%%%%%%%%%%%%%%%%
\begin{figure}[ht]
\begin{center}
\mbox{\psfig{file=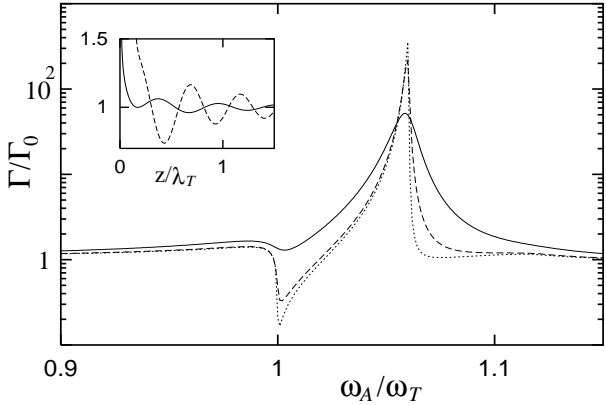,width=8.5cm}} 
\end{center} 
\caption{\label{fig:pl}
The rate of spontaneous decay of an excited atom 
near a planar dielectric half-space
is shown as a function of the transition frequency
for an $x$-oriented
transition dipole moment and a single-resonance
Drude--Lorentz-type dielectric
[\mbox{$\omega_{\rm P}$ $\!=$ $\!0.5\,\omega_T$}; 
\mbox{$\gamma/\omega_T$ $\!=$ $\!10^{-2}$} (solid line),
$10^{-3}$ (dashed line), and 
$10^{-4}$ (dotted line); \mbox{$z$ $\!=$ $\!0.05\,\lambda_T$}].
The inset illustrates the dependence on the distance 
of the decay rate [$\gamma/\omega_T\!=\!$ $10^{-3}$; 
$\omega_A/\omega_T\!=\!$ $0.9$ (solid line),  
$1.06$ (dashed line)].
}
\end{figure}
%%%%%%%%%%%%%%%%%%%%%%%%%%%%%%%%%%%%%%%%%%%%%%%%%%%%%%%%%%%%%%%%%%%%
Examples of the spontaneous decay rate are shown in 
Fig.~\ref{fig:pl} for dielectric matter of Drude--Lorentz type.
Note the strong absorption-assisted
enhancement of spontaneous decay that is observed
inside the band gap at \mbox{$\omega_A$ $\!\simeq$
$\!\sqrt{\omega_T^2\!+\!\frac{1}{2}\omega_P^2}$}. It
corresponds [for small values of $\varepsilon_I(\omega_A)$]
to the condition \mbox{$\varepsilon(\omega_A)$
$\!+$ $\!1$ $\!\simeq$ $\!0$}, which marks 
the position of the highest density of the 
surface-guided waves [cf. Eq.~(\ref{e4.5.7})].

Similarly, from Eq.~(\ref{e3.14}) together with
Eqs.~(\ref{e5.4}) -- (\ref{e5.6}), the line shift due
to the presence of the macroscopic body reads
\begin{eqnarray}
\label{e5.9}
\lefteqn{
     \delta\omega_A = {3\Gamma_0\over 32} 
     \left(1+{d_z^2\over d^2} \right)
     \left({ c \over \omega_Az}\right)^3 
}
\nonumber\\[.5ex]&&\times\,
     {|\varepsilon(\omega_A)|^2-1
     \over |\varepsilon(\omega_A)+1|^2}
     + O\!\left({c \over \omega_Az }\right)\!.    
\end{eqnarray}
In contrast to the decay rate, here the leading
(\mbox{$\sim$ $\!z^{-3}$}) term even appears when absorption is
disregarded [\mbox{$\varepsilon_{I}(\omega_A)$ $\!=$ $\!0$}].

%%%%%%%%%%%%%%%%%%%%%%%%%%%%%%%%%%%%%%%%%%%%%%%%%%%%%%%%%%%%%%%%%%%%%%
%%%%%%%%%%%%%%%%%%%%%%%%%%%%%%%%%%%%%%%%%%%%%%%%%%%%%%%%%%%%%%%%%%%%%%

\setcounter{equation}{0}
\section{SPHERICAL MICRO- \\ RESONATOR}
\label{sec:beyond}

In Section \ref{subsec:LFcorrection}, an atom in a microsphere
whose radius is much smaller than the wavelength of the atomic
transition was considered. If the radius is not small compared
with the wavelength, the cavity can act as a resonator. It is
well known that the spontaneous decay of an excited atom can be
strongly modified when it is placed in a microresonator 
[\citet{Hinds91,Haroche92,Meschede92,Meystre92,Berman94,Kimble98}].
There are typically two qualitatively different regimes:
the weak-coupling regime and the 
strong-coupling regime. In the weak-coupling regime the Markov
approximation applies and a monotonous exponential decay is observed,
the decay rate being enhanced or reduced compared
to the free-space value depending on whether the atomic transition
frequency fits a cavity resonance or not. The strong-coupling regime,
in contrast, 
is characterized by reversible Rabi oscillations where the energy of the
initially excited atom is periodically exchanged between the atom
and the radiation field. This usually requires that the emission
is in resonance with a high-quality cavity mode. 

%%%%%%%%%%%%%%%%%%%%%%%%%%%%%%%%%%%%%%%%%%%%%%%%%%%%%%%%%%%%%%%%%%%%
\begin{figure}[ht]
 \begin{center}
\mbox{\psfig{file=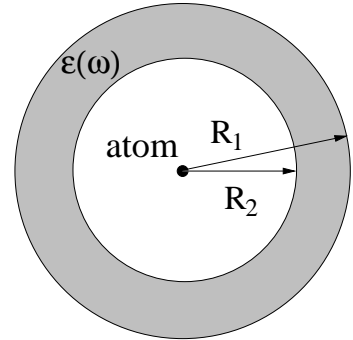,width=4.5cm}}
 \end{center}
\caption{\label{fig:slabho} Scheme of the spherical microresonator.}
\end{figure}
%%%%%%%%%%%%%%%%%%%%%%%%%%%%%%%%%%%%%%%%%%%%%%%%%%%%%%%%%%%%%%%%%%%%
%
Let us consider an excited atom placed at the 
center of a spherical three-layer structure (Fig.~\ref{fig:slabho}).
The outer layer ($r$ $\!>$ $\!R_1$) and the inner layer
\mbox{($0$ $\!\le$ $\!r$ $\!<$ $\!R_2$)} 
are vacuum, whereas the middle layer 
($R_2$ $\!\le$ $\!r$ $\!\le$ $\!R_1$), which plays the role of 
the resonator wall, is matter. In particular for a
Drude--Lorentz-type dielectric, the wall would be perfectly
reflecting in the band-gap zone,
provided that absorption could be disregarded. 
Restricting our attention to a true resonator,
we may assume that the condition
$R_2\omega_A/c\!\gg\! 1$ is satisfied.

%%%%%%%%%%%%%%%%%%%%%%%%%%%%%%%%%%%%%%%%%%%%%%%%%%%%%%%%%%%%%%%%%%%%

\subsection{Weak Coupling}

{F}rom Eq.~(\ref{e3.12}) together with the Green tensor 
for a spherical three-layer structure as given in
Appendix \ref{app:spherical}, 
the decay rate becomes [\citet{Ho00}] 
\begin{eqnarray}
\label{e6.1}
    \Gamma
\hspace{-1ex}&\simeq&\hspace{-1ex}
    \Gamma_0 \,{\rm Re}\! \left[
    \frac{n(\omega_A)-i\tan(\omega_AR_2/c)}
    {1-in(\omega_A)\tan(\omega_AR_2/c)} 
    \right]
\nonumber \\[.5ex] \hspace{-1ex}&=&\hspace{-1ex}
    \Gamma_0 \,
    n_R(\omega_A) [1+\tan^2(\omega_AR_2/c)]
\nonumber \\[.5ex] \hspace{-1ex}&&\hspace{-1ex}\times\,
    \Bigl\{
    [1+n_I(\omega_A) \tan(\omega_AR_2/c)]^2
\nonumber \\[.5ex] \hspace{-1ex}&&\hspace{1ex}
    +\,n_R^2(\omega_A) \tan^2(\omega_AR_2/c)
    \Bigr\}^{-1}\!\!.
\end{eqnarray}
Note that in Eq.~(\ref{e6.1}) it is assumed that 
\mbox{$\exp[-in_I(\omega_A)(R_1-R_2)\omega_A/c]$ $\!\ll$ $\!1$}
(thick cavity wall).

%%%%%%%%%%%%%%%%%%%%%%%%%%%%%%%%%%%%%%%%%%%%%%%%%%%%%%%%%%%%%%%%%%%%%%%%%%
\begin{figure}[ht]
\begin{center}
\mbox{\psfig{file=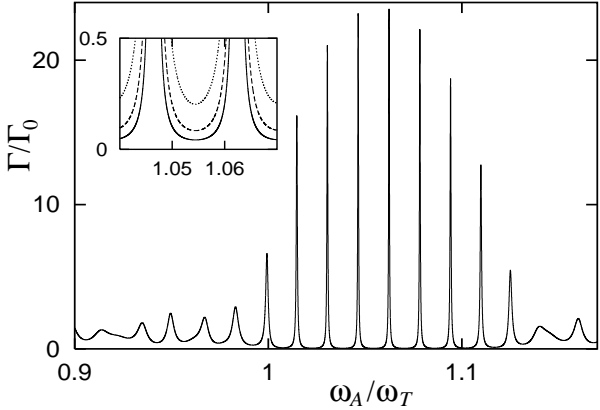,width=8.5cm}}
\end{center}
\caption{\label{fig:ho1}
The rate of spontaneous decay of an excited atom 
at the center of a spherical microresonator
is shown as a function of the transition frequency
for a single-resonance Drude--Lorentz-type dielectric
wall (\mbox{$R_2$ $\!=$ $\!30\,\lambda_T$};
\mbox{$R_1$ $\!-$ $\!R_2$ $\!=$ $\!\lambda_T$};
\mbox{$\omega_{\rm P}$ $\!=$ $\!0.5\,\omega_T$};
\mbox{$\gamma$ $\!=$ $\!10^{-2}\omega_T$}).
The curves in 
the inset correspond to \mbox{$\gamma/\omega_T$ $\!=$
$\!10^{-2}$} (solid line), $2\times 10^{-2}$ (dashed line),
and \mbox{$5\times 10^{-2}$ (dotted line)}.
[After \citet{Ho00}.]
}
\end{figure}
%%%%%%%%%%%%%%%%%%%%%%%%%%%%%%%%%%%%%%%%%%%%%%%%%%%%%%%%%%%%%%%%%%%%%%%%%%
The dependence on the transition frequency of the decay rate
is illustrated in Fig.~\ref{fig:ho1}. It is seen that
the decay rate very sensitively depends on
the transition frequency. Narrow-band enhancement of spontaneous decay
($\Gamma/\Gamma_0$ $\!>$ $\!1$) alternates with broadband inhibition
($\Gamma/\Gamma_0$ $\!<$ $\!1$). The frequencies at which the maxima of
enhancement are observed correspond to the resonance frequencies
of the cavity. Within the band gap the heights and widths of the
frequency intervals in which spontaneous decay is feasible are
essentially determined by material absorption. Outside the
band-gap zone the change of the decay rate is less pronounced,
because of the relatively large input-output coupling, the
(small) material absorption being of secondary importance. 

%%%%%%%%%%%%%%%%%%%%%%%%%%%%%%%%%%%%%%%%%%%%%%%%%%%%%%%%%%%%%%%%%%%%%%
\begin{figure}[ht]
\begin{center}
\mbox{\psfig{file=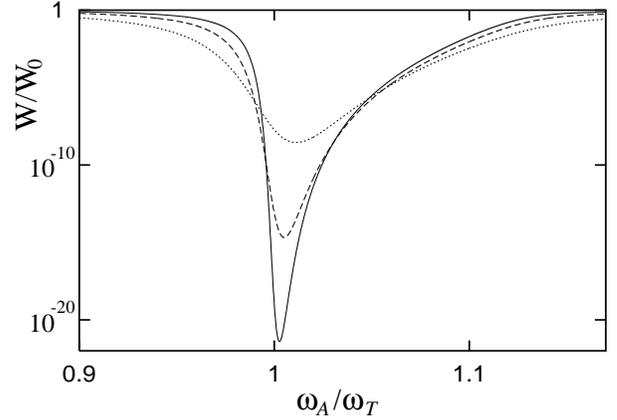,width=8.5cm}} 
\end{center} 
\caption{\label{fig3-1}
The amount of the (outside the resonator) available
radiation energy after spontaneous decay of an excited atom 
at the center of a spherical microresonator is shown as a function
of the transition frequency for a single-resonance
Drude--Lorentz-type dielectric wall
[\mbox{$\gamma/\omega_T\!=\!10^{-2}$} (solid line),  
$2\times 10^{-2}$ (dashed line), $5\times 10^{-2}$ (dotted line);
the other parameters are the same as in Fig.~\protect\ref{fig:ho1}].
[After \citet{Ho00}.]
}
\end{figure}
%%%%%%%%%%%%%%%%%%%%%%%%%%%%%%%%%%%%%%%%%%%%%%%%%%%%%%%%%%%%%%%%%%%%%%
The widths of the resonance lines are responsible for the damping
of the corresponding intracavity field (mode). There are two
damping mechanisms:
photon leakage to the outside of the cavity and 
photon absorption by the cavity-wall material.
The first mechanism is the dominant one outside bands in regions where
normal dispersion \mbox{(${\rm d}n_R/{\rm d}\omega$ $\!>$ $\!0$)} 
is observed, while the latter dominates inside band gaps where 
anomalous dispersion \mbox{(${\rm d}n_R/{\rm d}\omega$ $\!<$ $\!0$)} 
is observed. To illustrate this in more detail,
let us consider the total
amount of radiation energy observed outside the cavity and compare
it with the energy $W_0$ $\!=$ $\!\hbar\omega_{\rm A}$
emitted by an atom in free space.
Application of Eq.~(\ref{e3.29}) [together with
Eqs.~(\ref{e3.26}) and (\ref{e3.28})] yields [\citet{Ho00}]
\begin{equation}
\label{E41C}
        {W\over W_0}  
        \simeq \frac{ |{\cal A}^N_l(\omega_{\rm A})|^2}
        {1+{\rm Re}\, {\cal C}^N_l(\omega_{\rm A}) }\,,
\end{equation}
with ${\cal A}^N_l(\omega_{\rm A})$ and
${\cal C}^N_l(\omega_{\rm A})$ being given according
to Eqs.~(\ref{A.25}) and (\ref{A.26}) respectively.
Examples of the dependence of $W/W_0$ on the atomic transition 
frequency are plotted in Fig.~\ref{fig3-1}. It is seen
that inside the band gap most of the energy emitted by the
atom is absorbed by the cavity wall in the course of time, 
while outside the band gap the absorption is (for the chosen
values of $\gamma$) much less significant. Note that with
increasing value of $\gamma$ the band gap is smoothed a little bit,
and thus the fraction of light that escapes from the cavity
can increase.  

%%%%%%%%%%%%%%%%%%%%%%%%%%%%%%%%%%%%%%%%%%%%%%%%%%%%%%%%%%%%%%%%

\subsection{Strong Coupling}
\label{subsec:st_coupling}

When the coupling between the atom and a cavity resonance 
(mid-frequency $\omega_C$, line width $\Delta \omega_C$)
is so strong that the (weak-coupling) decay rate becomes comparable 
to the cavity line width, \mbox{$\Gamma_C$ $\!\gsim$ $\!\Delta \omega_C$}, 
the Markov approximation is no longer adequate. In this case,
the integral equation (\ref{e3.8}) or approximate equations of the
type (\ref{e6.3}) and (\ref{e6.5}) should be used in order to 
describe the temporal evolution of the (upper) atomic state.
For the configuration under investigation, the cavity line width
can be given by
\begin{equation}
\label{e6.4}
   \Delta\omega_C = \frac{c\Gamma_0}{R_2\Gamma_C} \,.
\end{equation}
Inside a band gap, $\Gamma_C$ is essentially
determined by material absorption. In particular, the
single-resonance Drude--Lorentz model reveals that  
\begin{eqnarray}
\label{e6.4a}
\Gamma_C
\hspace{-1ex}&\simeq&\hspace{-1ex}
\Gamma_0 \,\frac{n_I^2(\omega_C)+1}{n_R(\omega_C)}
\nonumber\\[.5ex]
\hspace{-1ex}&\simeq&\hspace{-1ex}
\Gamma_0 \,\frac{2\sqrt{(\omega_L^2-\omega_C^2)(\omega_C^2-\omega_T^2)}}
{\gamma\omega_C}
\qquad
\end{eqnarray}
($\gamma$ $\!\ll$ $\!\omega_T$, $\!\omega_P$,
$\!\omega_P^2/\omega_T$).
Below the band gap radiative losses dominate and $\Gamma_C$
reads 
\begin{equation}
\label{e6.4b}
\Gamma_C \simeq \Gamma_0 \,n_R(\omega_C)
   \simeq \Gamma_0 \,\sqrt{\frac{\omega_L^2-\omega_C^2}
   {\omega_T^2-\omega_C^2}}
\end{equation}
($n_R$ $\!\gg$ $\!n_I$).
%
%%%%%%%%%%%%%%%%%%%%%%%%%%%%%%%%%%%%%%%%%%%%%%%%%%%%%%%%%%%%%%%%%%%%%%%%
\begin{figure}[ht]
\begin{center}
\mbox{\psfig{file=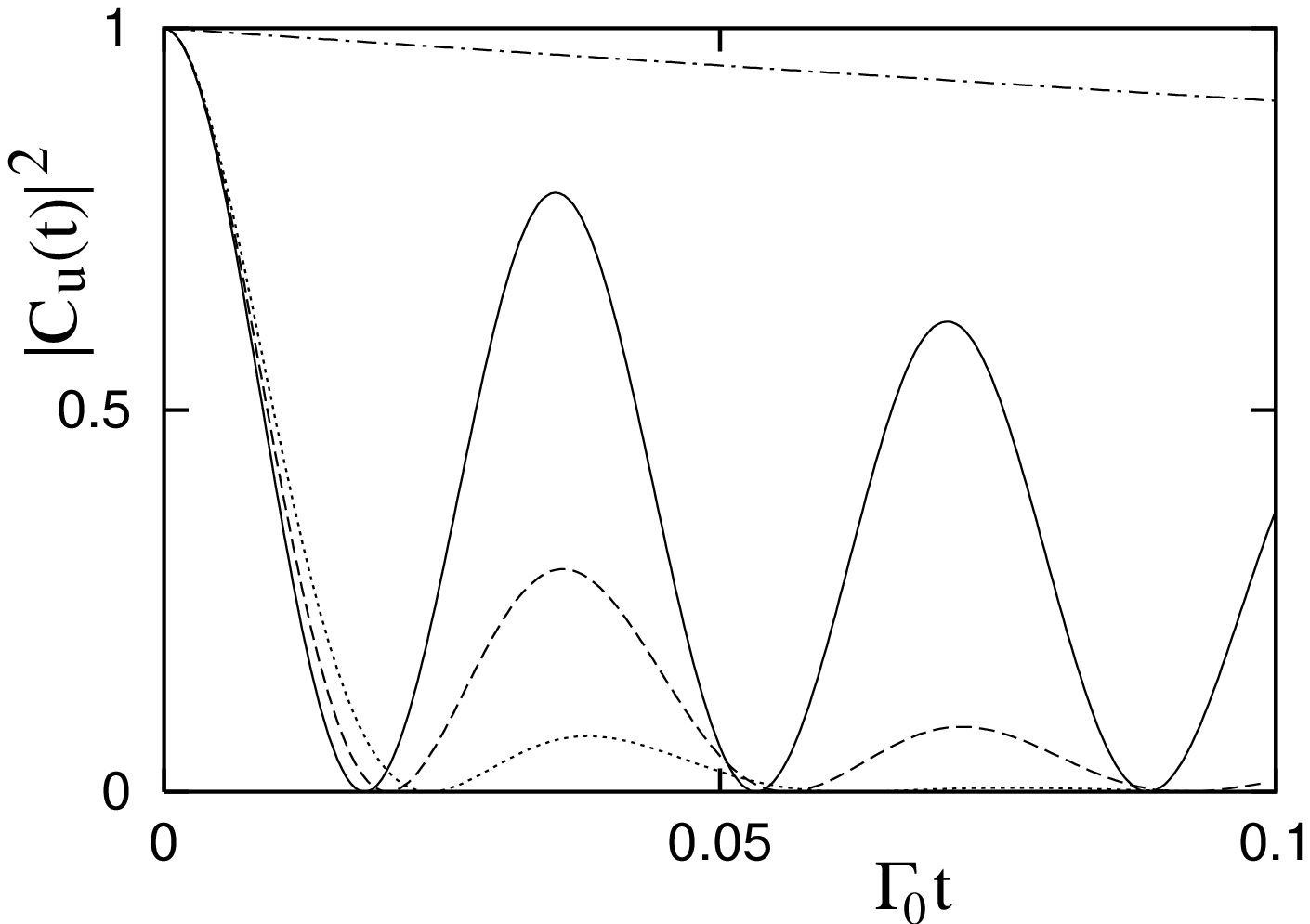,width=8.5cm}}
\end{center}
\caption{\label{fig:ho2}
The temporal evolution of the occupation probability
of the upper state of an initially excited atom 
at the center of a
spherical microresonator is shown for a single-resonance
Drude--Lorentz-type dielectric wall 
[\mbox{$R_2$ $\!=$ $\!30\,\lambda_T$};
\mbox{$R_1$ $\!-$ $\!R_2$ $\!=$ $\!\lambda_T$}; 
\mbox{$\omega_{\rm P}$ $\!=$ $\!0.5\,\omega_T$};
\mbox{$\omega_A$ $\!=$ $\!1.046448\,\omega_T$};
\mbox{$\Gamma_0\lambda_T/(2c)$ $\!=$ $\!10^{-6}$};
\mbox{$\gamma/\omega_T$ $\!=$ $\!10^{-4}$} (solid line),
$5\times 10^{-4}$ (dashed line), $10^{-3}$ (dotted line)].
For comparison, the exponential decay in free space is shown
(dashed-dotted line). [After \citet{Ho00}.]}
\end{figure}
%%%%%%%%%%%%%%%%%%%%%%%%%%%%%%%%%%%%%%%%%%%%%%%%%%%%%%%%%%%%%%%%%%%%%%%%%

Typical examples of the time evolution of
the upper-state occupation probability are shown in
Fig.~\ref{fig:ho2}. The curves are the exact (numerical) solutions
of the integral equation (\ref{e3.8}) [together with the kernel
function (\ref{e3.19})] for a (single-resonance) dielectric wall
of Drude--Lorentz type. 
The figure shows that with increasing value of the
intrinsic absorption constant $\gamma$ of the wall material
the Rabi oscillations become less pronounced. 
Clearly, larger values of
$\gamma$ mean enlarged absorption probability of
the emitted photon by
the cavity wall and thus reduced probability of 
atom-field energy interchange.

%%%%%%%%%%%%%%%%%%%%%%%%%%%%%%%%%%%%%%%%%%%%%%%%%%%%%%%%%%%%%%%%%%%%%%
%%%%%%%%%%%%%%%%%%%%%%%%%%%%%%%%%%%%%%%%%%%%%%%%%%%%%%%%%%%%%%%%%%%%%%

\setcounter{equation}{0}
\section{MICROSPHERE}
\label{sec:microsphere}

Light propagating in a dielectric sphere can be trapped by repeated
total internal reflections. When the round-trip optical path
fits integer numbers of the wavelength, whispering gallery (WG)
waves are formed, which combine extreme photonic confinement with 
very high quality factors [\citet{Collot93,Chang96,Gorodetsky96,
Vernooy98b,Uetake99}] -- properties that are crucial for cavity QED 
experiments [\citet{Lin92,Barnes96,
Lermer98,Vernooy98a,Fujiwara99,Yukawa99}] 
and certain optoelectronical applications [\citet{Chang96}].
WG waves are commonly classified by means of three numbers
[\citet{Collot93,Chang96}]: the angular-momentum number $l$,
the azimuthal number $m$, and the number $i$ of radial maxima of
the field inside the sphere. In the case of a uniform sphere,
the WG waves are \mbox{($2l$ $\!+$ $\!1$)}-fold degenerate, i.e.,
the \mbox{$2l$ $\!+$ $\!1$} azimuthal resonances
belong to the same frequency $\omega_{l,i}$.

A dielectric microsphere whose permittivity is
of Drude--Lorentz type does not only give
rise to WG waves, but can also feature surface-guided (SG) waves
inside band-gap regions.\footnote{For the dependence on frequency
   of the quality factors of WG and SG waves, see \citet{Ho01}.}
In contrast to WG
waves, each angular-momentum number $l$ is associated
with only one SG wave. 

If an excited atom is situated near a dielectric
microsphere, spontaneous decay sensitively depends
on whether or not the transition is tuned to a WG or an
SG resonance. Moreover, whereas WG waves typically suffer from
material absorption, the effect of material absorption on
SG waves is weak in general.
     
%%%%%%%%%%%%%%%%%%%%%%%%%%%%%%%%%%%%%%%%%%%%%%%%%%%%%%%%%%%%%%%%%%%

\subsection{Decay Rate}
\label{sec7.1}

Applying Eq.~(\ref{e3.17a}) together with the Green tensor
of a microsphere (Appendix \ref{app:spherical}),
the spontaneous decay rate for a (with respect to the
sphere) radially oriented dipole moment can be given by
\begin{eqnarray}
\label{e7.1}
\lefteqn{
      \Gamma^\perp =\Gamma_0 \biggl\{ 1 
      +{\textstyle{3\over 2}}
      \sum_{l=1}^\infty
      \biggl[
      l(l+1)(2l+1)
}
\nonumber\\&&\times\,
      {\rm Re}\biggl(\!{\cal B}^N_l\!(\omega_A)
      \biggl[ {h^{(1)}_l({k_Ar_A}) \over {k_Ar_A}} \biggr]^2 \biggr)
      \biggr]
      \biggr\},
\qquad      
\end{eqnarray}
and for a tangential dipole it reads
\begin{eqnarray}
\label{e7.2}
\lefteqn{
      \Gamma^\parallel = \Gamma_0 \biggl\{ 1
      + {\textstyle{3\over 4}} \sum_{l=1}^\infty
      \biggl[
      (2l+1)
}
\nonumber\\&&\hspace{-1.5ex}\times\,
      {\rm Re}\biggl(\!{\cal B}^M_l\!(\omega_A)
      \left[ h^{(1)}_l({k_Ar_A}) \right]^2
\nonumber\\&&\hspace{-1.5ex} 
      +\,{\cal B}^N_l\!(\omega_A)
      \biggl[ { \bigl[{k_Ar_A} h^{(1)}_l({k_Ar_A})\bigr]^\prime 
      \over {k_Ar_A} } \biggr]^2
      \biggr)
      \biggr]
      \biggr\}
\qquad      
\end{eqnarray}
%%%%%%%%%%%%%%%%%%%%%%%%%%%%%%%%%%%%%%%%%%%%%%%%%%%%%%%%%%%%%%%%%
\begin{figure}[ht]
\begin{center}
\mbox{\psfig{file=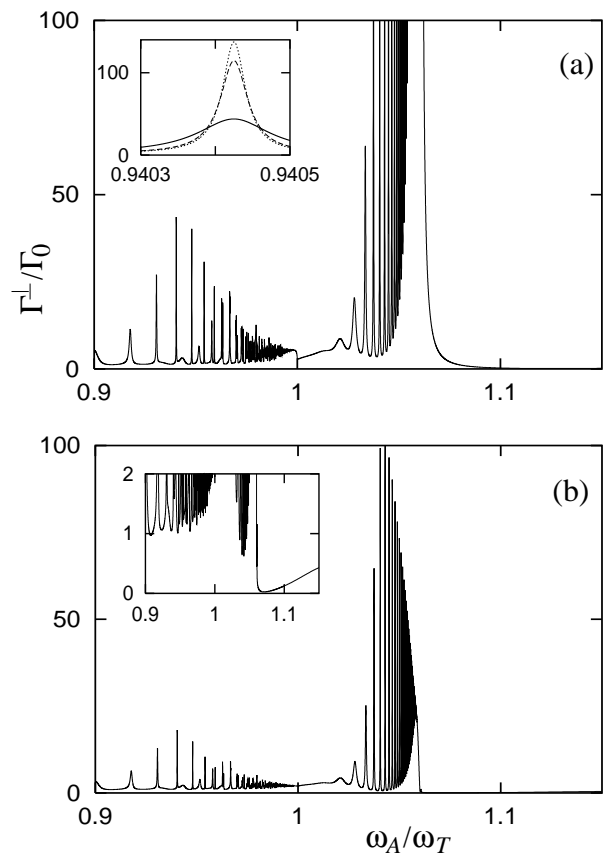,width=8.5cm}} 
\end{center} 
\caption{\label{f3}
The rate of spontaneous decay 
of an excited atom near a microsphere 
is shown as a function of the transition frequency
for a radially oriented transition dipole moment
and a single-resonance Drude--Lorentz-type dielectric
[\mbox{$R$ $\!=$ $\!2\,\lambda_T$};
\mbox{$\omega_P$ $\!=$ $\!0.5\,\omega_T$}; 
\mbox{$\gamma/\omega_T$ $\!=$ $\!10^{-4}$};
(a)
\mbox{$\Delta r$ $\!=$ $\!0.02\,\lambda_T$}; 
inset:  
\mbox{$\gamma/\omega_T$ $\!=$ $\!\!10^{-4}$} (solid line), 
$10^{-5}$ (dashed line), 
$10^{-6}$ (dotted line); 
(b) \mbox{$\Delta r$ $\!=$ $\!0.1\,\lambda_T$}].
[After \citet{Ho01}.] 
}
\end{figure}
%%%%%%%%%%%%%%%%%%%%%%%%%%%%%%%%%%%%%%%%%%%%%%%%%%%%%%%%%%%%%%%%%
($k_A$ $\!=$ $\omega_A/c$), where ${\cal B}^N_l\!(\omega_A)$
is defined according to Eq.~(\ref{eA.21}), and the prime indicates
\label{page1}
the derivative with respect to $k_Ar_A$ [\citet{Ho01}].\footnote{Equations
   (\ref{e7.1}) and (\ref{e7.2}) can be regarded as being the
   natural extension of the (classical) theory for nonabsorbing
   matter as given by \citet{Chew87} and \citet{Klimov96}. Note that
   when the formulas for nonabsorbing matter are given
   in terms of the Green tensor, without an explicit decomposition
   in real and imaginary parts, then for the real 
   permittivity the complex one can be substituted.\label{fnote2}}
Note that a radially oriented transition dipole moment only
couples to TM waves, whereas a tangentially oriented dipole
moment couples to both TM and TE waves.

It is worth noting that when
the atom is very close to the microsphere, then
the decay rates Eqs.~(\ref{e7.1}) and (\ref{e7.2}) reduce 
to exactly the same form as in Eq.~(\ref{e5.7}) for a planar
interface, with $z$ being now the distance between the atom and
the surface of the microsphere. Obviously, nonradiative decay,
which dominates in this case, does not respond sensitively
to the actual radiation-field structure. 

The dependence on the transition frequency of the decay rate,
as it is typically observed for not too small (large) values
of the atom-surface distance (material absorption),
is illustrated in Fig.~\ref{f3} for a radially
oriented transition dipole moment. Since
it mimics the single-quantum excitation spectrum of the
sphere-assisted (TM) radiation field, the figure reveals
that both the WG and SG field resonances can strongly
enhance the spontaneous decay. Material absorption
broadens the resonance lines at the expense of the heights,
and the enhancement is accordingly reduced
[see the inset in Fig.~\ref{f3}(a)]. Clearly, the sphere-assisted
enhancement of spontaneous decay decreases with increasing distance
between the atom and the sphere [compare Figs.~\ref{f3}(a) and (b)].

Figure \ref{f3} also reveals that SG waves
can give rise to a much stronger enhancement of
the spontaneous decay than WG waves.
In particular, with increasing angular-momentum number
the SG field resonance lines strongly overlap
and huge enhancement [e.g., of the order
of magnitude of $10^4$ for the parameters chosen in
Fig.~\ref{f3}(a)] can be observed for transition
frequencies inside a band gap.
When the distance between the atom and the sphere increases, then
the atom rapidly decouples from that part of the field. Thus,  
the huge enhancement of spontaneous decay rapidly reduces and
the interval in which inhibition of spontaneous decay is
typically observed, extends accordingly [see Fig.~\ref{f3}(b)]. 

%%%%%%%%%%%%%%%%%%%%%%%%%%%%%%%%%%%%%%%%%%%%%%%%%%%%%%%%%%%%%%%%%%%%%%%%

\subsection{Frequency Shift}
\label{sec7.2}

The sphere-assisted frequency shift calculated from
Eq.~(\ref{e3.14}) together with the Green tensor given
in Appendix \ref{app:spherical} reads
\begin{eqnarray}
\label{e7.3}
\lefteqn{
      \delta\omega_A^\perp = -  
      {3\Gamma_0\over 4} \sum_{l=1}^\infty
      \biggl[
      l(l+1)(2l+1)
}
\nonumber\\&&\times\, 
      {\rm Im}\biggl( {\cal B}^N_l\!(\omega_A)
      \biggl[ {h^{(1)}_l({k_Ar_A}) \over {k_Ar_A}} \biggr]^2
      \biggr)
      \biggr]
\qquad       
\end{eqnarray}
for a radially oriented transition dipole moment, and 
\begin{eqnarray}
\label{e7.4}
\lefteqn{
      \delta\omega_A^\parallel = -  
      {3\Gamma_0\over 8} \sum_{l=1}^\infty
      \biggl[
      (2l+1)
}
\nonumber\\&&\hspace{-2ex}\times\,      
      {\rm Im}\biggl( {\cal B}^M_l\!(\omega_A)
      \left[ h^{(1)}_l({k_Ar_A}) \right]^2
\nonumber\\&&\hspace{-2ex} 
      +\,{\cal B}^N_l\!(\omega_A)
      \biggl[ {\bigl[{k_Ar_A} h^{(1)}_l({k_Ar_A})\bigr]^\prime 
      \over {k_Ar_A}} \biggr]^2
      \biggr)
      \biggr]
\qquad      
\end{eqnarray}
%%%%%%%%%%%%%%%%%%%%%%%%%%%%%%%%%%%%%%%%%%%%%%%%%%%%%%%%%%%%%%%%%%
\begin{figure}[ht]
\begin{center}
\mbox{\psfig{file=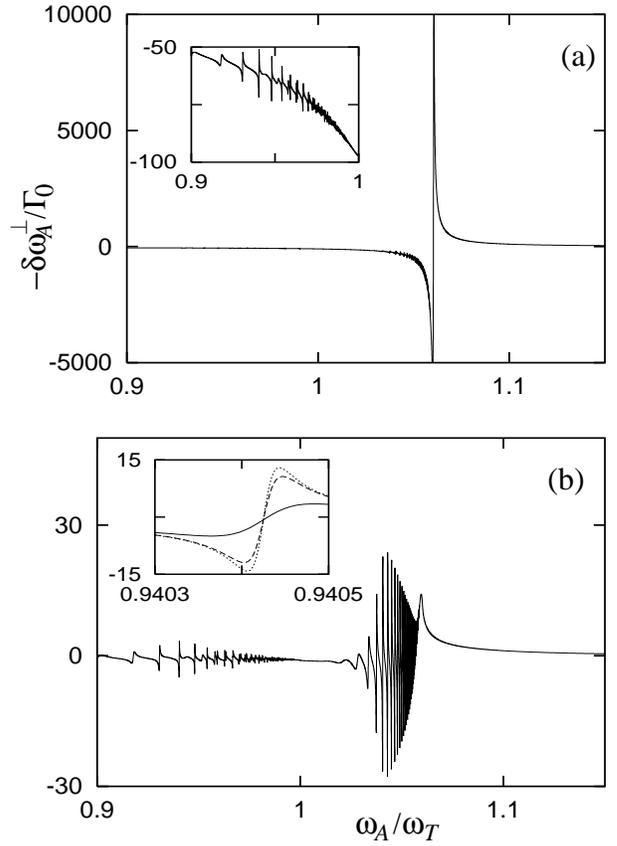,width=8.5cm}}
\end{center} 
\caption{
The frequency shift in spontaneous decay
of an excited atom near a microsphere 
is shown as a function of the transition frequency
for a radially oriented transition dipole moment
and a single-resonance Drude--Lorentz-type dielectric
[\mbox{$R$ $\!=$ $\!2\,\lambda_T$};
\mbox{$\omega_P$ $\!=$ $\!0.5\,\omega_T$}; 
\mbox{$\gamma/\omega_T$ $\!=$ $\!10^{-4}$};
(a) \mbox{$\Delta r$ $\!=$ $\!0.02\,\lambda_T$};
(b)\mbox{$\Delta r$ $\!=$ $\!0.1\,\lambda_T$};
inset:
\mbox{$\gamma/\omega_T$ $\!=$ $\!\!10^{-4}$} (solid line), 
$10^{-5}$ (dashed line), $10^{-6}$ (dotted line)].
[After \citet{Ho01}.] 
}
\label{f5}
\end{figure}
%%%%%%%%%%%%%%%%%%%%%%%%%%%%%%%%%%%%%%%%%%%%%%%%%%%%%%%%%%%%%%%%%%%
for a tangentially oscillating dipole.\footnote{Again,
   Eqs.~(\ref{e7.3}) and (\ref{e7.4}) could be obtained from the
   classical theory for nonabsorbing matter [\citet{Klimov96}];
   see footnote \ref{fnote2} on page \pageref{page1}.}
Note that the small quantum corrections that arise from the
second term in Eq.~(\ref{e3.14}) have been omitted.
For very small distance between the atom and the sphere,
Eqs.~(\ref{e7.3}) and (\ref{e7.4}) acquire 
the same form as for a planar interface, Eq. (\ref{e5.9}).

In Fig.~\ref{f5},
an example of the dependence on the transition frequency
of the frequency shift for a radially oriented dipole is
shown. It is seen that the field resonances can give rise to
noticeable frequency shifts in the very vicinities of the
corresponding resonance frequencies. Transition frequencies that are
lower (higher) than a resonance frequency are shifted to
lower (higher) frequencies. In close analogy to the behavior of
the decay rate, the frequency shift is more pronounced for SG
resonances than for WG resonances and can be huge for large
angular-momentum numbers when the lines of the SG field resonances
strongly overlap.  

The behavior of the frequency shift as shown in Fig.~\ref{f5}(b)
can already be seen in the single-resonance approximation
[\citet{Ching87}]. Let the atomic transition frequency $\omega_A$
be close to a resonance frequency $\omega_C$ of the microsphere
and assume that, in a first approximation, the effect from
the other resonances may be ignored. For a Lorentzian resonance
line of width $\Delta\omega_C$, from Eq.~(\ref{e3.13})
it then follows that
\begin{eqnarray}
\label{e7.5}
    \delta\omega_A
    \hspace{-1ex}&\simeq&\hspace{-1ex}
    -{{\cal P} \over 4\pi}\, 
    \!\! \int_{-\infty}^\infty \!\!{\rm d}\omega\, 
    {1\over \omega\!-\!\omega_A}
    {\Gamma_C \Delta\omega_C^2\over 
     (\omega\!-\!\omega_C)^2\!+\!\Delta\omega_C^2}  
\nonumber\\[.5ex] 
    \hspace{-1ex}&=&\hspace{-1ex}
    -{\Gamma_C \Delta\omega_C \over 2}              
    {\omega_A-\omega_C \over 
     (\omega_A-\omega_C)^2+\Delta\omega_C^2} \,,    
\end{eqnarray}    
where $\Gamma_C$ (which corresponds to the height of the line)
is the decay rate for \mbox{$\omega_A$
$\!=$ $\omega_C$}. In particular,
Eq.~(\ref{e7.5}) indicates that the frequency shift peaks at
half maximum  on both sides of the resonance line.
With increasing material absorption, the linewidth
$\Delta\omega_C$ increases while $\Gamma_C$ decreases,
and thus the absolute value of the frequency shift is
reduced, the distance between the maximum and the minimum
being somewhat increased. 
With decreasing distance between the atom and the microsphere
near-field effects become important and Eq.~(\ref{e7.5}) fails,
as it can be seen from a comparison of Figs. \ref{f5}(a) and (b).

%%%%%%%%%%%%%%%%%%%%%%%%%%%%%%%%%%%%%%%%%%%%%%%%%%%%%%%%%%%%%%%%%%%%%%%%

\subsection{Emitted-Light Intensity}
\label{sec7.3}

\subsubsection{Spatial distribution}
\label{sec7.3.1}

Substitution into Eq.~(\ref{e3.27}) of the expression
for the Green tensor (Appendix \ref{app:spherical}) yields 
(\mbox{$\theta_A$ $\!=$ $\!\phi_A$ $\!=$ $\!0$},
\mbox{$r_A$ $\!\le$ $\!r$})
\begin{eqnarray}
\label{e7.6}
\lefteqn{
     {\bf F}^\perp({\bf r},{\bf r}_A,\omega_A) =
     {k_A^3d \over 4\pi \varepsilon_0}
     \sum_{l=1}^\infty
     \biggl\{
     (2l+1)
}
\nonumber\\[.5ex]&&\hspace{-1ex} \times\, 
     {1\over {k_Ar_A}}
     \left[j_l({k_Ar_A}) + {\cal B}_l^N\!(\omega_A)
            h_l^{(1)}({k_Ar_A}) \right]
\nonumber\\[.5ex]&&\hspace{-1ex} \times\,
     \biggl[ {\bf e}_r \,l(l\!+\!1)\, {h_l^{(1)}({k_Ar})\over {k_Ar}}
     \,P_l(\cos\theta)
\nonumber\\[.5ex]&&\hspace{-1ex} 
     -\, {\bf e}_\theta \,{[{k_Ar}\,h_l^{(1)}({k_Ar})]'\over {k_Ar}}\,
     \sin\theta P_l'(\cos\theta)
     \biggr]
     \!\biggr\}
\qquad     
\end{eqnarray}
for a radially oriented transition dipole moment, and
\begin{eqnarray}
\label{e7.7}
\lefteqn{\hspace{-2ex}
     {\bf F}^\parallel({\bf r},{\bf r}_A,\omega_A) =
     {k_A^3d \over 4\pi \varepsilon_0}
     \sum_{l=1}^\infty
     \biggl\{
     {(2l+1)\over l(l+1)}
     \biggl[
     {\bf e}_r \cos \phi\,
}
\nonumber\\[.5ex]&&\hspace{-1ex}\times\, 
     \tilde{\cal B}_l^N l(l\!+\!1)\,{h_l^{(1)}({k_Ar})\over {k_Ar}}\,
     \sin\theta P_l'(\cos\theta)
\nonumber\\[.5ex]&&\hspace{-1ex} 
     + \,{\bf e}_\theta \cos \phi  
     \biggl(\tilde{\cal B}_l^M  h_l^{(1)}({k_Ar}) P_l'(\cos\theta)
\nonumber\\&&\hspace{-1ex}   
     + \,\tilde{\cal B}_l^N {[{k_Ar}\,h_l^{(1)}({k_Ar})]'\over {k_Ar}}\, 
     \tilde{P}_l(\cos\theta) 
     \biggr) 
\nonumber\\[.5ex]&&\hspace{-1ex} 
      -\,{\bf e}_\phi \sin \phi
     \biggl(\tilde{\cal B}_l^M  h_l^{(1)}({k_Ar}) \tilde{P}_l(\cos\theta)
\nonumber\\&&\hspace{-1ex}  
     +\,\tilde{\cal B}_l^N {[{k_Ar}\,h_l^{(1)}({k_Ar})]'\over
     {k_Ar}}\,P_l'(\cos\theta)
     \biggr)\! \biggr]\!
     \bigg\}
\quad      
\end{eqnarray}
for a tangentially oriented dipole in the $xz$-plane. Here
the abbreviating notations
\begin{eqnarray}
\label{e7.8}
\lefteqn{
     \tilde{\cal B}_l^N = 
     {1\over {k_Ar_A}}
     \Big\{ [{k_Ar_A}j_l({k_Ar_A})]'
}
\nonumber\\&& 
     +\, {\cal B}_l^N\!(\omega_A) 
            [{k_Ar_A}h_l^{(1)}({k_Ar_A})]'
     \Big\} ,
\end{eqnarray}
\begin{equation}     
\label{e7.9}
     \tilde{\cal B}_l^M = 
     j_l({k_Ar_A}) + {\cal B}_l^M\!(\omega_A) h_l^{(1)}({k_Ar_A}) ,
\end{equation}
\vspace{.1ex}
\begin{equation} 
\label{e7.10} 
     \tilde{P}_l (\cos\theta)=
     l(l+1)P_l(\cos\theta) - \cos\theta P'_l(\cos\theta)
\end{equation}
have been introduced. $|{\bf F}^{\perp(\|)}
({\bf r},{\bf r}_A,\omega_A)|^2$ determines, 
according to Eq.~(\ref{e3.26}), the spatial distribution
of the light emitted by a radially (tangentially) oriented dipole.    

%%%%%%%%%%%%%%%%%%%%%%%%%%%%%%%%%%%%%%%%%%%%%%%%%%%%%%%%%%%%%%%%%%%%%%%%
\begin{figure}[ht]
\begin{center}\hspace*{.5cm}
\parbox{\textwidth}{\psfig{file=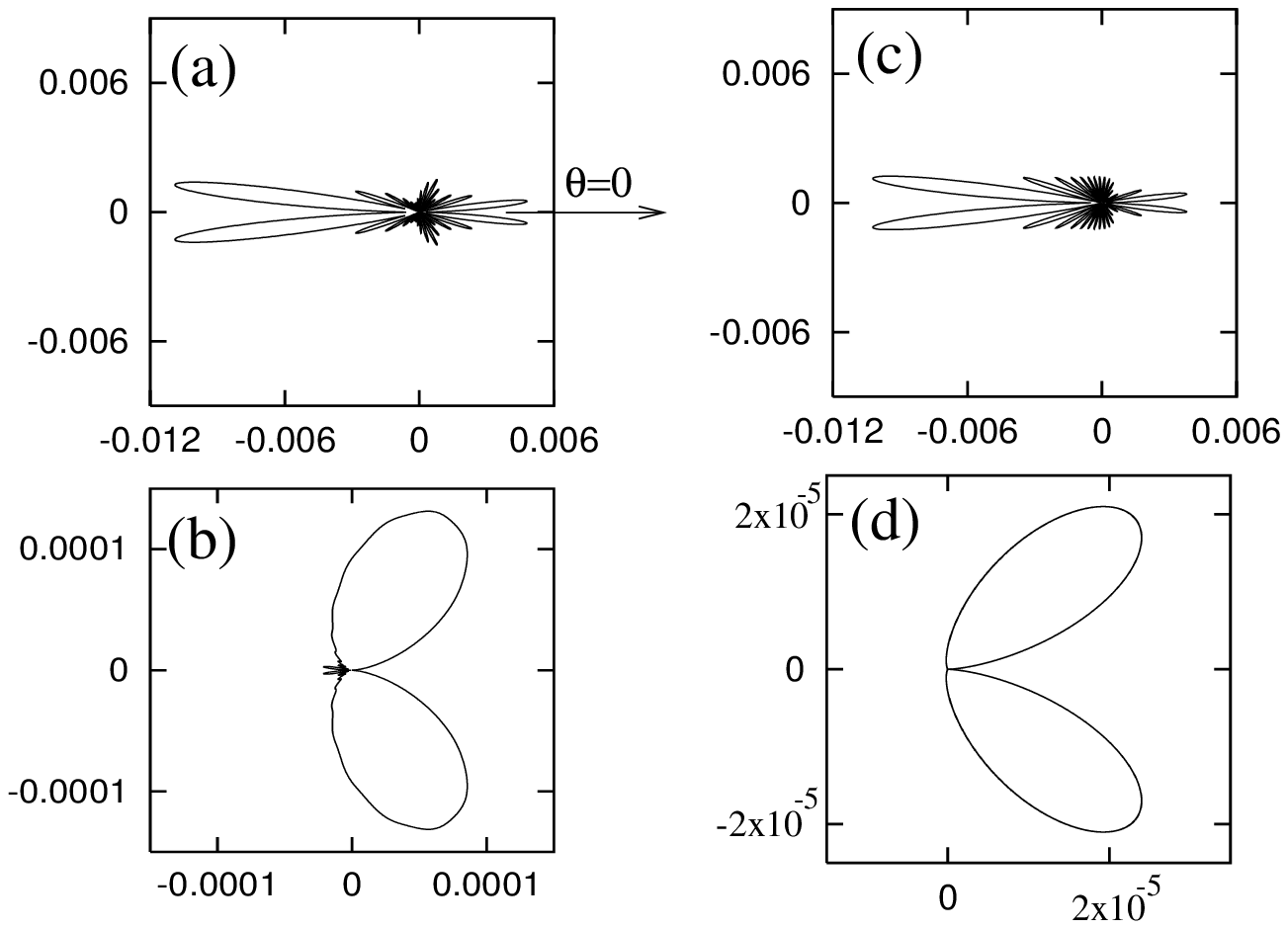,width=160mm}}
\end{center} 
\parbox{\textwidth}{
\caption{\label{f6}
      Polar diagrams of the far-field emission pattern 
      $|{\bf F}^\perp({\bf r},{\bf r}_A,\omega_A)|^2/
      (k_A^3d / 4\pi \varepsilon_0)^2$
      of a radially oscillating dipole
      near a single-resonance Drude--Lorentz-type dielectric
      microsphere 
      [\mbox{$R$ $\!=$ $\!2\,\lambda_T$};
      \mbox{$\omega_P$ $\!=$ $\!0.5\,\omega_T$}; 
      \mbox{$\gamma/\omega_T$ $\!=$ $\!10^{-4}$};
      \mbox{$\Delta r$ $\!=$ $\!0.02\,\lambda_T$};
      \mbox{$r$ $\!=$ $\!20\,\lambda_T$};
      \mbox{$\omega_A/\omega_T$ $\!=$ $\!0.94042$} (a), 
      $0.999$ (b), $1.02811$ (c), $1.06$ (d). [After \citet{Ho01}.]
      }
}
\end{figure}
%%%%%%%%%%%%%%%%%%%%%%%%%%%%%%%%%%%%%%%%%%%%%%%%%%%%%%%%%%%%%%%%%%%%%%%
Let us restrict our attention to a radially oriented transition
dipole moment. Examples of $|{\bf F}^\perp({\bf r},{\bf r}_A,\omega_A)|^2$
are plotted in Fig.~\ref{f6}. In this case, the far field is
essentially determined by $F^\perp_\theta$, as an inspection
of Eq.~(\ref{e7.6}) reveals. When the atomic transition frequency
coincides with the frequency of a WG wave of angular-momentum
number $l$ far from the band gap [Fig.~\ref{f6}(a)], then the 
corresponding $l$-term in the series (\ref{e7.6}) obviously
yields the leading contribution to the emitted radiation,
whose angular distribution is significantly determined by the term 
\mbox{$\sim$ $\!\sin\theta\, P'_l(\cos\theta)$}.
Thus, the emission pattern
has $l$ lobes in, say, the $yz$-plane, i.e., $l$
cone-shaped peaks around the $z$-axis, because of symmetry reasons.
The lobes near \mbox{$\theta$ $\!=$ $\!0$} and
\mbox{$\theta$ $\!=$ $\!\pi$} are the most 
dominant ones in general, because of
\begin{equation}
\label{e7.11}
     -\sin\theta P'_l(\cos\theta) 
     \sim (\sin\theta)^{-1/2} + O\bigr(l^{-1}\bigl)
\end{equation}
($0$ $\!<$ $\!\theta$ $\!\le$ $\!\pi$).
Note that the superposition of 
the leading term with the remaining terms  
in series (\ref{e7.6}) gives rise to some asymmetry 
with respect to the plane \mbox{$\theta$ $\!=$ $\!\pi/2$}.

When the atomic transition frequency approaches (from below)
a band gap (but is still outside it),
a strikingly different behavior is observed [Fig.~\ref{f6}(b)].
The emission pattern changes to a two-lobe 
structure similar to that observed in free space, but bent 
away from the microsphere surface, the emission 
intensity being very small.
Since near a band gap absorption losses dominate,
a photon that is resonantly emitted
is almost certainly absorbed and does not contribute
to the far field in general. If the photon is
emitted in a lower-order WG wave where radiative
losses dominate, it has a bigger chance to escape.
The superposition of
all these weak (off-resonant) contributions
just form the two-lobe emission pattern observed,
as it can also be seen from careful inspection of
the series (\ref{e7.6}).  

When the atomic transition frequency is inside 
a band gap and coincides with the frequency of 
a SG wave of low order such
that the radiative losses dominate, then the
emission pattern resembles that observed for resonant
interaction with a low-order WG wave [compare Figs.~\ref{f6}(a)
and (c)]. With increasing transition frequency the
absorption losses become substantial and eventually change the
%%%%%%%%%%%%%%%%%%%%%%%%%%
\newpage
\vspace*{134mm}
\noindent
emission pattern in a quite similar way as do below the band gap
[compare Figs.~\ref{f6}(b) and (d)]. Obviously,
the respective explanations are similar in the two cases.      

%%%%%%%%%%%%%%%%%%%%%%%%%%%%%%%%%%%%%%%%%%%%%%%%%%%%%%%%%%%%%%%%%%%%%%%%%%

\subsubsection{Radiative versus nonradiative decay}
\label{sec7.3.2}

Since the imaginary part of both the vacuum Green tensor $\mbb{G}^V$
and the scattering term $\mbb{G}^{R}$ is transverse, 
the decay rate (\ref{e3.12}) results from the coupling of the
atom to the transverse part of the electromagnetic field.
Nevertheless, the decay of the excited atomic state must not
necessarily be accompanied by the emission of a real photon,
but instead a matter quantum can be created, 
because of material absorption. To compare the two decay channels,
let us consider, according to Eq.~(\ref{e3.29}),
the fraction $W/W_0$ of the atomic (transition) energy that
is irradiated by an atom with a radially oriented 
transition dipole moment. Using Eqs.~(\ref{e3.26}) and
(\ref{e7.6}), one derives [\citet{Ho01}]
%%%%%%%%%%%%%%%%%%%%%%%%%%%%%%%%%%%%%%%%%%%%%%%%%%%%%%%%%%%%%%%%%%%%
\begin{figure}[ht]
\begin{center}
\mbox{\psfig{file=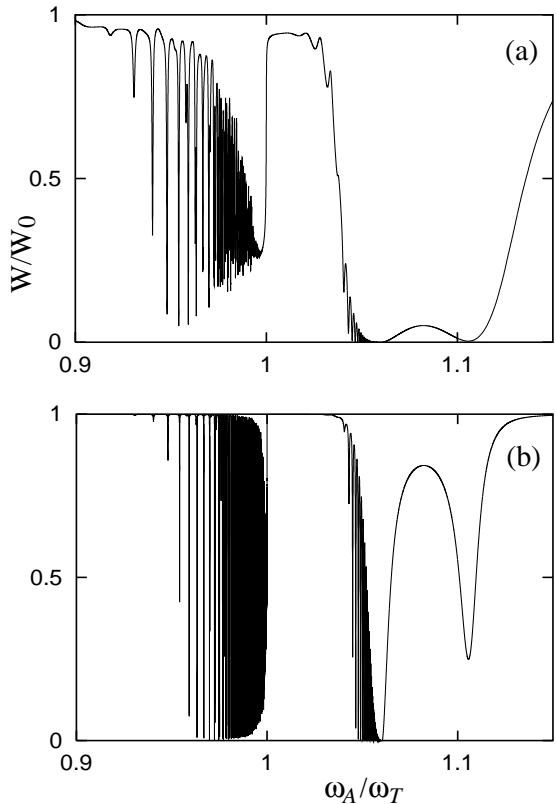,width=8.cm}} 
\end{center} 
\caption{\label{f8}
The fraction of emitted radiation energy in spontaneous decay
of an excited atom placed near a microsphere 
is shown as a function of the transition frequency
for a radially oriented transition dipole moment
and a single-resonance Drude--Lorentz-type dielectric
[\mbox{$R$ $\!=$ $\!2\,\lambda_T$};
\mbox{$\omega_P$ $\!=$ $\!0.5\,\omega_T$}; 
\mbox{$\gamma/\omega_T$ $\!=$ $\!10^{-4}$};
\mbox{$\Delta r$ $\!=$ $\!0.02\,\lambda_T$};
\mbox{$\gamma/\omega_T$ $\!=$ $\!10^{-4}$} (a),
$10^{-6}$ (b). [After \citet{Ho01}.] 
}
\end{figure}
%%%%%%%%%%%%%%%%%%%%%%%%%%%%%%%%%%%%%%%%%%%%%%%%%%%%%%%%%%%%%%%%%%%%
\begin{eqnarray}
\label{e7.12}
\lefteqn{          
{W\over W_0} = {3\Gamma_0\over 2 \Gamma^\perp}
      \sum_{l=1}^\infty 
     { l(l+1)(2l+1) \over ({k_Ar_A})^2}
}
\nonumber\\&&\times 
     \left|j_l({k_Ar_A}) \!+\! {\cal B}_l^N\!(\omega_A)
            h_l^{(1)}({k_Ar_A}) \right|^2\!\!.
\qquad\quad
\end{eqnarray}
Recall that \mbox{$W/W_0$ $\!=$ $\!1$} implies fully radiative decay, 
while \mbox{$W/W_0$ $\!=$ $\!0$} implies fully nonradiative one. 

The dependence of the ratio $W/W_0$ on the atomic
transition frequency is illustrated in Fig.~\ref{f8}.
The minima at the WG field resonance frequencies
indicate that the nonradiative decay is enhanced relative
to the radiative one. Obviously,
photons at these frequencies are captured inside the
microsphere for some time, and hence the probability
of photon absorption is increased. 
For transition frequencies inside a band gap, two regions
can be distinguished.
In the low-frequency region, where low-order 
SG waves are typically excited, radiative decay dominates. Here,
the light penetration depth into the sphere is small and
the probability of a photon being absorbed is small as well.  
With increasing atomic transition frequency
the penetration depth increases and the chance of 
a photon to escape drastically diminishes. As a result,
nonradiative decay dominates. Clearly, the strength of the
effect decreases with decreasing material absorption  
[compare Fig.~\ref{f8}(a) with (b)].

{F}rom the figure two well
pronounced minima of the
totally emitted light energy,
i.e., noticeable maxima of the energy transfer to the matter,  
are seen for transition frequencies inside the band gap.
The first minimum results from the overlapping high-order SG waves
that mainly underly absorption losses. The second one
is observed at the longitudinal resonance frequency of the medium.
It can be attributed to the atomic near-field interaction 
with the longitudinal component of the medium-assisted 
electromagnetic field, the strength of the longitudinal 
field resonance being proportional to $\varepsilon_{I}$.
Hence, the dip at the longitudinal frequency
of the emitted radiation energy reduces with decreasing
material absorption and may disappear when 
the atom is moved sufficiently away from the surface.

%%%%%%%%%%%%%%%%%%%%%%%%%%%%%%%%%%%%%%%%%%%%%%%%%%%%%%%%%%%%%%%%%%%%%%%%%

\subsubsection{Temporal evolution}
\label{sec7.3.3}

Throughout this section we have restricted our attention to the
weak-coupling regime where the excited atomic state decays
exponentially, Eq.~(\ref{e3.15}).
When retardation is disregarded, then the intensity of the
emitted light (at some chosen space point)
simply decreases exponentially, Eq.~(\ref{e3.26}).
To study the effect of retardation, the frequency integral
in the exact equation (\ref{e3.21}) must be performed
numerically in general. 

\begin{figure}[ht]
\begin{center}
\mbox{\psfig{file=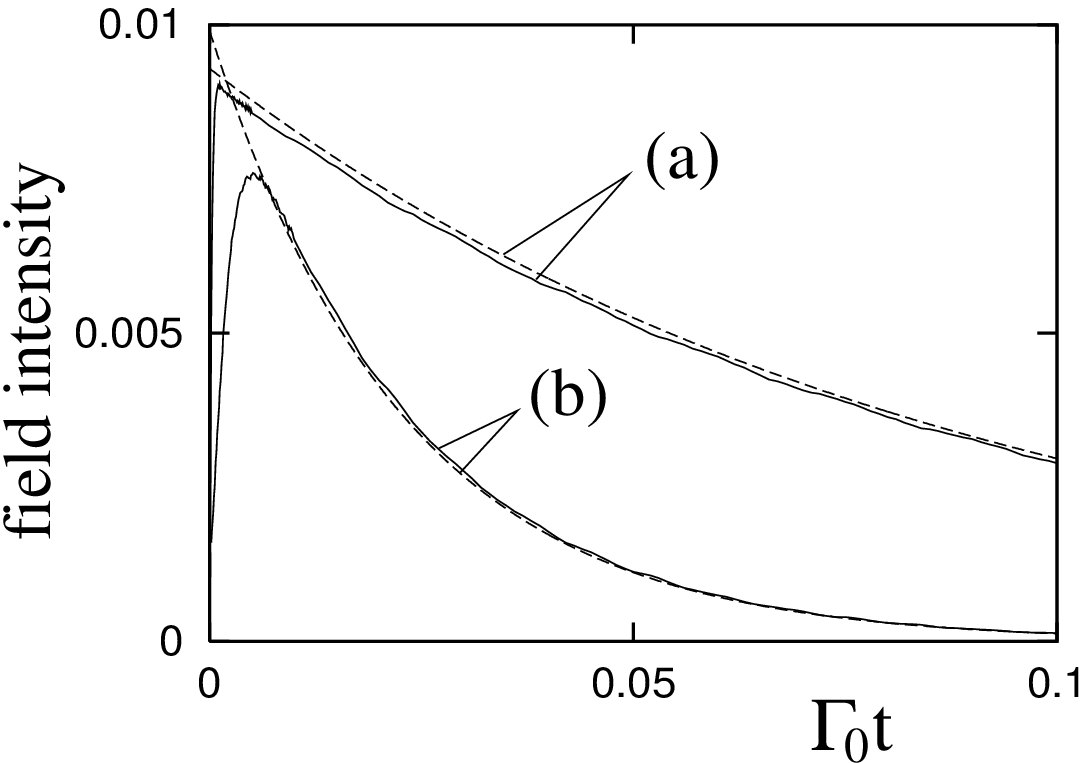,width=8.5cm}} 
\end{center} 
\caption{\label{f10}
Exact (solid lines) and approximate (dashed lines)
temporal evolution of the far-field intensity
$I({\bf r},t)/(k_A^3d/4\pi\varepsilon_0)^2$
at a fixed point of observation of a radially oscillating
dipole near a single-resonance Drude--Lorentz-type
dielectric microsphere 
[\mbox{$R$ $\!=$ $\!2\,\lambda_T$};
\mbox{$\omega_P$ $\!=$ $\!0.5\,\omega_T$}; 
\mbox{$\gamma/\omega_T$ $\!=$ $\!10^{-4}$};
\mbox{$\Delta r$ $\!=$ $\!0.02\,\lambda_T$};
\mbox{$r$ $\!=$ $\!20\,\lambda_T$};
\mbox{$\theta$ $\!=$ $\!3$};
\mbox{$\Gamma_0/\omega_T$ $\!=$ $\!10^{-7}$};
\mbox{$\omega_A$ $\!=$ $\!0.91779\,\omega_T$} (a),
$0.94042\,\omega_T$ (b). [After \citet{Ho01}.]
}
\end{figure}

Typical examples of the temporal evolution of the far-field
intensity are shown in Fig.~\ref{f10} for a radially oriented
transition dipole moment in the case when the atomic transition
frequency coincides with the frequency of a WG wave.
%%%%%%%%%%%%%%%%%%%%%%%%%%%%%%%%%%%%%%%%%%%%%%%%%%%%%%%%%%%%%%%%%%%%%%
%%%%%%%%%%%%%%%%%%%%%%%%%%%%%%%%%%%%%%%%%%%%%%%%%%%%%%%%%%%%%%%%%%%%%%%%
Whereas the long-time behavior of the intensity of the
emitted light is, with little error, exponential,
the short-time behavior (on a time scale given by the
atomic decay time) sensitively depends on the quality factor
[$Q$ $\!\sim$ $\!10^3$ in Fig.~\ref{f10}(a),
$Q$ $\!\sim$ $\!10^4$ in Fig.~\ref{f10}(b)].
The observed delay between the upper-state atomic population
and the intensity of the emitted light can be quite
large for a high-$Q$ microsphere, because the time that
a photon spends in the sphere increases with the $Q$\,value.  
Further, in the short-time domain some kink-like
fine structure is observed, which obviously reflects the
different arrival times associated with multiple reflections. 

%%%%%%%%%%%%%%%%%%%%%%%%%%%%%%%%%%%%%%%%%%%%%%%%%%%%%%%%%%%%%%%%%%%%%%%%

\subsection{Metallic Microsphere}
\label{sec7.4}

The permittivity of a metal (on the basis of the Drude model)
can be obtained by 
setting in Eq.~(\ref{e4.5})
the lowest resonance frequency $\omega_{T\alpha}$ equal
to zero. Thus, the results derived for the
band gap of a dielectric microsphere also apply,
for appropriately chosen values of
the corresponding
\mbox{$\omega_{P\alpha}$ $\!\equiv$ $\!\omega_P$}
and \mbox{$\gamma_\alpha$ $\!\equiv$ $\!\gamma$},
to a metallic sphere. In particular, the results
obtained in the nonretardation limit (\mbox{$c$ $\!\to$ $\!\infty$})
and for small sphere sizes (\mbox{$R$ $\!\ll$ $\!\lambda_P$})
[\citet{Ruppin82,Agarwal83}] are recovered.

%%%%%%%%%%%%%%%  F I G U R E %%%%%%%%%%%%%%%%%%%%%%
\begin{figure}[ht]
\begin{center}
\mbox{\psfig{file=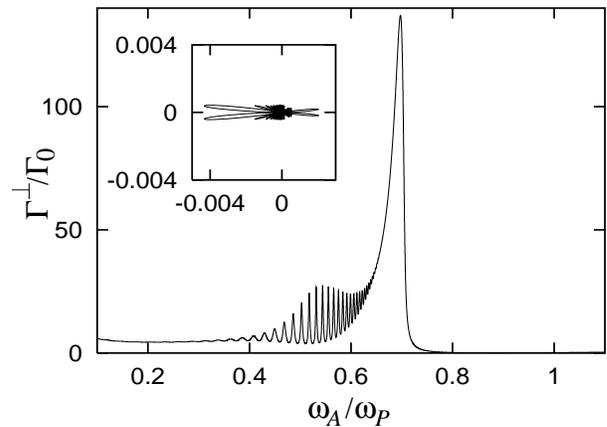,width=8.5cm}} 
\end{center} 
\caption{
The decay rate of an atom near a 
metallic microsphere is shown as a function of the transition
frequency for a radially oriented transition dipole moment
and a  single-resonance metal of Drude type
[\mbox{$R$ $\!=$ $\!5\lambda_P$};
\mbox{$\gamma/\omega_P$ $\!=$ $\!0.005$};
\mbox{$\Delta r$ $\!=$ $\!0.1\,\lambda_P$}.
Inset: polar diagram of the far-field emission pattern
$|{\bf F}^\perp({\bf r},{\bf r}_A,\omega_A)|^2/
(k_A^3\mu / 4\pi \epsilon_0)^2$
for \mbox{$r$ $\!=$ $\!50\,\lambda_P$} and   
\mbox{$\omega_A/\omega_P$ $\!=$ $\!0.5026$}.
[After \citet{Ho01}.]
}
\label{f15}
\end{figure}
%%%%%%%%%%%%%%%%%%%%%%%%%%%%%%%%%%%%%%%%%%%%%%%%%%%%%%%%%
The dependence of the decay rate on the transition
frequency of an excited atom near a metallic
microsphere is illustrated in Fig.~\ref{f15} for
\mbox{$R$ $\!>$ $\!\lambda_P$}. The inset shows 
the emission pattern for the case when the atomic
transition frequency coincides with the frequency of a
SG wave. Note that the SG field resonances seen in Fig.~\ref{f15}
obey, according to condition (\ref{e4.5.7}), the
relation \mbox{$\omega/\omega_P$ $\!<$ $\!1/\sqrt{2}$ $\!\simeq$
$\!0.71$}. When the radius of the microsphere becomes
substantially smaller than the wavelength $\lambda_P$,
then distinct
peaks are only seen  for a few lowest-order resonances
[\citet{Ruppin82,Agarwal83}]. It is worth noting that,
in contrast to dielectric matter, a large absorption 
in metals can substantially enhance the near-surface
divergence of the decay rate, Eq.~(\ref{e5.7}),
which is in agreement with experimental observations 
of the fluorescence from 
dye molecules near a planar metal surface
[\citet{Drexhage74}].

%%%%%%%%%%%%%%%%%%%%%%%%%%%%%%%%%%%%%%%%%%%%%%%%%%%%%%%%%%%%%%%%%%%%%
%%%%%%%%%%%%%%%%%%%%%%%%%%%%%%%%%%%%%%%%%%%%%%%%%%%%%%%%%%%%%%%%%%%%%%

\setcounter{equation}{0}
\section{\hspace{0ex}QUANTUM CORRELATIONS}
\label{sec:2atoms}

Let us finally address the problem of spontaneous decay
of an atom in the case when there is a second atom
that is resonantly dipole-dipole coupled to the first one.
Similarly to single-atom spontaneous decay, the
dipole-dipole interaction can be controlled by
the presence of macroscopic bodies.
%DGW Which papers refer to absorbing bodies?  
Various aspects of the problem have been discussed for 
bulk material [\citet{Knoester89}],
photonic crystals [\citet{Kurizki88,Kurizki90,John95,Bay97,Rupasov97}], 
optical lattices [\citet{Goldstein97b,Guzman98}],
planar cavities [\citet{Kobayashi95a,Agarwal98}]
(and unspecified %high-$Q$
cavities [\citet{Kurizki96,Goldstein97a}]),
and microspheres [\citet{Agarwal00}].
In particular,
resonant energy transfer realized through dipole-dipole 
interaction has been studied theoretically for bulk material 
[\citet{Juzeliunas94a,Juzeliunas94b}],
microspheres [\citet{Gersten84,Druger87,Leung88}],
and planar microcavities 
[\citet{Kobayashi95a,Kobayashi95b}], and experimentally for 
droplets [\citet{Folan85}] and planar microstructures
[\citet{Hopmeier99,Andrew00}], with potential for enhanced 
photon-harvesting systems and optical networks.

Interatomic interaction can give rise to nonclassical correlation
and may be used for entangled-state preparation,
which has been of increasing interest in the study of 
fundamental issues of quantum mechanics 
and with regard to application in quantum information processing.
The entanglement, which is very weak in free space, 
may be expected to be enhanced significantly
in resonator-like equipments.
Proposals have been made for entangling spatially 
separated atoms in Jaynes-Cummings systems 
through sequential or simultaneous strong atom-field coupling 
[\citet{Kudryavtsev93,Phoenix93,Cirac94,Freyberger96,
Gerry96,Plenio99,Beige00}].

Unfortunately, the Jaynes-Cummings 
model does not give any indication
of the influence on entanglement of the actual properties
(such as form and intrinsic dispersion and absorption)
of the bodies in a really used scheme.
{F}rom Section \ref{sec3.2} we know that not only the
evolution of a single atom but also the mutual evolution of
two atoms is fully determined by the Green tensor according
to the chosen configuration of macroscopic bodies. The
formalism thus renders it possible to examine
interatom quantum correlations established in
the presence of arbitrary macroscopic body.
%Let us consider the example of two atoms near a microsphere.

%%%%%%%%%%%%%%%%%%%%%%%%%%%%%%%%%%%%%%%%%%%%%%%%%%%%%%%%%%%%%%%%%%%%%%

\subsection{Entangled-state preparation}
\label{subsec:ent_gen}

%%%%%%%%%%%%%%%%%%%%%%%%%%%%%%%%%%%%%%%%%%%%%%%%%%%%%%%%%%%%%%%%%%%%%%%

\paragraph{Weak Coupling}

Let us consider, e.g., two atoms near a microsphere of the
type studied in Section \ref{sec:microsphere}.
To be more specific, let as assume that the (two-level) atoms
are of the same kind, that they are located at diametrically
opposite positions (\mbox{${\bf r}_A$ $\!=$ $\!-{\bf r}_B$}),
and that their transition dipole moments are radially
oriented.

Obviously, the conditions (\ref{e8.20c}) and (\ref{e8.20a})
are fulfilled for such a system, so that
from Eqs.~(\ref{e8.9}) and (\ref{e8.10}) together with
the Green tensor for a microsphere (Appendix \ref{app:spherical})
one then finds that 
\begin{eqnarray}
\label{e8.16}
\lefteqn{
      \Gamma_\pm^\perp = {\textstyle{3\over 2}} \Gamma_0
      \sum_{l=1}^\infty
      {\rm Re}\biggl\{
      {l(l+1)(2l+1) \over (k_Ar_A)^2}      
}
\nonumber\\&& 
      \times\,
      \left[ j_l(k_Ar_A)+ {\cal B}^N_l\!(\omega_A)
      h^{(1)}_l(k_Ar_A) \right] 
\nonumber\\&& 
      \times\,
      h^{(1)}_l(k_Ar_A) 
      \left[ 1\mp(-1)^l\right]
      \biggr\}.
\end{eqnarray}
When atom $A$ is initially in the
upper state and atom $B$ is accordingly in the lower state, then 
the two superposition states
$|+\rangle$ and $|-\rangle$, Eq.~(\ref{e8.14}),  are 
initially equally excited [\mbox{$C_+(0)$ $\!=$ $\!C_-(0)$ $\!=$
$\!2^{-\frac{1}{2}}$}].
If the atomic transition frequency coincides with a microsphere
resonance, the most significant contribution to the single-atom
decay rate $\Gamma^\perp$, Eq.~(\ref{e7.1}), comes
(for sufficiently small atom-surface distance) from the
corresponding term in the $l$-sum, i.e.,
\begin{eqnarray}
\label{e8.16b}
\lefteqn{
\Gamma^\perp
   \simeq {\textstyle{3\over 2}} \Gamma_0\,
      l(l+1)(2l+1)
}
\nonumber\\&& \times\,     
      {\rm Re} \biggl\{
      \biggl[{h^{(1)}_l(k_Ar_A) \over k_Ar_A}\biggr]^2
      {\cal B}^N_l\!(\omega_A) 
      \biggr\},
\quad      
\end{eqnarray}
and Eq.~(\ref{e8.16}) can be approximated as follows:
\begin{equation}
\label{e8.16c}
\Gamma_\pm^\perp \simeq \Gamma^\perp \left[1 \mp (-1)^l\right]. 
\end{equation}
Hence \mbox{$\Gamma_-$ $\!\gg$ $\!\Gamma_+$}
(\mbox{$\Gamma_+$ $\!\gg$ $\!\Gamma_-$})
if $l$ is even (odd), i.e., the state $|-\rangle$ ($|+\rangle$)
decays much faster than the state $|+\rangle$ ($|-\rangle$).

Consequently, there exists a time window,
during which the overall system is prepared in an entangled state
that is a superposition of the state with the atoms being
in the state $|+\rangle$ ($|-\rangle$)
and the medium-assisted field
being in the ground state, and all the states
with the atoms being in the lower state $|L\rangle$
and the medium-assisted field being in a single-quantum
Fock state. The window is opened
when the state $|-\rangle$ ($|+\rangle$) has already decayed
while the state $|L\rangle$ emerges, and
it is closed roughly after the lifetime of the state $|+\rangle$
($|-\rangle$). 
%%%%%%%%%%%%%%%%%%%%%%%%%%%%%%%%%%%%%%%%%%%%%%%%%%%%%%%%%%%%%%%%%%%%
\begin{figure}[ht]
\begin{center}
\mbox{\psfig{file=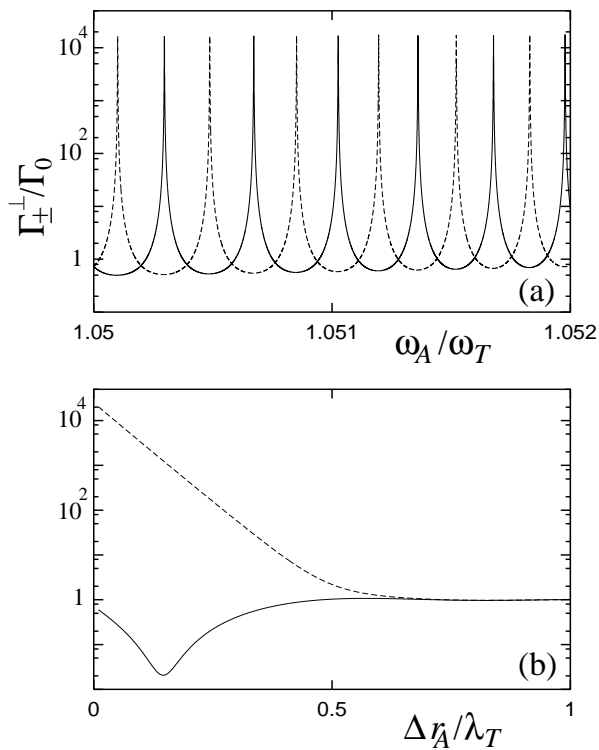,width=8.cm}}
\end{center} 
\caption{\label{fig:ent}
The dependence of the decay rates $\Gamma_+$ (solid line)
and $\Gamma_-$ (dashed line) on (a) the transition
frequency and (b) the distance of the atoms from
a microsphere is shown for two atoms at (with respect
to a sphere) diametrically
opposite positions, radially oriented
transition dipole moments, and a single-resonance
Drude--Lorentz-type dielectric
[\mbox{$R$ $\!=$ $\!10\,\lambda_T$};
\mbox{$\omega_P$ $\!=$ $\!0.5\,\omega_T$};
\mbox{$\gamma$ $\!=$ $\!10^{-6}\omega_T$};
\mbox{$\Delta r_B$ $\!=$ $\!\Delta r_A$
$\!\ge$ $\!10^{-2}\,\lambda_T$};
(a) \mbox{$\Delta r_A$ $\!=$ $\!0.02\,\lambda_T$};
(b) \mbox{$\omega_A$ $\!\simeq$ $\!1.0501\,\omega_T$}].
}
\end{figure}
%%%%%%%%%%%%%%%%%%%%%%%%%%%%%%%%%%%%%%%%%%%%%%%%%%%%%%%%%%%%%%%%%%%%%
As a result, the two atoms are also entangled to each other.
The state is a statistical mixture,
the density operator of which is
obtained from the density operator of 
the overall system by taking the trace with respect
to the medium-assisted field. Within approximation 
(\ref{e8.16c}) it takes the form of
\begin{eqnarray}
\label{e8.16a}
%\lefteqn{  
      \hat{\rho}_A \simeq 
      |C_\pm(t)|^2 |\pm\rangle\langle\pm|
%}
%\nonumber\\[.5ex] &&  
      + \left[1-|C_\pm(t)|^2 \right] 
        |L\rangle\langle L|,
\end{eqnarray}
where
\begin{equation}
\label{e8.16d}
C_\pm(t) \simeq 2^{-1/2} e^{-\Gamma_\pm t} .
\end{equation}
Applying the criterion suggested by \citet{Peres96},
it is not difficult to prove that the
state (\ref{e8.16a}) is indeed inseparable.
It is worth noting that the atoms become entangled 
within the weak-coupling regime, starting from
the state $|U_A\rangle$ (or $|U_B\rangle$) and the
vacuum field. In the language of (Markovian)
damping theory one would probably say that the two atoms are
coupled to the same dissipative system, which gives rise to
the quantum coherence. 

The frequency dependence of $\Gamma_\pm$ as given by
Eq.~(\ref{e8.16}) is illustrated in Fig.~\ref{fig:ent}(a)
for a frequency interval inside a band gap, and the
dependence on the atom-surface distance is illustrated
in Fig.~\ref{fig:ent}(b). 
We see that the values of $\Gamma_+$ and $\Gamma_-$ can
be substantially different from each other before they
tend to the free-space rate $\Gamma_0$ as the distance
from the sphere becomes sufficiently large. In particular, the
decay of one of the states $|+\rangle$ or $|-\rangle$ can
strongly be suppressed [see the minimum value of
$\Gamma_+$ in Fig.~\ref{fig:ent}(b)] at the expense
of the other one, which rapidly decays.  
Note that $\Gamma_+$ also differs from $\Gamma_-$ for
two atoms in free space [\citet{DeVoe96}]. However,
the difference that occurs by mediation of the microsphere
is much larger. For example, at the distance for
which in Fig.~\ref{fig:ent}(b) $\Gamma_+$ attains the
minimum the ratio 
\mbox{$\Gamma_-/\Gamma_+$ $\!\simeq$ $\!67000$} is
observed, which is to be compared with the free-space
ratio \mbox{$\Gamma_-/\Gamma_+$ $\!\simeq$ $\!1.0005$}.
The effect may become even more pronounced for larger microsphere
sizes and lower material absorption, i.e., sharper microsphere
resonances. Needless to say that it is not only
observed for SG waves considered in
Fig.~\ref{fig:ent}, but also for WG waves.  

%%%%%%%%%%%%%%%%%%%%%%%%%%%%%%%%%%%%%%%%%%%%%%%%%%%%%%%%%%%%%%%%%%%%%%%%%

\paragraph{Strong Coupling}

Entangled-state preparation in the weak-coupling regime
has the advantage that it could routinely be achieved experimentally. 
However, the value of $|C_+(t)|^2$ in Eq.~(\ref{e8.16a})
(or the value of $|C_-(t)|^2$ in the corresponding
equation for the state $|-\rangle$) is always less than  
$1/2$. In order to achieve a higher degree of entanglement,
the strong-coupling regime is required.

Let us assume that the two atoms are initially in the
ground state and the medium-assisted field is excited.
The field excitation can be achieved, for example,
by coupling an excited atom $D$ to the microsphere and
then making sure that the atomic excitation is
transferred to the field (cf. Section \ref{subsec:basic}).
If the atom $D$ strongly interacts with the field,
the excitation transfer can be controlled  
by adjusting the interaction time.
Another possibility would be measuring the state populations
and discarding the events where the atom
is found in the upper state.
Here we restrict our attention 
to the first method and assume that all three
atoms $D$, $A$, and $B$ strongly interact with the same
microsphere resonance (of mid-frequency $\omega_C$ and line
width $\Delta\omega_C$). According to Eq.~(\ref{e6.5}),
the upper-state probability amplitude $C_{U_D}(t)$ of
atom $D$ reads
\begin{equation}
\label{e8.19} 
       C_{U_D}(t) = e^{-\Delta\omega_C (t+\Delta t)/2}
       \cos [\Omega_D (t+\Delta t)/2], 
\end{equation}
with $\Omega_D$ being given according to Eq.~(\ref{e6.6}). For
\begin{equation}
\label{e8.22}
       \Delta t =  \pi/\Omega_D ,
\end{equation}
the initially (i.e., at time \mbox{$t$ $\!=$ $\!-\Delta t$}) excited
atom $D$ is at time $t$ $\!=$ $\!0$ in the lower state
[\mbox{$C_{U_D}(0)$ $\!=$ $\!0$}].

{F}rom the preceding subsection we know that
when the resonance angular-momentum number $l$ is odd (even),
then the state $|+\rangle$ ($|-\rangle$) ``feels'' a sharply
peaked high density of medium-assisted field states, so that a 
strong-coupling approximation of the type (\ref{e6.2}) applies.
The state $|-\rangle$ ($|+\rangle$), in contrast, ``feels'' a flat
one and the (weak-coupling) Markov approximation applies.
Assuming atom $A$ (or $B$) 
is at the same position
as was atom $D$, from Eqs.~(\ref{e8.17}), (\ref{e8.19}),
and (\ref{e8.22}) we then find that
\begin{equation}
\label{e8.21}
       C_\pm(t) \simeq - e^{-\Delta\omega_C( t +\pi/\Omega_D)/2}
          \sin(\Omega_\pm t/2)
\end{equation}
[with $\Omega_\pm$ according to Eq.~(\ref{e6.6})] and
\begin{equation}          
\label{e8.21a} 
       C_\mp(t) \simeq 0
\end{equation}
(the sign of $C_-(t)$ in Eq. (\ref{e8.21}) is reversed
if atom $B$ is at the same position as was atom $D$).
Note that $\Omega_\pm$ $\!=$ $\!2^{\frac{1}{2}}\Omega$,
because of Eq.~(\ref{e8.16c}). 
The two-atom entangled state is again of the form
given in Eq.~(\ref{e8.16a}), but now the weight of the
state $|+\rangle$ ($|-\rangle$) can reach values larger $1/2$,
provided that the resonance linewidth $\Delta\omega_C$ is small enough.

%%%%%%%%%%%%%%%%%%%%%%%%%%%%%%%%%%%%%%%%%%%%%%%%%%%%%%%%%%%%%%%%%%%%%%%%%

\subsection{Violation of Bell's inequality}
\label{subsec:Bell_ineq}

Violations of Bell's inequalities provide support 
to quantum mechanics versus local (hidden-variables)
theories [\citet{Bell65,Clauser69}]. Despite
   outstanding progress\footnote{For recent experiments
   using photons, see \citet{Weihs98,Kuzmich00}, and using
   trapped ions, see \citet{Rowe01}.}
in the test of Bell's inequalities, a decisive experiment to
rule out any local realistic theory is yet to be
performed [\citet{Vaidman01}], and the problem continues to
attract much attention. Though entangled states of spatially
separated atoms in a cavity  have been observed [\citet{Hagley97}],
a test of Bell's inequalities for such a system has yet to
be realized. 

The Bell's inequality for spin systems can be written in the form of
[\citet{Bell65,Clauser69}]
\begin{eqnarray}
\label{e8.23}
\lefteqn{         
       B_S=|E(\theta_1,\theta_2)-E(\theta_1,\theta'_2)
}
\nonumber\\[.5ex] &&  
       +\,E(\theta'_1,\theta_2)+E(\theta'_1,\theta'_2)|
           \le 2,
\qquad           
\end{eqnarray}
where
\begin{equation}
\label{e8.24}
       E(\theta_1,\theta_2) = 
       \bigl\langle \hat{\sigma}_A^{\theta_1} 
       \hat{\sigma}_B^{\theta_2} \bigr\rangle,      
\end{equation}
\begin{equation}
\label{e8.25}
\qquad\quad
       \hat{\sigma}_A^{\theta} = 
       \cos\theta \,\hat{\sigma}_A^x
       + \sin\theta \,\hat{\sigma}_A^y .
\end{equation}
When the atomic state $|u_A,u_B\rangle$ is not populated, as 
it is the case for a state of the type (\ref{e8.16a}), it is
not difficult to prove that
\begin{equation}
\label{e8.25a}
E(\theta_1,\theta_2) = E(\theta_1-\theta_2,0).
\end{equation}
%%%%%%%%%%%%%%%%%%%%%%%%%%%%%%%%%%%%%%%%%%%%%%%%%%%%%%%%%%%%%%%%%%%%%%
\begin{figure}[htb]
\begin{center}
\mbox{\psfig{file=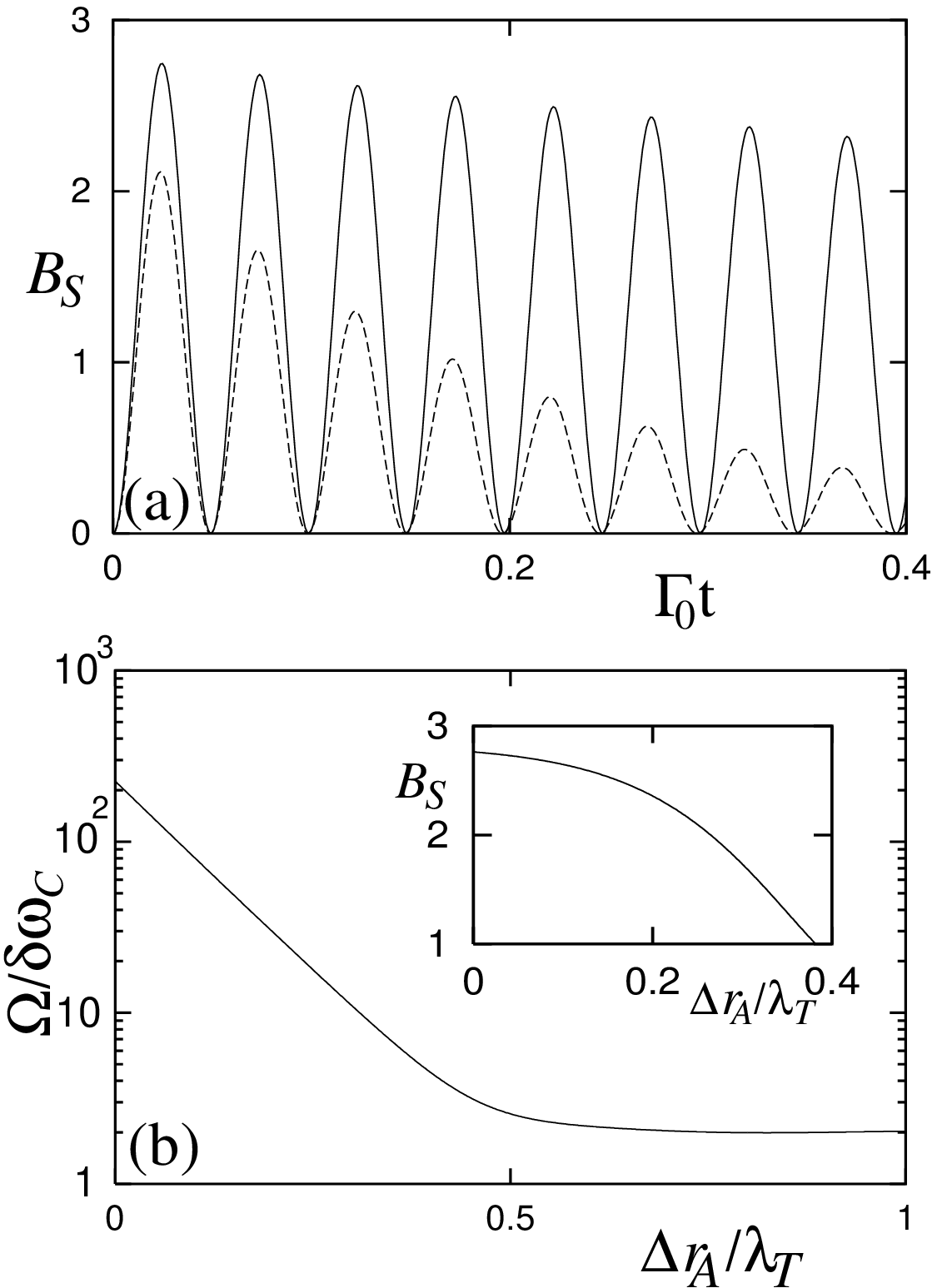,width=7.8cm}} 
\end{center} 
\caption{\label{fig:Bell_ineq}
The dependence on time of $B_S$ is shown for
two atoms at (with respect to a microsphere) diametrically
opposite positions, radially oriented transition dipole
moments, and a single-resonance Drude--Lorentz-type
dielectric
[\mbox{$R$ $\!=$ $\!10\,\lambda_T$};
\mbox{$\omega_P$ $\!=$ $\!0.5\,\omega_T$};
\mbox{$\Delta r_B$ $\!=$ $\!\Delta r_A$ $\!=$ $\!0.02\,\lambda_T$};
\mbox{$\omega_A$ $\!=$ $\!1.0501\,\omega_T$};
\mbox{$\Gamma_0$ $\!=$ $\!10^{-6}\,\omega_T$}; 
$\Omega_D$ $\!=$ $\!\Omega$;
\mbox{$\gamma/\omega_T$ $\!=$ $\!10^{-6}$} (solid line),
$10^{-5}$ (dashed line)].
(b) $\Omega/\Delta\omega_C$ versus $\Delta r_A$
for \mbox{$\gamma/\omega_T$ $\!=$ $\!10^{-6}$}
(\mbox{$\Delta r_A$ $\!\ge$ $\!10^{-3}\,\lambda_T$}).
The inset shows the %corresponding
variation of the first maximum value of $B_S$ in (a).                                   }
\end{figure}
%%%%%%%%%%%%%%%%%%%%%%%%%%%%%%%%%%%%%%%%%%%%%%%%%%%%%%%%%%%%%%%%%%%%%%
Let us choose
\begin{equation}
\label{e8.25b}
\theta = \theta_1 - \theta_2 =
   \theta_2 - \theta'_1 = \theta'_1 - \theta'_2\,.
\end{equation}
The inequality (\ref{e8.23}) thus simplifies to
\begin{equation}
\label{e8.26}
       B_S=|3E(\theta,0) - E(3\theta,0)|
           \le 2 .
\end{equation}

An entangled state of the type (\ref{e8.16a}) can 
only give rise to a violation of the Bell's inequality if
\mbox{$|C_+(t)|^2$ $\!\ge$ $\!2^{-\frac{1}{2}}$ $\!\simeq$ $\!0.71$}
[\citet{Beige00}], which cannot be achieved in the
weak-coupling regime, Eq.~(\ref{e8.16d}). It can be achieved,
in contrast, in the strong-coupling regime, Eq.~(\ref{e8.21}), where
\begin{eqnarray}
\label{e8.27}
\lefteqn{
       E(\theta,0) = \cos\theta\,|C_\pm(t)|^2
}
\nonumber\\[.5ex]&&\hspace{-1ex}       
       = \cos\theta\,
        e^{-\Delta\omega_C(t+\pi/\Omega_D)}
          \sin^2\bigl(\Omega t/\sqrt{2}\bigr).
\qquad          
\end{eqnarray}
Substitution of this expression into in Eq. (\ref{e8.26}) yields, 
on choosing \mbox{$\theta$ $\!=$ $\!\pi/4$}, 
\begin{equation}
\label{e8.28}
       B_S = 2\sqrt{2}\,e^{-\Delta\omega_C(t+\pi/\Omega_D)}
       \sin^2\bigl(\Omega t/\sqrt{2}\bigr),
\end{equation}
which clearly shows that $B_S$ $\!>$ $\!2$ becomes possible
as long as \mbox{$\Delta\omega_C(t$ $\!+$ $\!\pi/\Omega_D)$
$\!\ll$ $\!1$}.
Examples of the temporal evolution of $B_S$ for two atoms
near a dielectric microsphere are shown in 
Fig.~\ref{fig:Bell_ineq}(a).
In Figure~\ref{fig:Bell_ineq}(b) the dependence of 
the ratio $\Omega/\Delta\omega_C$ on the distance
of the atoms from the sphere is plotted. The
strong-coupling regime can be observed for
distances for which \mbox{$\Omega/\Delta\omega_C$ $\!\gg$ $\!1$}
is valid. The inset reveals that the maximum value of $B_S$ 
decreases with increasing atom-surface distance
and reduces below the threshold value of $2$ still 
in the strong-coupling regime.

%%%%%%%%%%%%%%%%%%%%%%%%%%%%%%%%%%%%%%%%%%%%%%%%%%%%%%%
%%%%%%%%%%%%%%%%%%%%%%%%%%%%%%%%%%%%%%%%%%%%%%%%%%%%%%%%

\section{SUMMARY}
\label{sec:concl}

We have studied spontaneous decay in the presence of 
dispersing and absorbing macroscopic bodies,
basing on quantization of the (macroscopic) electromagnetic
field in arbitrary linear, causal media. 
The formalism covers both
weak and strong couplings and enables one to include 
the material absorption and dispersion in a consistent way, 
without restriction to a particular frequency domain. It
replaces the standard concept of orthogonal-mode decomposition,
which requires real permittivities and thus does not allow
for material absorption, with a source-quantity representation
in terms of the classical Green tensor and appropriately
chosen bosonic-field variables. All relevant information 
about the bodies such as form and intrinsic dispersion and absorption
properties are contained in the Green tensor.

The formalism has been applied to study spontaneous decay of
a single atom in 
the presence of various absorbing and dispersing macroscopic 
bodies, including open configurations such as bulk and planar
half space media, and closed configurations such as a
spherical cavity or a microsphere. Absorption can
noticeably influence spontaneous decay. So, the decay rate
in absorbing bulk material takes a much more complicated form
than one would expect from the simple product form that is
commonly used for nonabsorbing matter.
The decay rate of an atom located very near a
planar surface shows that due to material absorption 
the decay rate drastically rises as the atom approaches
the surface of the body, because of near-field assisted
(nonradiative) energy transfer from the atom to the medium.
In fact, this is valid for an atom that is sufficiently
near an arbitrary body, because for short enough 
atom--surface distances, any curved surface can be
approximated  by a planar one.
 
Spontaneous decay can strongly be influenced by
field resonances that can appear due to the presence of
macroscopic bodies, depending on whether the
atomic transition frequency is tuned to a field
resonance or not.      
In particular, the decay process can be mainly radiative or
nonradiative, which depends on whether 
the radiative losses due to input-output coupling 
or the losses due to material absorption dominate. In particular,
to understand what happens when the atomic transition
frequency is inside a band-gap zone of a body,
inclusion in the study of material absorption is
necessary.

Finally, spontaneous decay of (two) dipole-dipole coupled
atoms in the presence of macroscopic bodies
offers the possibility of entangled state preparation
and verification of the violation of Bell's inequalities.
Whereas entangled states can already be prepared
in the weak-coupling regime, violation of 
Bell's inequalities requires the strong-coupling regime,
the ultimate limits being given by material absorption.

\section*{ACKNOWLEDGMENTS}
We thank S. Scheel, E. Schmidt, and A. Tip for
discussions. H.T.D. gratefully acknowledges support from 
the Alexander von Humboldt Stiftung. This work was supported 
by the Deutsche Forschungsgemeinschaft.

%%%%%%%%%%%%%%%%%%%%%%%%%%%%%%%%%%%%%%%%%%%%%%%%%%%%%%%%%%%%%%%%%%%%%
%%%%%%%%%%%%%%%%%%%%%%%%%%%%%%%%%%%%%%%%%%%%%%%%%%%%%%%%%%%%%%%%%%%%%

\begin{appendix}
\renewcommand{\theequation}{\Alph{section}.\arabic{equation}}
\setcounter{equation}{0}
\section{THE GREEN TENSOR}
\label{appA}

%%%%%%%%%%%%%%%%%%%%%%%%%%%%%%%%%%%%%%%%%%%%%%%%%%%%%%%%%%%%%%%%%%%%%%%%%%

\subsection{Bulk Medium}
\label{app:bulk}

For bulk material the Green tensor reads as
\mbox{($\mbb{\rho}$ $\!=$ ${\bf r}$ $\!-$ $\!{\bf r}'$)}
\begin{eqnarray}
\label{se2.1}
      \mbb{G}({\bf r},{\bf r}',\omega) 
=\left[\mbb{\nabla}^r\otimes\mbb{\nabla}^r \!+\! \mbb{I}q^2(\omega) \right]
\frac{{\rm e}^{iq(\omega)\rho}}
{4\pi q^2(\omega)\rho}\,,
\end{eqnarray}
where
\begin{eqnarray}
\label{se2.1a}
q(\omega)
\hspace{-1ex}&=&\hspace{-1ex}
\sqrt{\varepsilon(\omega)} \,\omega/c.
\end{eqnarray}
It can be decomposed into a longitudinal and
a transverse part, 
\begin{equation}
\label{se2.1b}
\mbb{G}({\bf r},{\bf r}',\omega) = \mbb{G}^\|({\bf r},{\bf r}',\omega)
   + \mbb{G}^\perp({\bf r},{\bf r}',\omega),
\end{equation}
where
\begin{eqnarray}
\label{se2.2a}
\lefteqn{  
    \mbb{G}^\|({\bf r},{\bf r}',\omega)
    = -\frac{1}{4\pi q^2}
    \bigg[ \frac{4\pi}{3} \delta(\mbb{\rho}) \mbb{I}
}
\nonumber\\[.5ex]&&\hspace{8ex}
   +\left( \mbb{I}
   -\frac{3\mbb{\rho}\otimes\mbb{\rho}}{\rho^2} \right)
   \frac{1}{\rho^3}
   \bigg]
\qquad   
\end{eqnarray}
and
\begin{eqnarray}
\label{se2.2b}
\lefteqn{   
        \mbb{G}^\perp({\bf r},{\bf r}',\omega)
        = \frac{1}{4\pi q^2} \bigg\{
        \left( \mbb{I} -\frac{3\mbb{\rho}
        \otimes\mbb{\rho}}{\rho^2} \right)
        \frac{1}{\rho^3}
}
\nonumber \\ &&\hspace{-1ex} 
        +\,q^3 \bigg[\! \left(\! \frac{1}{q\rho}\! +\!\frac{i}{(q\rho)^2}
        \!-\!\frac{1}{(q\rho)^3}\! \right) \mbb{I}
\nonumber \\ &&\hspace{-1ex} 
        \!-\!\left(\! \frac{1}{q\rho}
        \!+\!\frac{3i}{(q\rho)^2}\! -\!\frac{3}{(q\rho)^3}\! \right)
        \frac{\mbb{\rho}\otimes\mbb{\rho}}{\rho^2}\! \bigg]
        {\rm e}^{iq\rho}\! \bigg\}.
\qquad        
\end{eqnarray}
In particular, from Eq.~(\ref{se2.2b}) it follows that
\begin{eqnarray}
\label{se2.2c}
     {\rm Im}\,\mbb{G}^\perp({\bf r},{\bf r},\omega)
\hspace{-1ex}&=&\hspace{-1ex}
   \lim_{{\bf r}'\to{\bf r}}
   {\rm Im}\,\mbb{G}^\perp({\bf r},{\bf r}',\omega)
\nonumber \\
\hspace{-1ex}&=&\hspace{-1ex}
   \frac{\omega}{6\pi c}\,n_R(\omega) \mbb{I}.
\end{eqnarray}

%%%%%%%%%%%%%%%%%%%%%%%%%%%%%%%%%%%%%%%%%%%%%%%%%%%%%%%%%%%%%%%%%%%%%%%%%%

\subsection{Spherical Multilayers}
\label{app:spherical}

The Green tensor of a spherical structure, 
consisting of ${\cal N}$ concentric layers, enumerated 
from outward in (the outmost layer labeled as layer 1, 
the innermost layer as layer ${\cal N}$),
can be decomposed into two parts
\begin{equation} 
\label{A2.1}
\mbb{G}({\bf r},{\bf r'},\omega) 
= \mbb{G}^{(s)}({\bf r},{\bf r'},\omega) \delta_{fs} 
+ \mbb{G}^{(fs)}({\bf r},{\bf r'},\omega),
\end{equation}
where $\mbb{G}^{(s)}({\bf r},{\bf r'},\omega)$ represents 
the contribution of the direct waves from the source in an
unbounded space, and $\mbb{G}^{(fs)}({\bf r},{\bf r'},\omega)$
is the scattering part that describes the contribution of the
multiple reflection \mbox{($f$ $\!=$ $\!s$)} and transmission
\mbox{($f$ $\!\neq$ $\!s$)} due to the presence of
the surfaces of discontinuity ($f$ and $s$, respectively,
refer to the regions where are the field and source points
${\bf r}$ and ${\bf r}'$). In Eq.~(\ref{A2.1}), $\mbb{G}^{(s)}$
is nothing but the bulk-material Green tensor (\ref{se2.1}).
In the (local) spherical coordinate systems it reads as [see, e.g.,
\citet{Li94}] 
\begin{eqnarray}
\label{A2.2}
\lefteqn{
       \mbb{G}^{(s)}({\bf r},{\bf r}',\omega)
       = \frac{{\bf e}_r\otimes{\bf e}_r}{k_s^2}\, \delta(r-r')
}
\nonumber\\[.5ex]&&\hspace{-2ex}
       +\,{ik_s\over 4\pi} \sum_{e\atop o} 
       \sum_{l=1}^\infty \sum_{m=0}^l
       \biggl\{ 
       {2l\!+\!1\over l(l\!+\!1)}
       {(l\!-\!m)!\over(l\!+\!m)!}\,(2\!-\!\delta_{0m})
\nonumber\\[.5ex]&&\hspace{-2ex}
       \times
       \left[ {\bf M}^{(1)}_{{e \atop o}lm} ({\bf r},k_s) \otimes
       {\bf M}_{{e \atop o}lm} ({\bf r}',k_s) \right.
\nonumber\\[.5ex]&&\hspace{2ex}       
       \left. +\,{\bf N}^{(1)}_{{e \atop o}lm} ({\bf r},k_s) \otimes
       {\bf N}_{{e \atop o}lm} ({\bf r}',k_s)
        \right]
       \biggr\}
\qquad
\end{eqnarray}
if \mbox{$r$ $\!\ge$ $\!r'$}, and
\mbox{$\mbb{G}^{(s)}({\bf r},{\bf r}',\omega)$  $\!=$
$\!\mbb{G}^{(s)}({\bf r}',{\bf r},\omega)$}
if \mbox{$r$ $\!<$ $\!r'$}. The scattering part 
$\!\mbb{G}^{(fs)}({\bf r},{\bf r}',\omega)$ in
Eq.~(\ref{A2.1}) reads [\citet{Li94}]
\begin{eqnarray}
\label{A2.3}
\lefteqn{
       \mbb{G}^{(fs)}({\bf r},{\bf r'},\omega)
}
\nonumber\\&&\hspace{-3ex}
       = {ik_s\over 4\pi} \sum_{e\atop o} 
       \sum_{l=1}^\infty \sum_{m=0}^l
       \biggl( 
       {2l\!+\!1\over l(l\!+\!1)}
       {(l\!-\!m)!\over(l\!\!+m)!}\,(2\!-\!\delta_{0m})
\nonumber\\[.5ex]&&\hspace{-2ex}
       \biggl\{ 
       (1-\delta_{f{\cal N}}) 
       {\bf M}^{(1)}_{{e \atop o}lm} ({\bf r},k_f)
       \otimes
       \bigl[
       {\bf M}_{{e \atop o}lm} ({\bf r}',k_s)
\nonumber\\[.5ex]&&\hspace{-4ex}\times\,        
       (1-\delta_{s1})
       {\cal A}^M_l (\omega)
%\nonumber\\[.5ex]&&\hspace{1ex}
       +\,{\bf M}^{(1)}_{{e \atop o}lm} ({\bf r}',k_s)
       (1-\delta_{s{\cal N}}) {\cal B}^M_l (\omega)
       \bigr]        
\nonumber\\[.5ex]&&\hspace{-2ex}
        +\,(1-\delta_{f{\cal N}}) 
        {\bf N}^{(1)}_{{e \atop o}lm} ({\bf r},k_f)
        \otimes
        \bigl[
        {\bf N}_{{e \atop o}lm} ({\bf r}',k_s)
\nonumber\\[.5ex]&&\hspace{-4ex}\times\,
        (1-\delta_{s1})
        {\cal A}^N_l (\omega)
%\nonumber\\[.5ex]&&\hspace{0ex}
        +\,{\bf N}^{(1)}_{{e \atop o}lm} ({\bf r}',k_s)
        (1-\delta_{s{\cal N}}) {\cal B}^N_l (\omega)
        \bigr]
\nonumber\\[.5ex]&&\hspace{-2ex}
        +\,(1-\delta_{f1}) 
        {\bf M}_{{e \atop o}lm} ({\bf r},k_f)
        \otimes
        \bigl[
        {\bf M}_{{e \atop o}lm} ({\bf r}',k_s)
\nonumber\\[.5ex]&&\hspace{-4ex}\times\,
        (1-\delta_{s1}) {\cal C}^M_l (\omega)
%\nonumber\\&&\hspace{0ex}
        +\,{\bf M}^{(1)}_{{e \atop o}lm} ({\bf r}',k_s)
        (1-\delta_{s{\cal N}}) {\cal D}^M_l (\omega)
        \bigr]
\nonumber\\[.5ex]&&\hspace{-2ex}
        +\,(1-\delta_{f1}) 
        {\bf N}_{{e \atop o}lm} ({\bf r},k_f)
        \otimes
        \bigl[
        {\bf N}_{{e \atop o}lm} ({\bf r}',k_s)
\nonumber\\[.5ex]&&\hspace{-4ex}\times\,
        (1-\delta_{s1}) {\cal C}^N_l (\omega)
\nonumber\\&&\hspace{-0ex}
        +\,{\bf N}^{(1)}_{{e \atop o}lm} ({\bf r}',k_s)
        (1-\delta_{s{\cal N}}) {\cal D}^N_l (\omega)
        \bigr]
    \biggr\}
    \!\biggr),
\end{eqnarray}
where
\begin{equation}
\label{A2.3a}
    k_{f(s)}=\sqrt{\varepsilon_{f(s)}(\omega)}\,{\omega\over c}\,, 
\end{equation}
and ${\bf M}$ and ${\bf N}$ represent TM- and TE-waves, respectively,
\begin{eqnarray} 
\label{A2.4}
\lefteqn{\hspace{-7ex}
       {\bf M}_{{e \atop o}nm}({\bf r},k) 
%}
%\nonumber\\[.5ex]&& \hspace{-2ex}
       \!=\! \mp {m \over \sin\theta} j_n(kr)
       P_n^m(\cos\theta) {\sin\choose\cos} (m\phi) {\bf e}_{\theta}
}
\nonumber\\&& \hspace{-2ex}
        - j_n(kr) \frac{dP_n^m(\cos\theta)}{d\theta} 
       {\cos\choose\sin} (m\phi) {\bf e}_{\phi}\,,
\end{eqnarray}
\begin{eqnarray}
\label{A2.5}
\lefteqn{\hspace{-4ex}
       {\bf N}_{{e \atop o}nm}({\bf r},k)
%}        
%\nonumber\\&& \hspace{-2ex}
       \!=\! {n(n\!+\!1)\over kr} j_n(kr)
        P_n^m(\cos\theta) {\cos\choose\sin} (m\phi) {\bf e}_r
}
\nonumber\\&&\hspace{-2ex}
        + {1\over kr} \frac{d[rj_n(kr)]}{dr} 
       \Biggl[
       \frac{dP_n^m(\cos\theta)}{d\theta} 
       {\cos\choose\sin} (m\phi) {\bf e}_{\theta}
\nonumber \\&&\hspace{-2ex}
       \mp {m \over \sin\theta}
       P_n^m(\cos\theta) {\sin\choose\cos} (m\phi) {\bf e}_{\phi}
       \Biggr], 
\end{eqnarray}
with $j_n(x)$ being the spherical Bessel function of the first kind
and $P_n^m(x)$ being the associated Legendre function. 
The superscript ${(1)}$ in Eq.~(\ref{A2.3}) indicates that 
in Eqs.~(\ref{A2.4}) and (\ref{A2.5}) the spherical Bessel function 
$j_n(x)$ has to be replaced by the first-type
spherical Hankel function $h^{(1)}_n(x)$.

The coefficients ${\cal A}_l^{M,N}$, ${\cal B}_l^{M,N}$, 
${\cal C}_l^{M,N}$, and ${\cal D}_l^{M,N}$ 
are to be found from the coupled recurrence equations
\begin{eqnarray}
\label{A2.6}
\lefteqn{ 
      \left(
       \begin{array}{cc}
       {\cal A}_{(f+1)s}^{M,N}+\delta_{(f+1)s} & {\cal B}_{(f+1)s}^{M,N}\\
       {\cal C}_{(f+1)s}^{M,N}                      & {\cal D}_{(f+1)s}^{M,N}
       \end{array}
       \right) 
}\nonumber\\[.5ex]&&
       = \left(
       \begin{array}{cc}
      1/T_{Ff}^{M,N}                  & R_{Ff}^{M,N} /T_{Ff}^{M,N} \\
      R_{Pf}^{M,N} /T_{Pf}^{M,N} & 1/T_{Pf}^{M,N} 
      \end{array}
      \right)
\nonumber\\[.5ex]&&\hspace{2ex}
      \times
      \left(
      \begin{array}{cc}
      {\cal A}_{fs}^{M,N} & {\cal B}_{fs}^{M,N} \\
      {\cal C}_{fs}^{M,N} & {\cal D}_{fs}^{M,N}+\delta_{fs} 
      \end{array}
      \right) ,
\end{eqnarray}
\begin{equation}
\label{A2.6a}
      {\cal A}_{{\cal N}s}^{M,N} ={\cal B}_{{\cal N}s}^{M,N} 
      ={\cal C}_{1s}^{M,N} ={\cal D}_{1s}^{M,N} = 0   
\end{equation}
[with $f$ and $s$ being taken according to Eq.~(\ref{A2.3})].
Here, the coefficients are redefined as 
\begin{eqnarray}
\label{A2.6b}
&&      {\cal A}_l^{M,N} \equiv {\cal A}_{fs}^{M,N},  \hspace{1ex}
      {\cal B}_l^{M,N} \equiv {\cal B}_{fs}^{M,N},  
\nonumber\\
&&      {\cal C}_l^{M,N} \equiv {\cal C}_{fs}^{M,N},  \hspace{1ex}
      {\cal D}_l^{M,N} \equiv {\cal D}_{fs}^{M,N},
\qquad      
\end{eqnarray}
and
\begin{eqnarray}
\label{A2.7}
\lefteqn{\hspace{-6ex}
       R^M_{Pf} = 
       \frac 
       { k_{f+1} H'_{(f+1)f}H_{ff} \!-\! k_f H'_{ff}H_{(f+1)f} }
       { k_{f+1} J_{ff}H'_{(f+1)f} \!-\! k_f J'_{ff}H_{(f+1)f} }\,,
}
\\[.5ex] && \hspace{-10ex}
\label{A2.8}
       R^M_{Ff} = 
       \frac 
       { k_{f+1} J'_{(f+1)f} J_{ff} \!-\! k_f J'_{ff}J_{(f+1)f} }
       { k_{f+1} J'_{(f+1)f} H_{ff} \!-\! k_f J_{(f+1)f} H'_{ff} }\,,
\\[.5ex] &&\hspace{-10ex}
\label{A2.9}
       R^N_{Pf} = 
       \frac 
       { k_{f+1} H_{(f+1)f} H'_{ff} \!-\! k_f H_{ff}H'_{(f+1)f} }
       { k_{f+1} J'_{ff} H_{(f+1)f} \!-\! k_f J_{ff}H'_{(f+1)f} }\,,
\\[.5ex] && \hspace{-10ex}
\label{A2.10}
       R^N_{Ff} = 
       \frac 
       { k_{f+1} J_{(f+1)f} J'_{ff} \!-\! k_f J_{ff}J'_{(f+1)f} }
       { k_{f+1} J_{(f+1)f} H'_{ff} \!-\! k_f J'_{(f+1)f}H_{ff} }\,,
\end{eqnarray}
\begin{eqnarray}
\lefteqn{\hspace{-3ex}
\label{A2.11}
       T^M_{Pf} =
       \frac 
       { k_{f+1} [ J_{(f+1)f}H'_{(f+1)f} 
       \!-\! J'_{(f+1)f}H_{(f+1)f} ]}
       { k_{f+1} J_{ff}H'_{(f+1)f} \!-\! k_f J'_{ff}H_{(f+1)f} }\,,
}
\nonumber\\[.5ex]&&       
\\&& \hspace{-7ex}
\label{A2.12}
       T^M_{Ff} = 
       \frac 
       { k_{f+1}[J'_{(f+1)f} H_{(f+1)f} 
       \!-\! J_{(f+1)f} H'_{(f+1)f} ] }   
       { k_{f+1} J'_{(f+1)f} H_{ff} \!-\! k_f J_{(f+1)f} H'_{ff} }\,,
\nonumber\\[.5ex]&&
\\&& \hspace{-7ex}
\label{A2.13}
       T^N_{Pf} = 
       \frac 
       { k_{f+1} [J'_{(f+1)f} H_{(f+1)f} 
       \!-\! J_{(f+1)f}H'_{(f+1)f} ] }
       { k_{f+1} J'_{ff} H_{(f+1)f} \!-\! k_f J_{ff}H'_{(f+1)f} }\,,
\nonumber\\[.5ex]&&
\\&& \hspace{-7ex}
\label{A2.14}
       T^N_{Ff} = 
       \frac 
       { k_{f+1}[ J_{(f+1)f} H'_{(f+1)f} 
       \!-\! J'_{(f+1)f}H_{(f+1)f} ] }
       { k_{f+1} J_{(f+1)f} H'_{ff} \!-\! k_f J'_{(f+1)f}H_{ff} }\,,
\nonumber\\[.5ex]&&       
\end{eqnarray}
with
\begin{eqnarray}
\label{A2.15}
       J_{il} \hspace{-1ex}&=&\hspace{-1ex} j_n(k_iR_l) ,
\\[.5ex]
\label{A2.16}
       H_{il} \hspace{-1ex}&=&\hspace{-1ex} h^{(1)}_n(k_iR_l) ,
\\[.5ex]
\label{A2.17}
       J'_{il} \hspace{-1ex}&=&\hspace{-1ex} {1\over \rho} 
       { d[\rho j_n(\rho)] \over d\rho } 
       \bigg|_{\rho=k_iR_l} ,
\\[.5ex]
\label{A2.18}
       H'_{il}\hspace{-1ex}&=&\hspace{-1ex} {1\over \rho} 
       { d[\rho h^{(1)}_n(\rho)] \over d\rho } 
       \bigg|_{\rho=k_iR_l}.
\end{eqnarray}

The coefficients in the first and the last layers can be found 
immediately from Eqs. (\ref{A2.6}) and (\ref{A2.6a}). The 
rest can be obtained by again using recurrence equations 
(\ref{A2.6}).

%%%%%%%%%%%%%%%%%%%%%%%%%%%%%%%%%%%%%%%%%%%%%%%%%%%%%%

\subsubsection{Two-layered medium} 
\label{app:2l}

For a sphere of
radius $R$ (including the special cases of an empty sphere 
in an otherwise homogeneous medium and a material sphere in vacuum)
we have [\citet{Li94}]
\begin{eqnarray}
\label{A2.19}
\lefteqn{
       \mbb{G}^{(11)}({\bf r},{\bf r'},\omega)
}
\nonumber\\[.5ex]&&\hspace{-2ex}
       = {ik_1\over 4\pi} \sum_{e\atop o} 
       \sum_{l=1}^\infty \sum_{m=0}^l 
       {2l+1\over l(l+1)} {(l-m)!\over(l+m)!}\,(2\!-\!\delta_{0m})
\nonumber\\[.5ex]&&\hspace{-2ex}\times 
       \Bigl[ {\cal B}^M_l (\omega) 
       {\bf M}^{(1)}_{{e \atop o}lm} ({\bf r},k_1) \otimes
       {\bf M}^{(1)}_{{e \atop o}lm} ({\bf r}',k_1)
\nonumber\\[.5ex]&&\hspace{-2ex}
       +\; {\cal B}^N_l (\omega)
       {\bf N}^{(1)}_{{e \atop o}lm} ({\bf r},k_1) \otimes
       {\bf N}^{(1)}_{{e \atop o}lm} ({\bf r}',k_1) \Bigr]  
\label{A2.20}
\end{eqnarray}
\mbox{($r,r'$ $\!>$ $\!R$)},
\begin{eqnarray}
\lefteqn{
       \mbb{G}^{(22)}({\bf r},{\bf r'},\omega)
}
\nonumber\\[.5ex]&&\hspace{-2ex}
       ={ik_2\over 4\pi} \sum_{e\atop o} 
       \sum_{l=1}^\infty \sum_{m=0}^l 
       {2l+1\over l(l+1)} {(l-m)!\over(l+m)!}\,(2\!-\!\delta_{0m})
\nonumber\\[.5ex]&&\hspace{-2ex}\times 
       \Bigl[ {\cal C}^M_l (\omega)
       {\bf M}_{{e \atop o}lm} ({\bf r},k_2) \otimes
       {\bf M}_{{e \atop o}lm} ({\bf r}',k_2)
\nonumber\\[.5ex]&&\hspace{-2ex}
       +\; {\cal C}^N_l (\omega)
       {\bf N}_{{e \atop o}lm} ({\bf r},k_2) \otimes
       {\bf N}_{{e \atop o}lm} ({\bf r}',k_2) \Bigr]   
\end{eqnarray}
\mbox{($r,r'$ $\!<$ $\!R$)}, where
\begin{eqnarray}
\label{eA.21}
       {\cal B}^{M,N}_l(\omega)
       \hspace{-1ex}&=&\hspace{-1ex} - R^{M,N}_{F1} ,
\\[.5ex]
\label{eA.22}
       {\cal C}^{M,N}_l(\omega)
       \hspace{-1ex}&=&\hspace{-1ex} - R^{M,N}_{F1}
       \frac{T^{M,N}_{F1} R^{M,N}_{P1}}{T^{M,N}_{P1}} \,.
\qquad       
\end{eqnarray}

%%%%%%%%%%%%%%%%%%%%%%%%%%%%%%%%%%%%%%%%%%%%%%%%%%%%%%

\subsubsection{Three-layered medium}
\label{app:3l}

%WWW%
For three-layered media of radii $R_1$ and $R_2$ ($R_1>R_2$), 
the spherical cavity presented in Fig. \ref{fig:slabho} 
in particular, we have that [\citet{Li94}]
\begin{eqnarray}
\label{A.23}
\lefteqn{\hspace{-4ex}
       \mbb{G}^{(13)}({\bf r},{\bf r'},\omega)
       = {ik_3\over 4\pi} \sum_{e\atop o} 
       \sum_{n=1}^\infty \sum_{l=0}^n 
}
\nonumber\\[.5ex] 
       &&\hspace{-7ex}\times
       \Biggl\{ \frac{2n\!+\!1}{n(n\!+\!1)} 
       \frac{(n\!-\!l)!}{(n\!+\!l)!}  
       (2\!-\!\delta_{0l})
\nonumber \\[.5ex] 
       &&\hspace{-7ex}\times 
       \biggl[ {\cal A}^M_l (\omega)
       {\bf M}^{(1)}_{{e \atop o}nl} ({\bf r},k_1) \otimes
       {\bf M}_{{e \atop o}nl} ({\bf r}',k_3)
\nonumber \\[.5ex] 
       &&\hspace{-7ex}
       +\,{\cal A}^N_l (\omega)
       {\bf N}^{(1)}_{{e \atop o}nl} ({\bf r},k_1) \otimes
       {\bf N}_{{e \atop o}nl} ({\bf r}',k_3) \biggr]\Biggr\}
\quad       
\end{eqnarray}
($r<R_2,\ r'>R_1$),
\begin{eqnarray}
\label{A.24}
\lefteqn{\hspace{-4ex}
       \mbb{G}^{(33)}({\bf r},{\bf r'},\omega)
       = {ik_3\over 4\pi} \sum_{e\atop o} 
       \sum_{n=1}^\infty \sum_{l=0}^n 
}
\nonumber\\[.5ex] 
       &&\hspace{-7ex}\times
       \Biggl\{ \frac{2n\!+\!1}{n(n\!+\!1)} 
       \frac{(n\!-\!l)!}{(n\!+\!l)!}  
       (2\!-\!\delta_{0l})
\nonumber \\[.5ex] 
       &&\hspace{-7ex}\times 
       \biggl[ {\cal C}^M_l (\omega)
       {\bf M}_{{e \atop o}nl} ({\bf r},k_1) \otimes
       {\bf M}_{{e \atop o}nl} ({\bf r}',k_3)
\nonumber \\[.5ex] 
       &&\hspace{-7ex}
       +\,{\cal C}^N_l (\omega)
       {\bf N}_{{e \atop o}nl} ({\bf r},k_1) \otimes
       {\bf N}_{{e \atop o}nl} ({\bf r}',k_3) \biggr]\Biggr\}
\quad        
\end{eqnarray}
($r,\ r'>R_1$), where
\begin{eqnarray}
\label{A.25}
\lefteqn{\hspace{-4ex}
       {\cal A}^{M,N}_l (\omega)
       \!=\! \frac { T_{F1}^{M,N} T_{F2}^{M,N} T_{P1}^{M,N} }
       { T_{P1}^{M,N} \!+\! T_{F1}^{M,N} R_{P1}^{M,N} R_{F2}^{M,N} }\,, 
}
\\[.5ex] &&\hspace{-7ex}
\label{A.26}
       {\cal C}^{M,N}_l (\omega)
        \!=\! { {\cal A}^{M,N}_l \over T_{P2}^{M,N}}
       \left[ {R_{P2}^{M,N} \over T_{F1}^{M,N}}  
       +{R_{P1}^{M,N} \over T_{P1}^{M,N}}\right].
\quad        
\end{eqnarray}

\end{appendix}

%%%%%%%%%%%%%%%%%%%%%%%%%%%%%%%%%%%%%%%%%%%%%%%%%%%%%%%%%%%%%%%%%%%%%%%%%
%%%%%%%%%%%%%%%%%%%%%%%%%%%%%%%%%%%%%%%%%%%%%%%%%%%%%%%%%%%%%%%%%%%%%%%%%

\end{document}